\documentclass[usenatbib]{mn2e}
\usepackage{graphicx}
\usepackage{natbib}
\usepackage{amsmath}
\usepackage{color}
\usepackage{dsfont}
\def\apjl{{ApJ}}                % Astrophysical Journal, Letters
\def\apjs{{ApJS}}               % Astrophysical Journal, Supplement
\def\ao{{Appl.~Opt.}}           % Applied Optics
\def\aap{{A\&A}}                % Astronomy and Astrophysics
\begin{document}

\title[ISW Ray-Tracing]{Full-sky map of the ISW and Rees-Sciama effect from Gpc simulations}
\author[Cai et al.]{Yan-Chuan Cai$^{1,2}$, Shaun Cole$^{1}$, Adrian Jenkins$^{1}$, Carlos
  S. Frenk$^{1}$ \\
$^{1}$Institute for
Computational Cosmology, Durham University, South Road, Durham, DH1
3LE, UK\\
$^{2}$Dept. of Physics and Astronomy, University of Pennsylvania, Philadelphia, PA 19104}
\maketitle
\begin{abstract}
We present a new method for constructing maps of the secondary
temperature fluctuations imprinted on the cosmic microwave background
(CMB) radiation by photons propagating through the evolving cosmic
gravitational potential. Large cosmological N-body simulations are
used to calculate the complete non-linear evolution of the peculiar
gravitational potential.  Tracing light rays back through the past
lightcone of a chosen observer accurately captures the temperature
perturbations generated by linear (the integrated Sachs-Wolfe or ISW
effect) and non-linear (the Rees-Sciama or RS effect) evolution.
These effects give rise to three kinds of non-linear features in the
temperature maps. (a) In overdense regions, converging flows of matter
induce cold spots of order 100~Mpc in extent which can dominate over
the ISW effect at high redshift, and are surrounded by hot rings. (b)
In underdense regions, the RS effect enhances ISW cold spots which can
be surrounded by weak hot rings. (c) Transverse motions of large lumps
of matter produce characteristic dipole features, consisting of
adjacent hot and cold spots separated by a few tens of Megaparsecs.
These non-linear features are not easily detectable; they modulate the
ISW sky maps at about the 10~per~cent level. The RS effect causes the
angular power spectrum to deviate from linear theory at $l \sim 50$
and generates non-Gaussianity, skewing the one-point distribution
function to negative values. Cold spots of similar angular size, but
much smaller amplitude than the CMB cold spot reported by Cruz et
al. are produced. Joint analysis of our maps and the corresponding
galaxy distribution may enable techniques to be developed to detect
these non-linear, non-Gaussian features. Our maps are available at
http://astro.dur.ac.uk/$\sim$cai/ISW.
\end{abstract}

\section{Introduction}
The integrated Sachs-Wolfe (ISW) effect \citep{Sachs67} arises from
the decay of the large-scale potential fluctuations as they are
traversed by cosmic microwave background (CMB) photons and induces
secondary temperature perturbations in the CMB radiation.  It occurs
in both open cosmological models ($\Omega_{\rm{m}}<1$) and models containing a
cosmological constant ($\Omega_\Lambda>0$) or dark energy.  It
provides an independent way of measuring the dynamical effect
of dark energy, but it is challenging to detect: at large scales,
the ISW signal suffers from cosmic variance, as there are too few
independent modes, while at somewhat smaller scales, it is entangled
with the Rees-Sciama (RS) effect \citep{Rees68}, which arises from the
non-linear evolution of gravitational potential perturbations. Since
it is impossible to evade the cosmic variance at large scales, the
information at small scales becomes valuable. To extract
cosmological information from such scales  requires disentangling
the ISW and RS effects.

Studies using N-body simulations have found that the combined ISW plus
RS power spectra of CMB temperature fluctuations
deviate from linear theory expectation at very large
scales, $k \sim 0.1$ $h$~Mpc$^{-1}$ at $z=0$, with the scale of this
transition becoming even larger at higher redshift.  This discovery
that the non-linear effect was more important at higher redshift was
made by \citet{Cai09}
and later confirmed by \citet{Smith09}.
Recent studies using both perturbation theory and
fitting formulae indicate that the combined ISW and RS effect makes a
non-trivial contribution to the overall CMB non-Gaussianity as
measured by the bispectrum \citep[e.g.][]{Verde02,Giovi03,
Boubekeur09, Mangilli09}. The primordial-lensing-RS correlation
contribution may yield an effective non-Gaussianity parameter, $f_{\rm
NL}$, of 10 \citep{Mangilli09}. All these results suggest that the RS
effect is a very important supplement to the ISW effect even at very
large scales.

Observational evidence for a possible contribution from the RS effect
at large scales has come from a combined study of the SDSS LRG samples
and the CMB \citep{Granett09}. These authors report a $4$-$\sigma$
detection of the ISW and RS signal at the scale of 4~degrees, somewhat
higher than expected from the standard flat $\Lambda$CDM universe.
There have also been attempts to attribute an extreme cold spot in the
CMB to the
ISW and RS effects \citep[e.g.][]
{Martinez-Gonzalez90b,Martinez-Gonzalez90a,
Rudnick07,Inoue06,Inoue07,Tomita08, Masina09b, Masina09a}.  However, on
theoretical grounds such an explanation seems unlikely because the
estimated sizes of the non-linear structures responsible for the cold
spot are typically $>100$~Mpc, which seems too large to occur in a
$\Lambda$CDM universe which assumes Gaussian initial conditions \citep[see
also][for discussion of non-Gaussianity.] {Cruz05,McEwen05,
Cruz06,McEwen06, Cruz07,McEwen08}. However, it is still unclear whether
or not the combined ISW and RS effects can generate cold spots of a
few degrees with the right amplitudes and whether such large-scale
non-Gaussianity can arise from large-scale structure.  A
full understanding of the ISW and Rees-Sciama effect could be crucial in
explaining the oddities of these observations.  Meanwhile, the
increase in sensitivity of forthcoming CMB experiments opens
possibilities of exploiting CMB temperature fluctuations down to
arcmin scales (Planck\footnote{www.sciops.esa.int/PLANCK/},
ACT\footnote{http://www.physics.princeton.edu/act/},
SPT\footnote{http://pole.uchicago.edu/} and
APEX-SZ\footnote{http://bolo.berkeley.edu/apexsz}), at which the ISW
and RS effect may also entangle with other large-scale astrophysics of
interest, i.e. lensing \citep[e.g.][]{Verde02, Nishizawa08,Mangilli09}
and the Sunyaev-Zel'dovich (SZ) effect \citep{Sunyaev72}
\citep[e.g.][]{Cooray02a, Fosalba03,Bielby09}. To disentangle all
these effects is the key to thoroughly exploiting the information
encoded in these upcoming CMB measurements.

N-body simulations are the ideal tool for investigating the phenomena
discussed above since they treat the non-linear regime accurately and
permit the construction of a full sky map of the ISW and RS effects
and the full underlying 3-dimensional light-cone.
Maps of the ISW effect have been
constructed from both simulations and observations
\citep{Barreiro08,Granett09}. Most of them simply adopt the linear
approximation, using only the density field to estimate the time
derivative of the potential. We will show in this paper that these
linear maps are far from accurate.  There are also maps constructed
from ray-tracing through simulations \citep{Tuluie95,Puchades06}, but
these are limited by small simulation box sizes and  are
not adequate to explore very large-scale structures. A full sky map
of the RS effect has been constructed using a constrained
high-resolution hydrodynamical simulation to model the RS effect
in the very local universe \citep{Maturi07a}.  This map is useful for
understanding the RS effect from within the radial distance of $110$
Mpc, which is a very small volume. Maps from the ray-tracing of large
cosmological volumes are still missing.

In this paper, we develop a new method
of constructing a full sky light-cone of the time derivative of
the potential, $\dot{\Phi}$, using a large N-body simulation.
Our method of computing $\dot{\Phi}$ is fully non-linear and so
should model the complete RS effect as well as the ISW component.
Our Gpc box size simulation provides a sufficient number of
independent large scale modes to investigate the ISW effect fully.
We ray-trace through the light-cone to produce maps of temperature
fluctuations induced by the ISW and RS effects. Our maps
cover a large range of scales and cosmic time
with high accuracy and allow us to investigate
the ISW and RS effects. The maps will also be a valuable
source for understanding CMB secondary non-Gaussianity arising from
large-scale structure. They may also prove useful for
disentangling lensing and SZ effects from the ISW
and RS effects.

The paper is organized as follows. In \S2, we present the basic
physics and the mathematical description of the ISW and RS effects. In
\S3, we describe our method of computing CMB temperature perturbations
from our N-body simulation and ray-tracing to produce full sky maps
from light-cone data.  In \S4, we identify and discuss three characteristic
non-linear features of the temperature perturbations.  Full sky maps
are presented in \S5.  Finally, in \S6, we discuss our results and
draw conclusions.

\section{the ISW and Rees-Sciama effect}
\label{ISWRS}

In a $\Lambda$CDM universe, dominance of the cosmological
constant, $\Lambda$, causes the expansion factor of the universe, $a$,
to grow at a faster rate than the linear growth of density
perturbations, $\delta$. Consequently, the cosmological constant
has the direct dynamical effect of causing gravitational potential
perturbations, $\Phi \propto -\delta/a$, to decay. The ISW effect is
caused by the change in energy of CMB photons as they traverse these
linearly evolving potentials.  A CMB photon passing through an overdense
region, or cluster, will gain more energy falling into the potential
well than it later looses climbing out of the evolved shallower
potential well. Therefore, overdense regions correspond to hot regions in a
linear ISW map. The converse is true for a photon passing through an
underdense region. Here, the potential fluctuation is positive and the
CMB photon looses more energy climbing the potential hill than it
subsequently regains from its descent.
Therefore, underdense regions appear cold in a linear ISW map.
Non-linear growth of the density perturbations modifies this picture,
producing additional temperature perturbations -- the RS effect.
In overdense regions, the accelerated non-linear growth of structure
acts to increase the depth of the potential wells resulting in a
reduction in the CMB temperature, partially cancelling the ISW effect.
In contrast, in underdense regions the RS effect enhances the ISW
effect as saturation of the density contrast in voids further suppresses
growth of the gravitational perturbation. We will discuss other
situations in which the RS effect makes a significant contribution
and analyze the morphology of the resulting features in the sky maps in
\S~\ref{SkyMap}.

The net induced ISW plus RS, hereafter ISWRS, temperature fluctuation
along a direction $\hat n$ can be written as an integral of the time
derivative of the gravitational potential, $\dot{\Phi}$, from the last
scattering surface to the present \citep[i.e.][]{Sachs67, Martinez-Gonzalez90a},
\begin{equation}\label{eq1}
\Delta T (\hat n) = \frac{2}{c^2}\bar{T_0}\int_{t_{\rm L}}^{t_0}
\dot{\Phi}(t,\hat n) \, dt,
\end{equation}
where $t$ is cosmic time, $t_{\rm L}$ the age of the universe at the
last scattering surface, $t_0$ the present age,
$\dot{\Phi}$  the time derivative of
the gravitational potential, $\bar{T_0}$  the mean CMB temperature
and $c$  the speed of light. This is equivalent to the integral
over radial comoving distance, $r$,
\begin{equation}\label{eq1b}
\Delta T (\hat n)
=  \frac{2}{c^3}\bar{T_0}\int_{0}^{r_{\rm L}}
\dot{\Phi}(r,\hat n)\, a \, dr,
\end{equation}
where $r_{L}$ is the comoving distance to the last scattering surface
and $a$ the expansion factor.

To compute $\dot{\Phi}$ accurately from our simulation we make use of
the Poisson equation expressed in comoving coordinates,
$\bigtriangledown^2\Phi(\vec{x},t)=4\pi G
\bar{\rho}(t) a^2\delta(\vec{x}, t)$, which can be written in Fourier
space as
\begin{equation}\label{eq2}
\Phi(\vec{k},t)=-\frac{3}{2}\left(\frac{H_0}{k}\right)^2\Omega_{\rm
  m}\frac{\delta(\vec{k},t)}{a} .
\end{equation}
Here $\bar{\rho}(t)$ is the mean density of the universe, $\delta$
the density contrast $\delta \equiv (\rho-\bar{\rho})/\bar{\rho}$,
$H_0$ and $\Omega_{\rm m}$ the present values of the Hubble
and matter density parameters
and $G$ is the gravitational constant. Taking the time
derivative yields
\begin{equation}\label{eq3}
\dot{\Phi}(\vec{k},t)=\frac{3}{2}\left(\frac{H_0}{k}\right)^2\Omega_{\rm m}
\left[\frac{\dot{a}}{a^2}\delta(\vec{k},t)-\frac{\dot\delta(\vec{k},t)}{a}\right]
.
\end{equation}
Combining this with the Fourier space form of the continuity equation,
$\dot\delta(k,t)+i\vec k\cdot \vec p(\vec k,t)=0$, we have
\begin{equation}\label{eq4}
\dot{\Phi}(\vec{k},t)=\frac{3}{2}\left(\frac{H_0}{k}\right)^2\Omega_{\rm m}
\left[\frac{\dot{a}}{a^2}\delta(\vec{k},t)+\frac{i\vec{k}\cdot\vec{p}(\vec{k},t)}{a}\right],
\end{equation}
where $\vec p(\vec k,t)$ is the Fourier transform of the
momentum density divided by the mean mass
density, $\vec p(\vec x,t)=[1+\delta(\vec x,t)]\vec v(\vec x,t)$.
Equation~(\ref{eq4}) enables us to compute $\dot{\Phi}$ (including
the contributions of both the linear ISW effect and the non-linear RS
effects) to high accuracy using the density and momentum fields of our
simulation. We will refer to the results obtained
using equation~(\ref{eq4}) as the ISWRS.

We wish to contrast these ISWRS predictions with the corresponding
results from linear theory.  In the linear regime, $\dot\delta(\vec
k,t)=\dot D(t)\delta(\vec k,z=0)$, where $D(t)$ is the linear growth
factor. Substituting this into equation~(\ref{eq3}) yields
\begin{equation}\label{eq5}
\dot{\Phi}(\vec{k},t)=\frac{3}{2}\left(\frac{H_0}{k}\right)^2\Omega_{\rm m}
\frac{\dot{a}}{a^2}\delta(\vec k,t)[1-\beta(t)],
\end{equation}
where $\beta(t)$ denotes the linear growth rate
$\beta(t)\equiv{d \ln D(t)}/{d \ln a}$.
This equation represents the
conventional way of modelling the ISW effect and uses only the
information from the density field. It is equivalent to assuming that
the velocity field is related to the density field by the linear
approximation
$\vec{p}(\vec{k},t)=i{\dot\delta(k,t)\vec k}/{k^2}\approx
i\beta(t)\delta(k,t) (\dot a/a )\vec k/k^2$. We will refer to results obtained
using this linear approximation for the velocity field as
LAV.
Note that with this LAV approximation, $\dot \Phi$ simply scales
with time according to the ISW linear growth factor,
$G(t)=\dot{a}D(t)[1-\beta(t)]/a^2$.

In the LAV approximation the density field, $\delta$, directly
determines the potential field, $\Phi$,  and its derivative,
$\dot{\Phi}$.
Hence the morphology of the $\dot{\Phi}$ field is determined directly
by the form of the density field, $\delta$, with overdense regions
($\delta>0$) corresponding to positive regions of $\dot{\Phi}$
and underdense regions corresponding to negative
regions of $\dot{\Phi}$. In contrast, for the exact calculation, represented by equation~(\ref{eq4}),
this correspondence is broken and the dynamics of the density field
play an additional direct role in determining $\dot{\Phi}$.
This is fully discussed in \S~\ref{slices}.

\section{Constructing full sky maps}
\label{method}
In this section, we describe our method of constructing full sky maps
of the ISWRS effect using our large simulation.

\subsection{The Gpc simulation}
To investigate the ISWRS effect, we need a simulation
of sufficiently large volume
to include the very large scale
perturbation modes (hundreds of Megaparsecs) necessary to verify convergence
with linear theory.  At the same time, we need sufficiently high resolution to
investigate the effects of small-scale non-linearity on the structures
resolved in the sky maps.  Moreover, for high accuracy ray-tracing,
we need the redshift spacing of the simulation outputs to be small
enough that interpolation between neighbouring redshifts does not
introduce significant systematic errors.

The N-body simulation we employ follows  $2200^3$-particles in a
$1~h^{-1}$Gpc periodic box in a $\Lambda$CDM cosmological
model with $\Omega_\Lambda=0.74$, $\Omega_{\rm m}=0.26$, $\Omega_{\rm
b}=0.044$, $\sigma_8=0.8$ and $H_0=71.5$~km~s$^{-1}$~Mpc$^{-1}$,
chosen
for consistency with recent CMB and large scale structure data
\citep{Sanchez09}.  The simulation, run on the COSMA supercomputer at
Durham, was designed in order to make mock galaxy catalogues for
forthcoming surveys (e.g. Pan-STARRS1 and EUCLID) and, as such, resolves
the dark matter halos of luminous galaxies
(Baugh et al.  in prep).  The simulation has a softening length
(Plummer equivalent) of $0.023~h^{-1}$Mpc which provides more than
adequate resolution for our purposes.  The initial conditions were set
up at redshift $z=49$ and the simulation was run
using GADGET \citep{Springel05} with
output at 50 snapshots, 48 of which lay between $z=0$ and $z=10$,
where we mainly focus our analysis. The redshift intervals between
neighbouring simulation outputs correspond to about $100~h^{-1}$~Mpc
in radial comoving distance. We have verified that this is adequate to
ensure that the errors induced by interpolating $\dot \Phi$ between
snapshots are less than $2\%$ at Megaparsec scales and even less at larger
scales.

\subsection{Map construction}

Our method of constructing the ISWRS sky maps has two stages.
First, at each output redshift, we construct an estimate of
$\dot \Phi$ on a cubic grid. In the second stage, we propagate
light rays from the observer and, as we move along each ray, we
interpolate $\dot \Phi$ from the grids and accumulate the
$\Delta T$ defined by the integral of equation~(\ref{eq1b})
along the past light-cone of the observer. Finally, we use HEALPix
\citep{Gorski05} to visualize the $\Delta T$ map in spherical
coordinates. We now describe these steps in more detail.

\subsubsection{Cartesian Grids}

We construct the density field, $\delta(\vec x)$, by assigning the
mass of each dark matter particle to a 3D mesh of cubic grid cells
using the cloud-in-cell assignment scheme \citep{Hockney81}.  We apply
the same scheme to accumulate the momentum density field,
$\vec p(\vec x)$, by assigning the vector momentum of each particle to
a Cartesian grid.  Then we perform four independent Fast Fourier
Transforms to compute the Fourier space versions of the density
field, $\delta(\vec k)$, and the three components of the momentum field, $\vec
p(\vec k)$. These are then combined using equation~(\ref{eq4}) to yield the
$\dot{\Phi}$ field in Fourier space.  Finally, we perform an inverse
Fourier transform to obtain $\dot{\Phi}(\vec x)$ in real space on a
cubic grid.  We repeat the final two operations using equation~(\ref{eq5})
to obtain the alternative LAV $\dot{\Phi}$ field.  This whole process
is then repeated for each of the 50 simulation outputs.

For our $1~h^{-1}$~Gpc simulation, we have used a grid of $1000^3$
cells each of $1~h^{-1}$~Mpc on a side.  The resolution of the N-body
simulation is significantly greater and would warrant using a finer
grid. However, this would be very demanding of both machine memory and
hard disk space and is not necessary to resolve
structures over $10~h^{-1}$~Mpc accurately, which is the smallest scale of interest in this paper.

\subsubsection{Ray tracing through the light-cone}

The next step is to choose a location for the observer and propagate
rays through the simulation. We choose to place the observer at the
corner of the simulation box at Cartesian coordinate (0,0,0). We then
used HEALPix \citep{Gorski05} to generate the directions of
$3\,145\,728$ evenly distributed rays, corresponding to an angular
pixel scale of $(6.87')^2$ which is sufficient to resolve upto a
spherical harmonic scale of $l\sim 1000$.  For each ray, we accumulate the
integral given by equation~(\ref{eq1b}) by taking fixed discrete steps in
comoving radial distance. At the location of each step, we find the two
output snapshots that bracket the lookback time at this distance.
Using the Cartesian grids at each of these outputs we use the
cloud-in-cell assignment scheme \citep{Hockney81}
to obtain the values
of $\dot{\Phi}_1$ and $\dot{\Phi}_2$ at the chosen position on the ray
at these two lookback times. Note that if the position along the ray
lies outside the simulation box, we use the periodic
boundary conditions to map the location back into the box.
Finally, to estimate $\dot{\Phi}$ at the
lookback time corresponding to the position on the ray, we linearly
interpolate $\dot{\Phi}(t)/G(t)$ with comoving radial distance, $r$,
using
\begin{equation}
(r_2-r_1)\frac{\dot{\Phi}}{G}=(r_2-r)\frac{\dot{\Phi}_1}{G_1}+
(r-r_1)\frac{\dot{\Phi}_2}{G_2} .
\end{equation}
Here $G$, $G_1$ and $G_2$ are the linear growth factors for $\dot\Phi$
at the lookback time corresponding to the position on the light ray and
the two neighbouring outputs that bracket it.  The values of $r$,
$r_1$ and $r_2$ are the corresponding comoving radial
distances.  This interpolation scheme guarantees that we recover the
linear theory result exactly when $\dot{\Phi}$ is evolving according
to linear theory.

The maps that we present in Section~\ref{SkyMap} are constructed and analysed using
the HEALPix package \citep{Gorski05}. We show maps corresponding to
the contribution of the integral in equation~(\ref{eq1b}) over different
finite intervals of comoving radial distance. If we were to integrate over a
comoving distance larger than the $1~h^{-1}$Gpc size of our
simulation, the assumed periodic boundary conditions would
create artifacts in the maps. For instance, in directions
corresponding to the principal axes of the simulation volume the light ray
would pass through the same location  every $1~h^{-1}$Gpc.
The contributions to $\Delta T(\hat n)$ from these replicas would add
coherently and lead to larger fluctuations along these special
directions than on average.  In similar applications some authors
choose to rotate, flip and shift the replicated simulation cube.
While this avoids the artificial coherence between one replica and the next,
it generates other problems including discontinuities of $\dot \Phi$
at cube boundaries. We have chosen to evade this difficulty by mainly focusing
on generating and analyzing maps with a maximum radial depth equal to the
simulation cube size of $1~h^{-1}$Gpc. It may still remain the case
that the same structure is seen in more than one direction, but
this does not affect the mean power spectrum of the map and only produces
non-Gaussian features on angular scales approaching the angular size
subtended by the simulation cube. Hence, the power spectra and small
scale features of the maps are not affected by the
periodic nature of the simulation.\footnote{In fact, we find that if we ignore
this problem and produce power spectra directly from maps projected
over a depth of $7000~h^{-1}$Mpc or more, they do not differ significantly from
those built up from the power spectra of successive $1000~h^{-1}$Mpc
slices. Hence, at least for angular power spectra, our simulation size
is sufficiently large and the evolution of $\dot \Phi$ sufficiently rapid
that the superposition of periodic replicas
is not a concern.}

It is also important to establish that the resolution of the maps
is not compromised by the spatial and temporal resolution of
the simulation or our choices of size of grid cell and integration
step. The pixel size of our HEALpix maps is $6.87'$ and matches the
linear size of our Cartesian grid cells at an angular diameter
distance of $500~h^{-1}$~Mpc. Thus, we would expect that beyond
$500~h^{-1}$~Mpc our sky maps are not lacking any small scale features
due to the limited resolution of our cubical grids. We have tested
this using a smaller simulation of the same resolution but grided
using $0.5$ rather than $1~h^{-1}$~Mpc cells. This test indicates that
at $l=100$, the larger, default, choice of cell size suppresses the
power in the maps made from the innermost $250~h^{-1}$Mpc by just 10\%.
As we ray-trace further in the radial direction, the accuracy improves,
and is within $10\%$ for $l\sim1000$ for maps extending to 1000
$h^{-1}$Mpc.
Additional tests have shown that it is this cell size
which is the limiting factor in the resolution of our maps.
For instance, we have checked that the maps are essentially unchanged
if we make the integration step size smaller than our default choice
or if we have more closely spaced simulation outputs.
The full resolution of the N-body simulation could be exploited by
using a finer mesh but this is unnecessary for our purposes.

\section{The Effects of Non-linearity}
\label{slices}

% Full area slices
\begin{figure*}
\begin{center}
\resizebox{\hsize}{!}{
\includegraphics{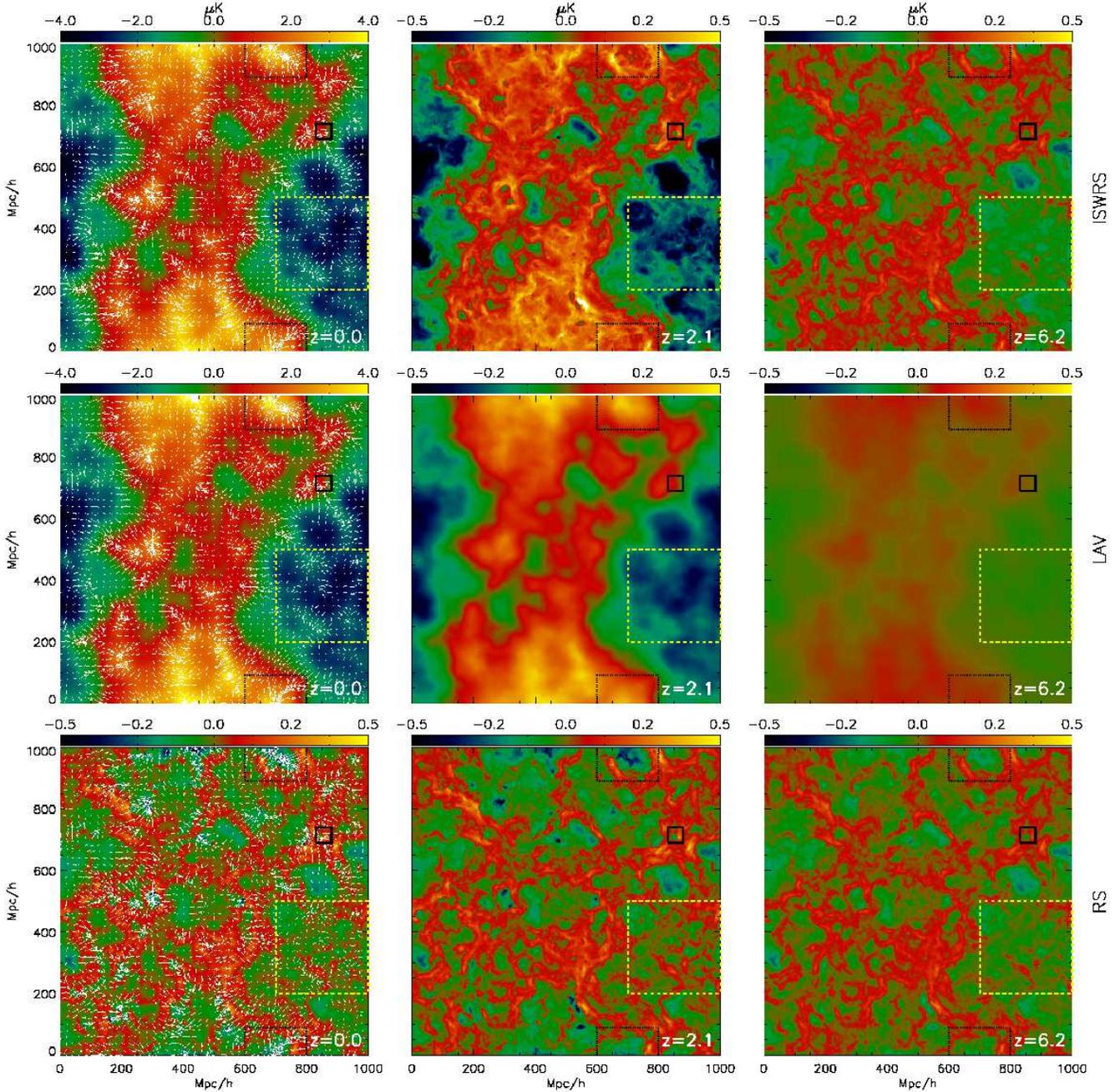}}
\caption{\label{FullBox100} Maps of $\Delta T$ generated from a slab of
thickness $\Delta r=100 h^{-1}$~Mpc taken from our Gpc
simulation. From left to right, they are maps at $z=$ 0.0,
2.1, 6.2 respectively.  From top to bottom, they are maps
constructed using equation~(\ref{eq4}), which includes the ISW and
Rees-Sciama effects (ISWRS), maps constructed using the linear
approximation for the velocity field, equation~(\ref{eq5}), (LAV), and
residual maps of the top panels minus the middle panels, leaving
essentially the Rees-Sciama (RS) contribution to the temperature
fluctuations.  Note the individual temperature scales for each panel.
At $z=0$, we also show the momentum field, averaged over the same
slice of the simulation, by the overplotted arrows.  The three square
boxes indicate the three regions we show in detail in
Fig.~\ref{Dipole} (solid-black box), Fig.~\ref{ColdInHot}
(dotted-black box) and Fig.~\ref{ColdInVoid} (dashed-yellow box).}
\end{center}
\end{figure*}

% Radial tracks
\begin{figure}
\resizebox{\hsize}{!}{
\includegraphics{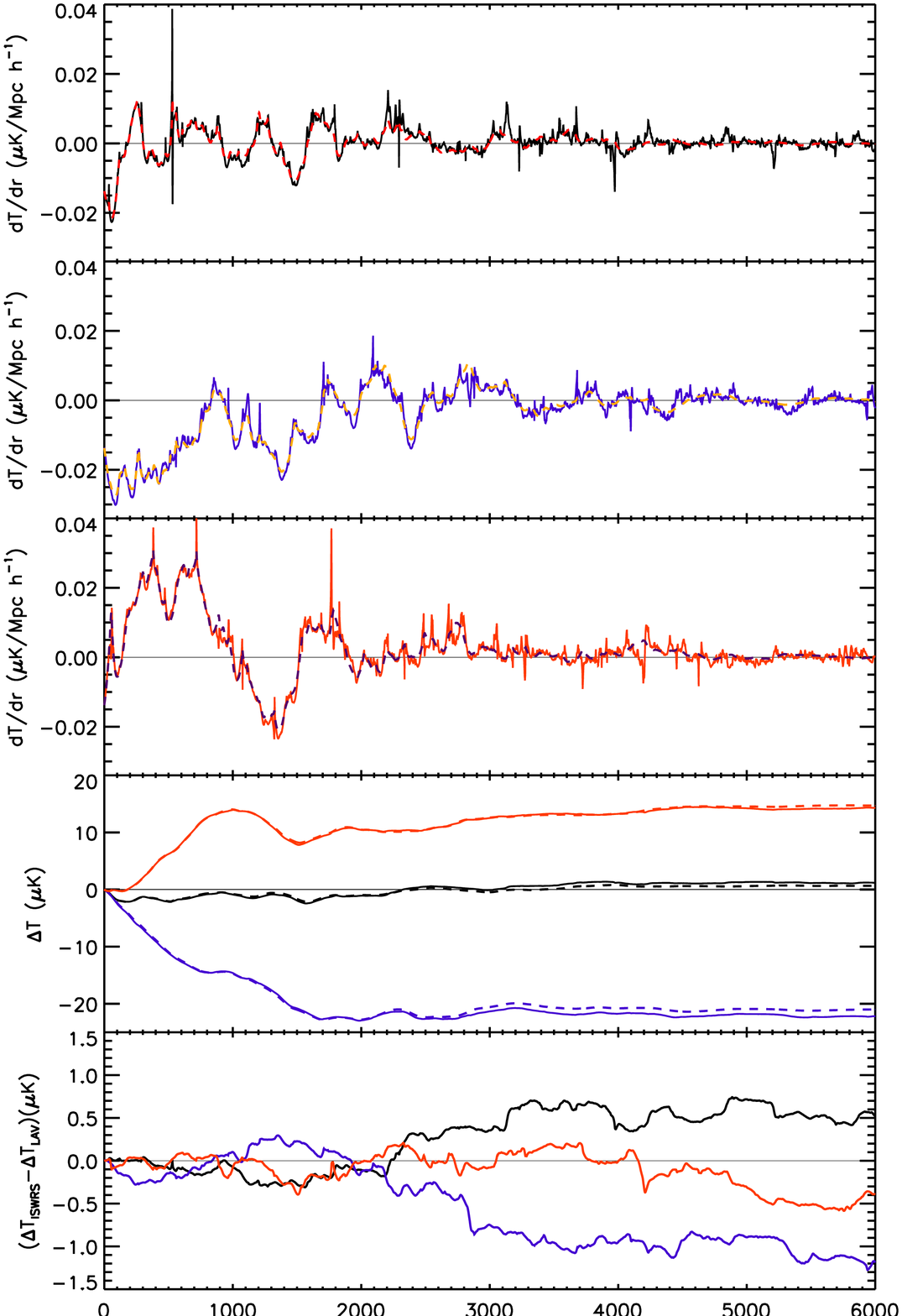}}
\\
\\
\caption{\label{RayTraceAll} Temperature perturbations along three light
rays, shown in black, blue and red from $z=0$ to $z=5.8$,
corresponding to the comoving distance from $r_c=0$ to
$r_c=6000~h^{-1}$~Mpc. The top three panels show the temperature
perturbations per unit comoving distance along each  radial direction.
Solid lines show the results of the full ISW and Rees-Sciama effects
(ISWRS) and dashed lines the results of the linear approximation for
the velocity field (LAV). The fourth panel from the top shows the
accumulated temperature perturbations along the radial direction for
ISWRS (solid lines) and LAV (dash lines). The bottom panel shows
differences between these accumulated  ISWRS and LAV temperature
perturbations.}
\end{figure}

In this section, we compare our full non-linear estimates of the ISWRS
effect with those of the LAV approximation. We both quantify and
characterise the features that are generated by non-linear
gravitational evolution and elucidate the physical processes by which
they are generated.

Fig.~\ref{FullBox100} shows the temperature maps at three different
redshifts that result from computing the contribution to the ISWRS
integral equation~(\ref{eq1b}) from a $100 h^{-1}$~Mpc thick slice of the Gpc
simulation.  The top row shows the full ISWRS calculation in which $\dot
\Phi$ is computed using equation~(\ref{eq4}), while the middle row shows
the result of using the LAV approximation for $\dot \Phi$ given by
equation~(\ref{eq5}). The LAV differs slightly from the standard ISW contribution to the
temperature fluctuations because to compute the ISW one should really use
the linear theory prediction for the density field in
equation~(\ref{eq5}), whereas for the LAV we use the actual non-linear
density field, but assume the velocity field is related to the density
as in linear theory.  However, the difference between the LAV and the
true ISW contribution is extremely small, as we shall see when we
compare their power spectra in Section~\ref{SkyMap}.  Consequently, the bottom
row of panels, which are the residual produced by subtracting the LAV
maps from the corresponding ISWRS maps, are essentially the RS
contribution to the temperature fluctuations.

Let us consider first  the LAV maps. It is clear that the hot (red/yellow) and
cold (blue/green) regions have a large coherence scale, approaching the
size of the simulation box. The coherence scale is much larger than
that of the underlying density field simply because of the $k^{-2}$
factor in equation~(\ref{eq3}), which arises from the Poisson equation
relating $\Phi$ and $\delta$. As explained at the beginning of
Section~\ref{ISWRS}, hot and cold spots in these maps are induced by the
decay of the gravitational potential in overdense and underdense
regions
respectively
caused by the dynamical effect of the cosmological constant,
$\Omega_\Lambda$.  In the higher redshift slices, the
amplitude of the LAV/ISW fluctuations are greatly reduced. This is
easily understood: as redshift increases $\Omega_\Lambda$ becomes
smaller and linear theory predicts density perturbations grow as
$\delta \propto a$ and so the corresponding potential perturbations, $\Phi
\propto -\delta/a$, are constant. For $\dot \Phi=0$ there are no ISW
fluctuations as CMB photons traversing such a potential well will
loose just as much energy leaving the perturbation as they gain on
entering it.

The RS contribution to the temperature fluctuations, shown in the
bottom row of Fig.~\ref{FullBox100}, have a much shorter coherence
scale than the ISW fluctuations. Their evolution with redshift is much
more gradual than for the ISW fluctuations. In the redshift $z=0$
slice they are a minor contribution to the overall ISWRS map (top
row), but they become a significant contribution at redshift $z=2.1$,
generating small scale structure within the smooth large-scale ISW
fluctuations. At $z=6.2$, the RS fluctuations are almost completely
dominant and while they are on a smaller scale than the ISW
fluctuations, they still produce filamentary structures which can be
seen to be one hundred to a few hundreds of Megaparsecs across.
These panels give a visual confirmation of the two-point statistics
presented in \citet{Cai09}, which demonstrated that both the importance and
physical scale of the RS effect become greater with increasing
redshift.

% Zoom in on a dipole
\begin{figure*}
\begin{center}
\resizebox{\hsize}{!}{
\includegraphics{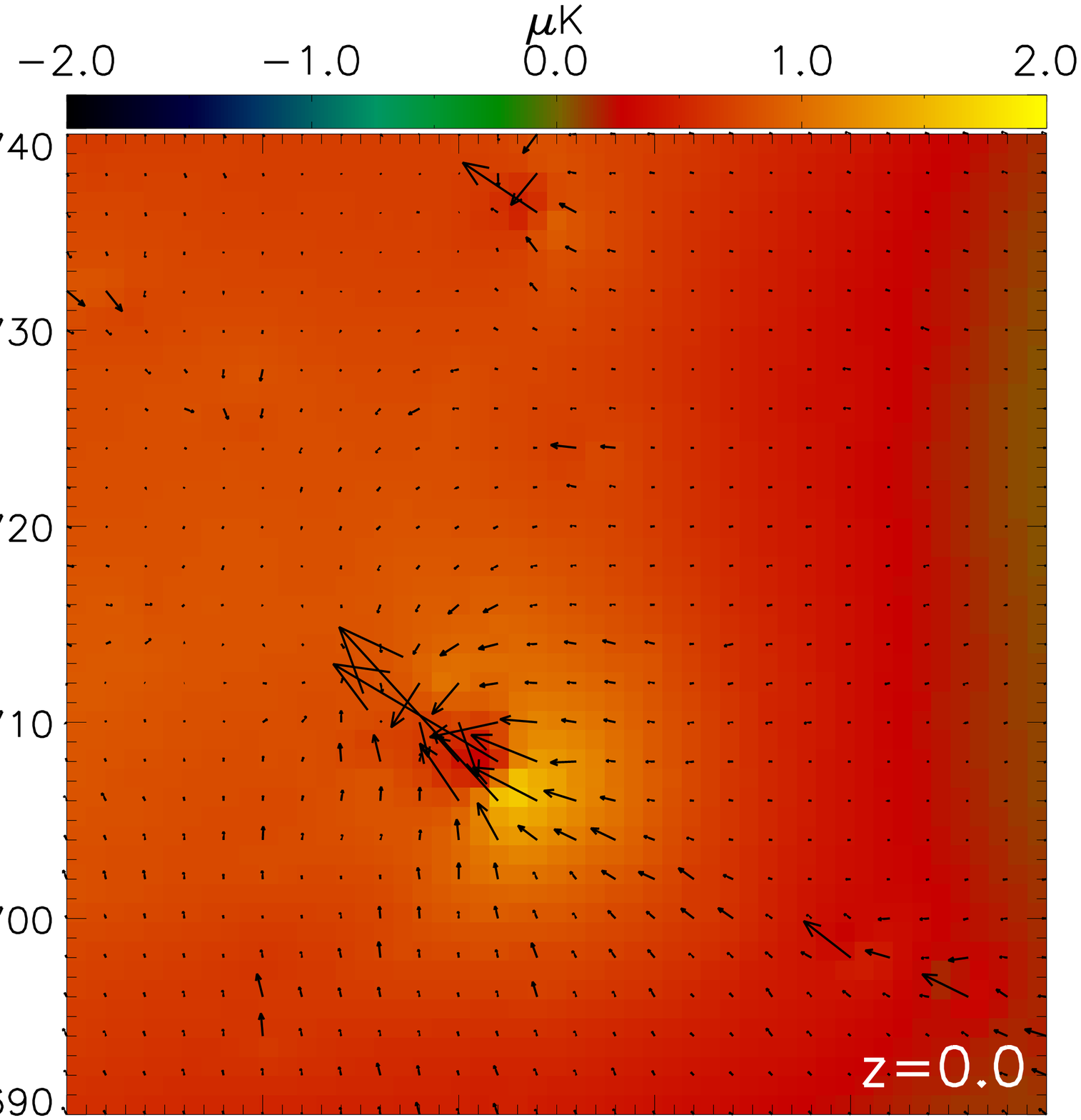}
\includegraphics{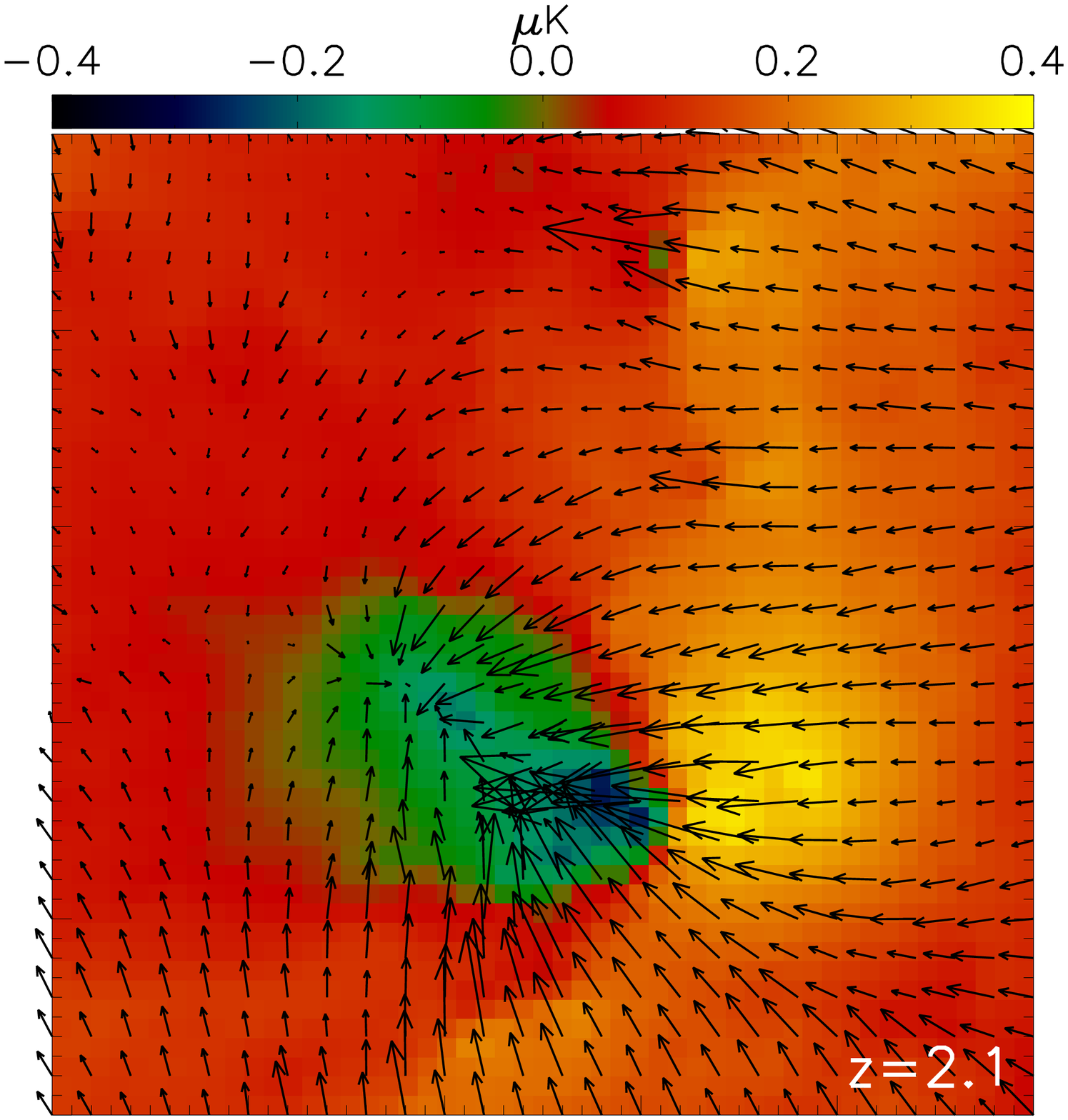}
\includegraphics{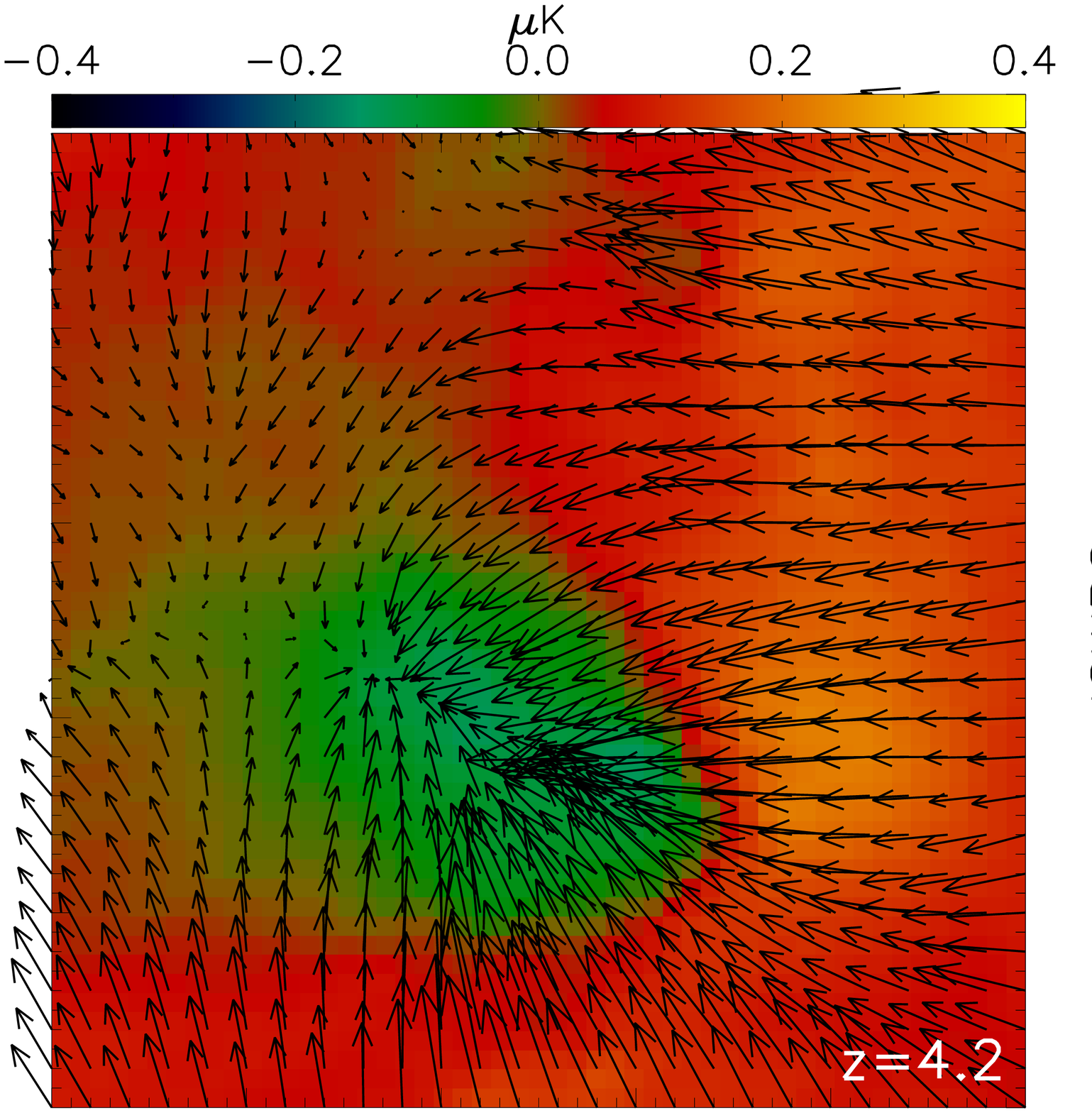}}
\resizebox{\hsize}{!}{
\includegraphics{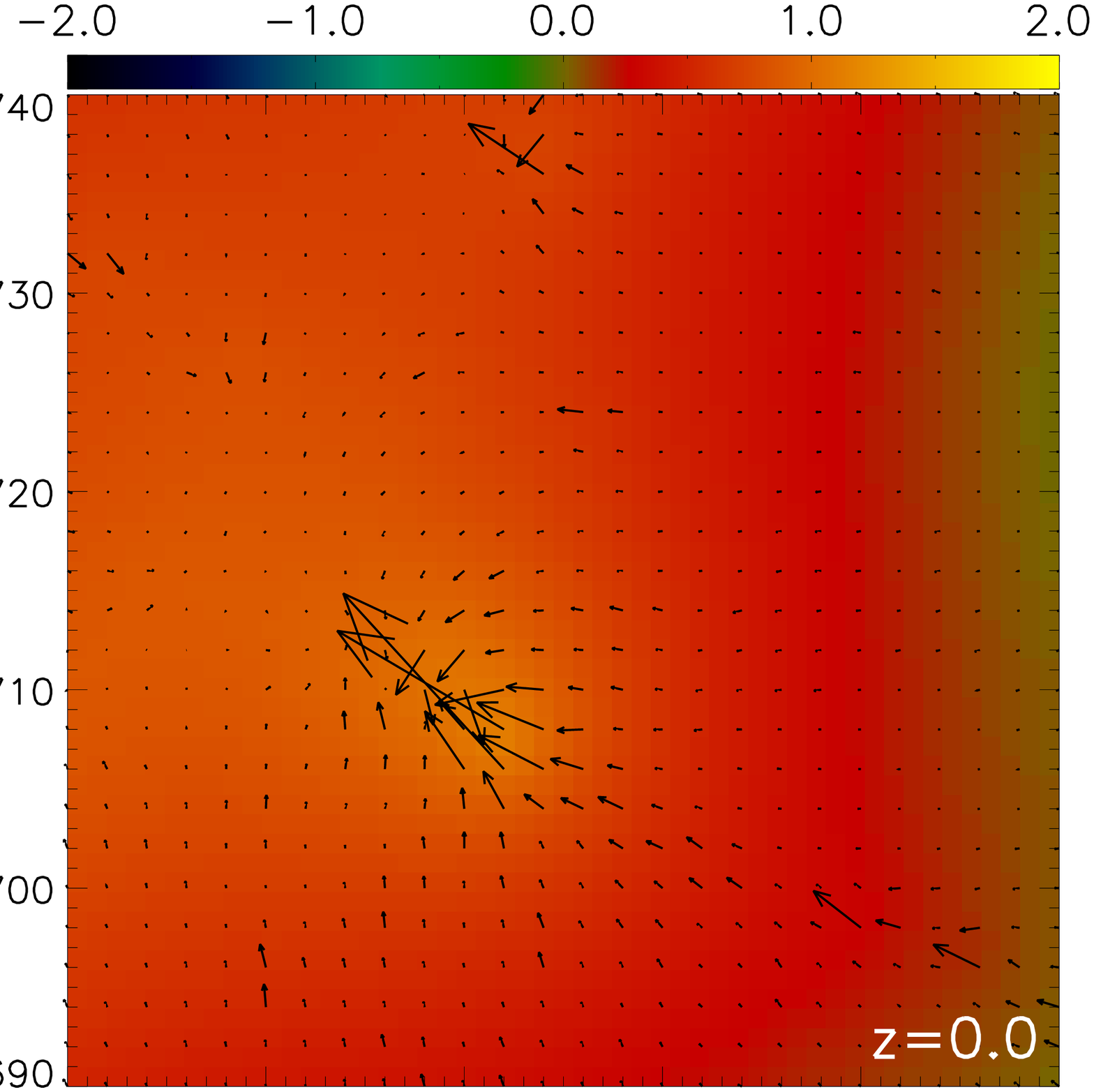}
\includegraphics{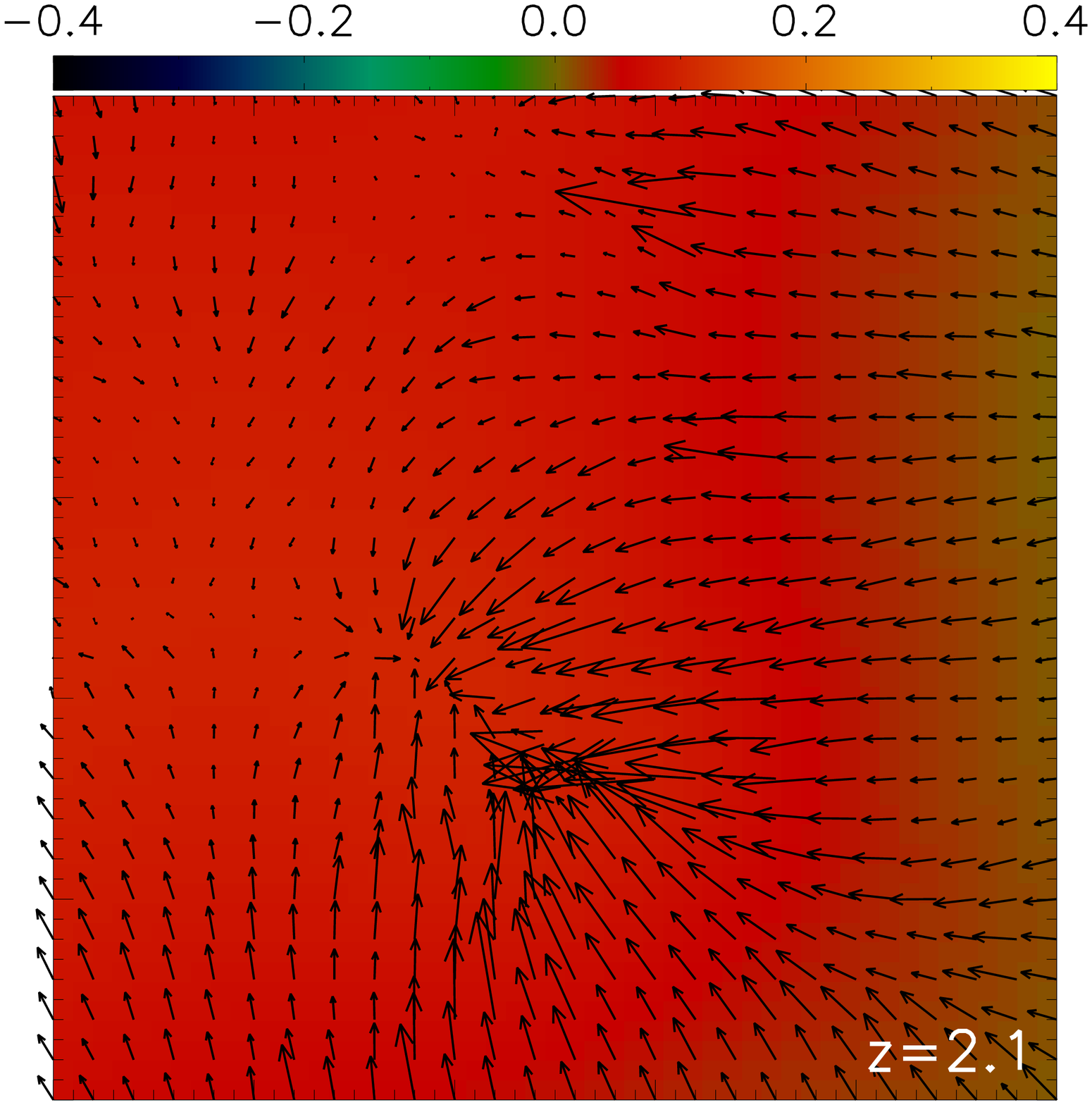}
\includegraphics{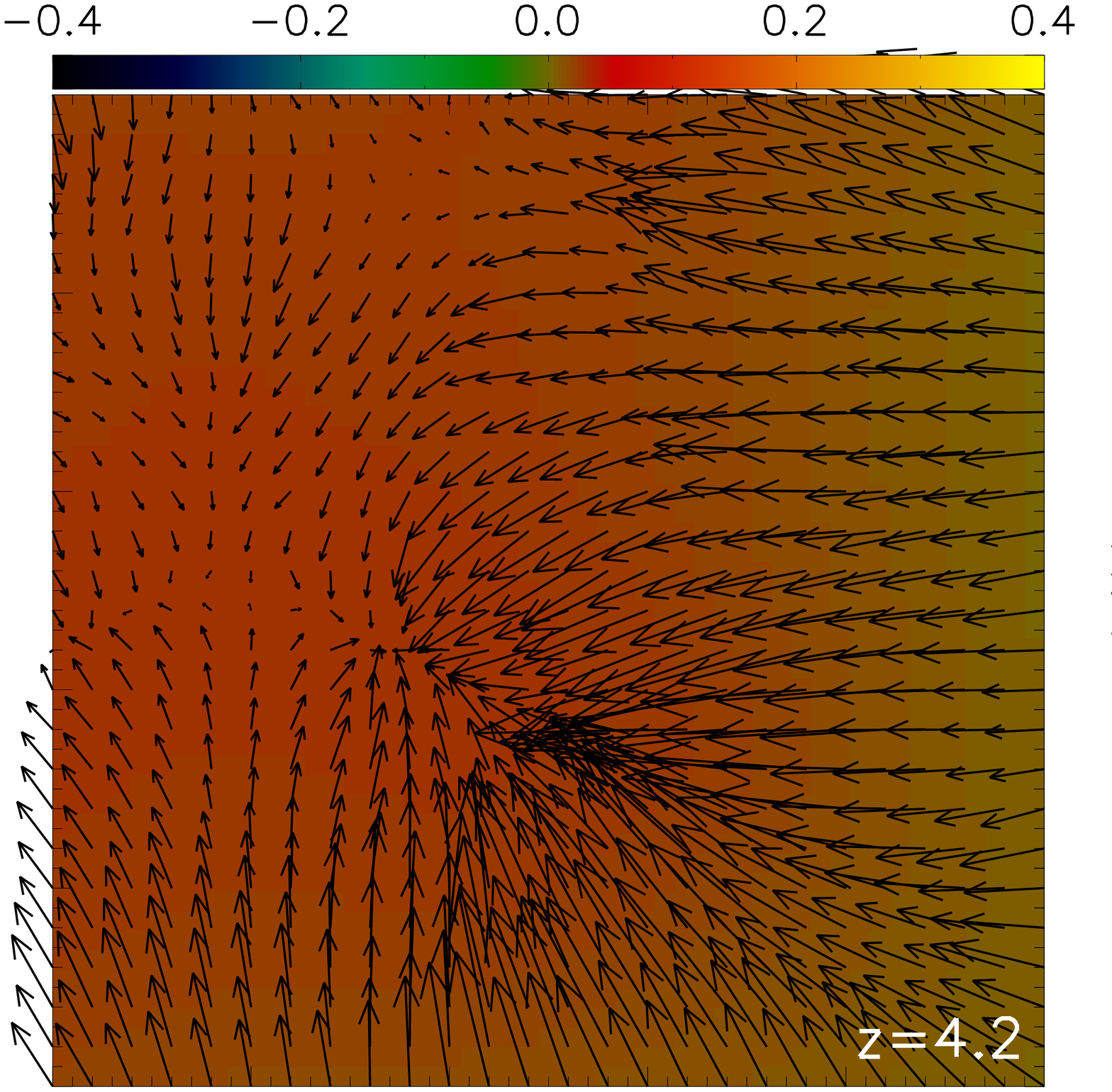}}
\resizebox{\hsize}{!}{
\includegraphics{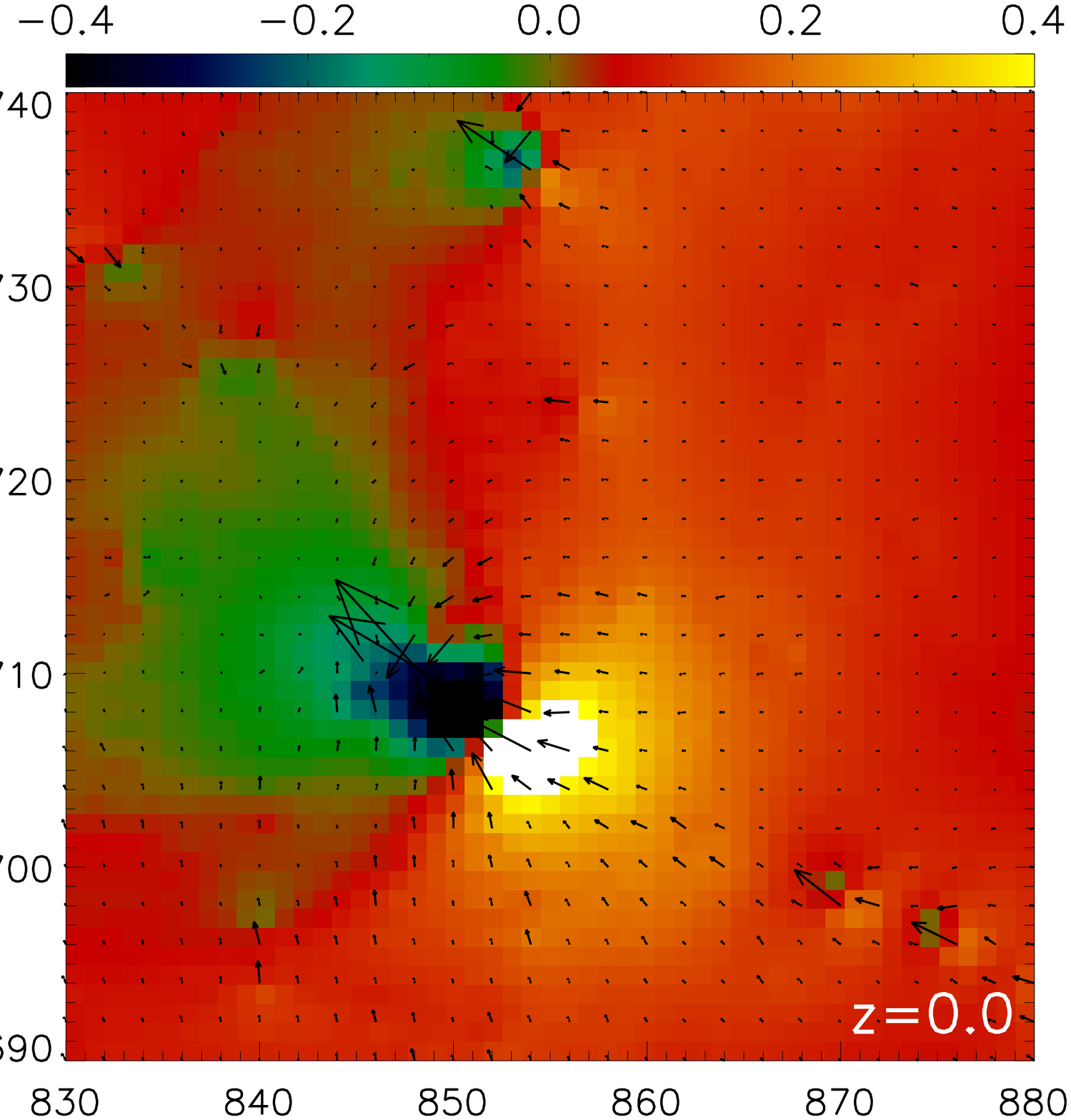}
\includegraphics{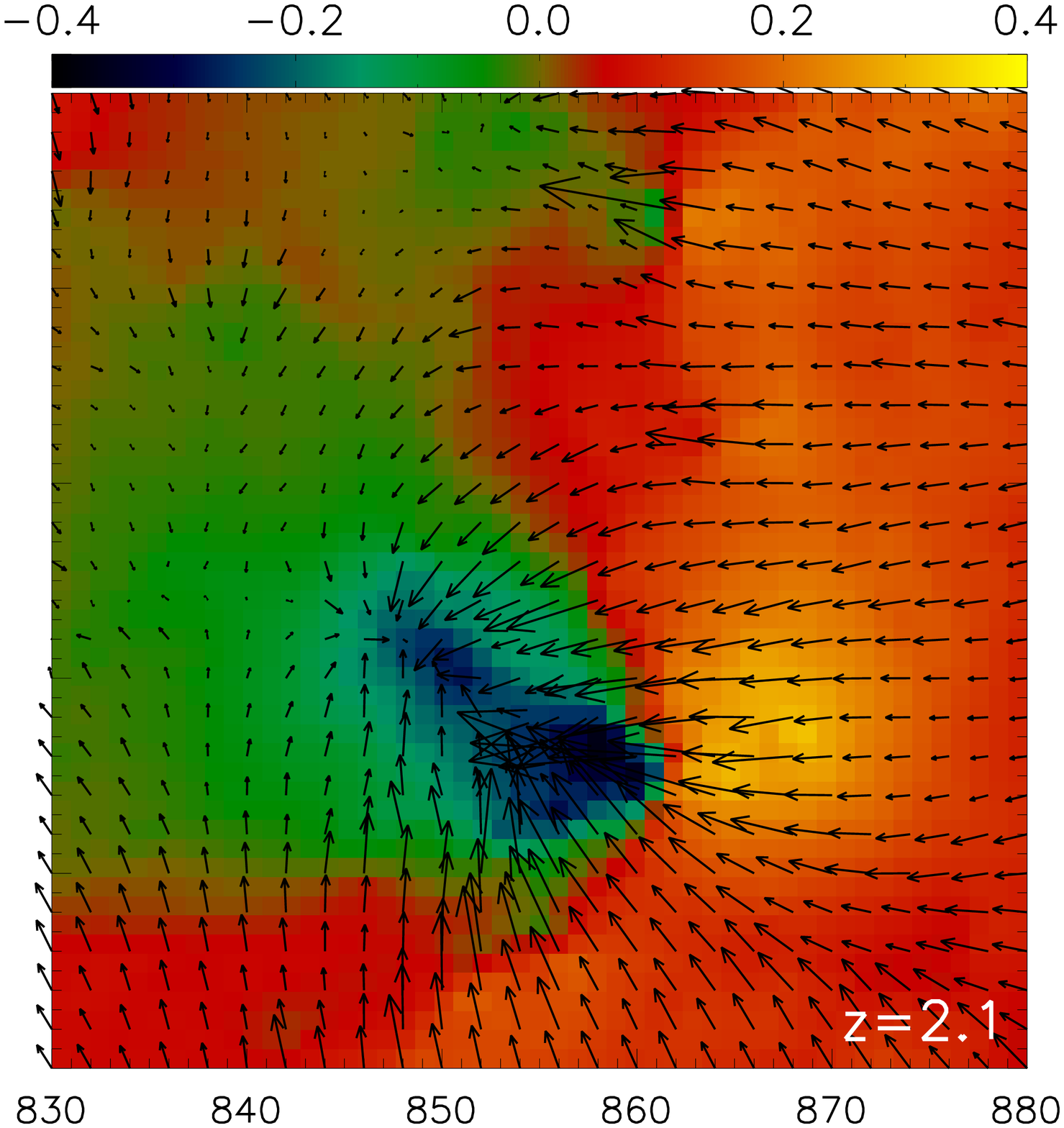}
\includegraphics{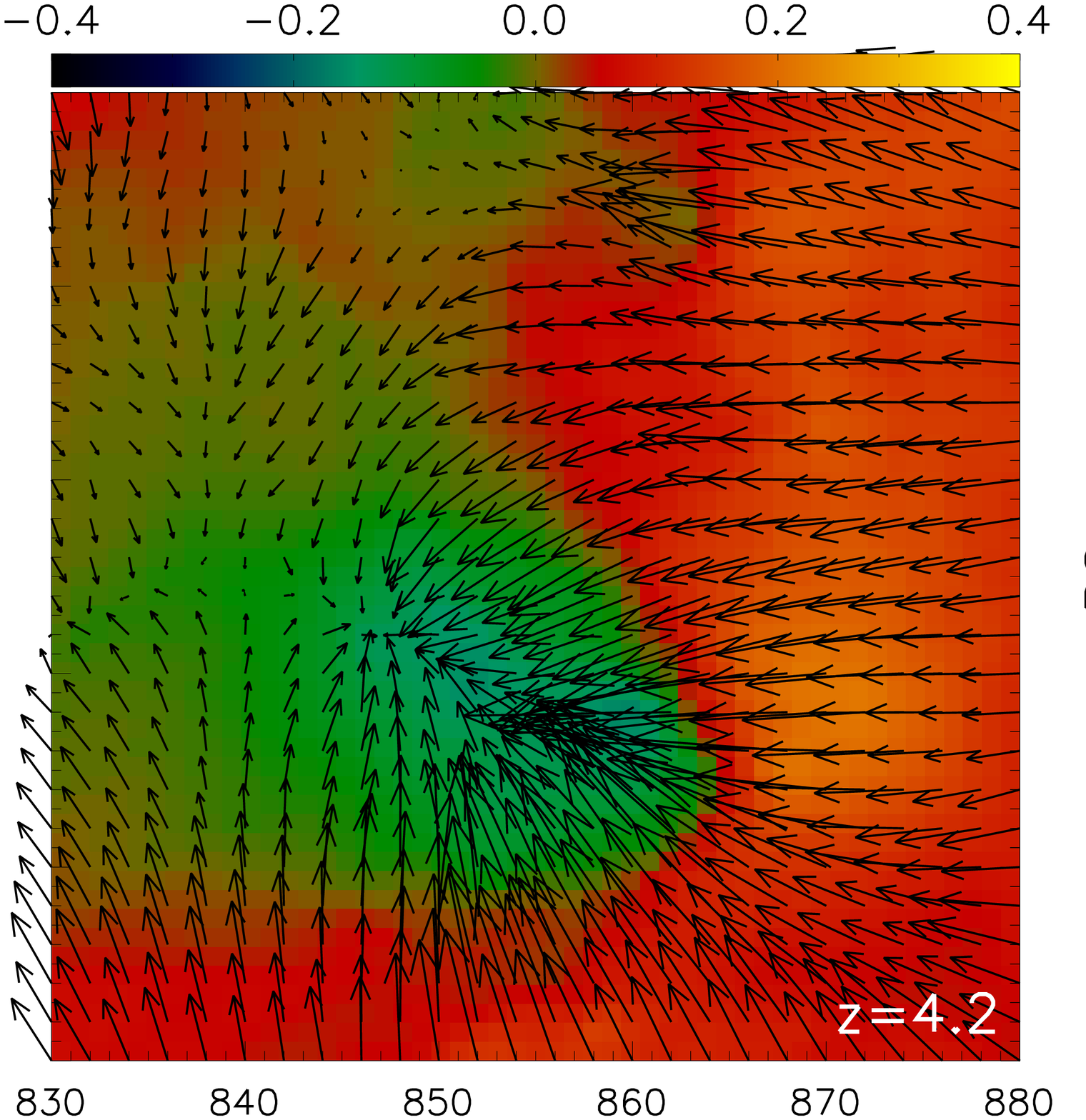}}
\caption{\label{Dipole} Maps of $\Delta$T in a slice of thickness
$\Delta r$=50$h^{-1}$~Mpc and an area of
$50\times50~[h^{-1}$~Mpc]$^2$. The $x$ and $y$ labels give  the exact
coordinates of this set of maps in the full-box maps shown in
Fig.~\ref{FullBox100}. From left to right, they are
maps at $z=$ 0.0, 2.1, 4.2 respectively. From the top to the bottom,
they are maps constructed using
equation~(\ref{eq4}), which includes the ISW and Rees-Sciama effect
(ISWRS), maps with linear approximation for the velocity
field, equation~(\ref{eq5}), (LAV), and residual maps of the top
panels minus the middle panels, which is essentially the
Rees-Sciama (RS) contribution. The overplotted arrows show the
projected momentum field of the slice.}
\end{center}
\end{figure*}

% zoom in of radial tracks
\begin{figure*}
\begin{center}
\advance\leftskip 0.40cm
\resizebox{\hsize}{!}{
\includegraphics{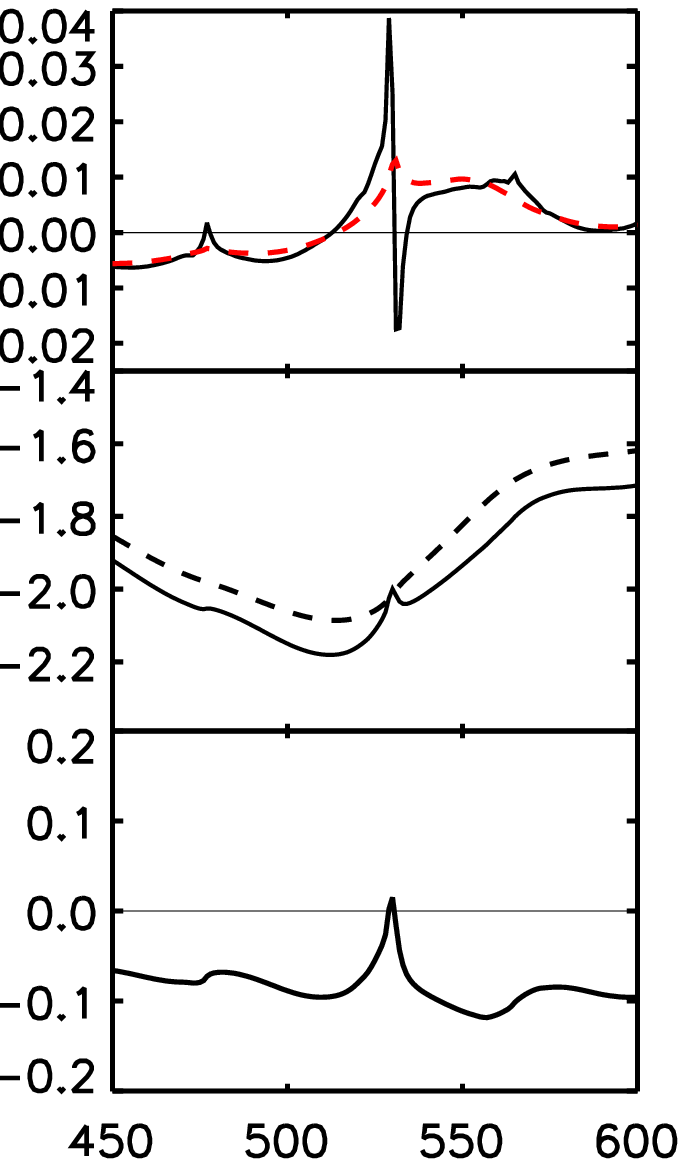}
\includegraphics{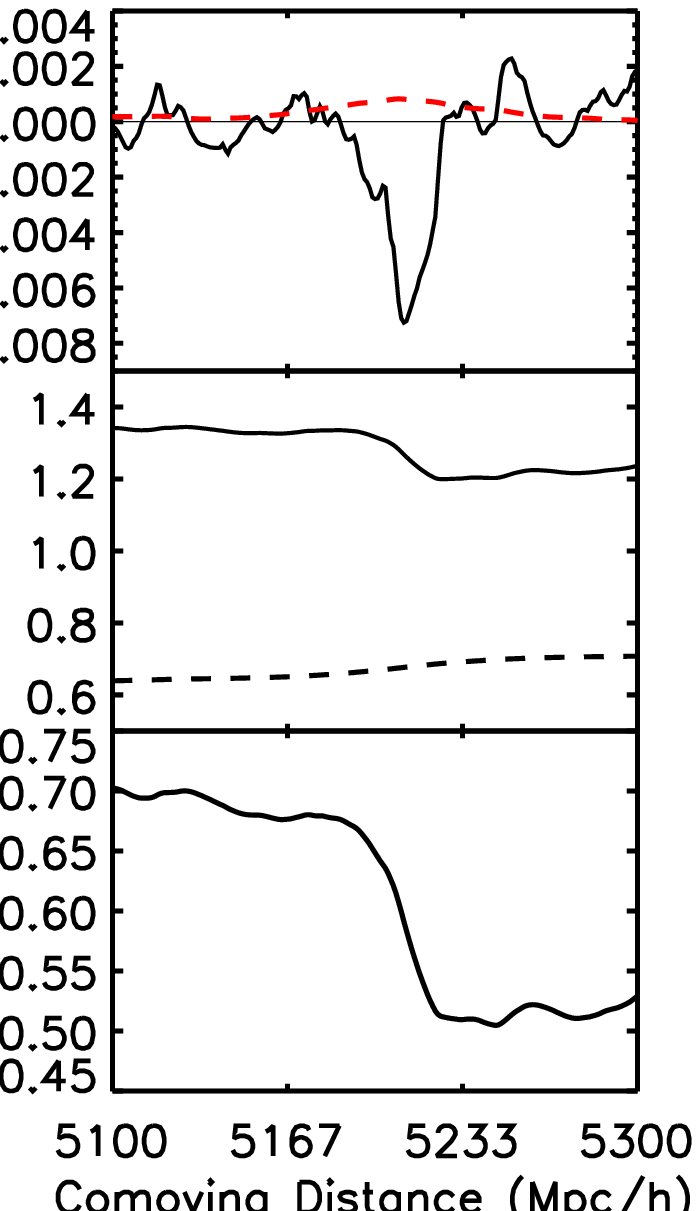}
\includegraphics{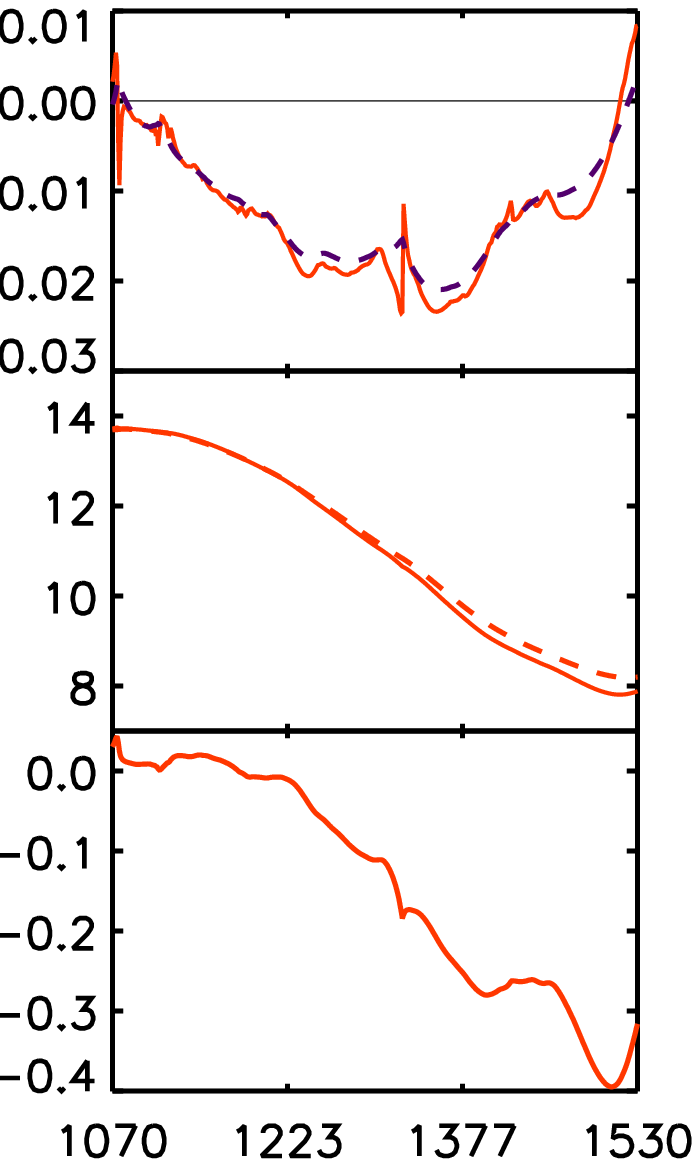}}
\caption{\label{RayTraceDetail} Segments of the three light rays shown in
Fig.~\ref{RayTraceAll}, with the distances shown on the $x$-axis. The top
panels show the temperature perturbations per unit comoving distance
in these chosen
directions. Solid lines show the results of the full ISW and
Rees-Sciama effects (ISWRS) and dashed lines those from the linear
approximation for the velocity field (LAV). The middle panels show the
accumulated temperature perturbations along the radial direction for
ISWRS (solid lines) and LAV (dashed lines). The bottom panels shows the
accumulated difference between the ISWRS and LAV perturbation.  The
left panel is an example of a dipole, the middle panel of a convergent
flow and the right panel of a divergent flow around an empty void.}
\end{center}
\end{figure*}

%Zoom in of convergent flow
\begin{figure*}
\begin{center}
\resizebox{\hsize}{!}{
\includegraphics{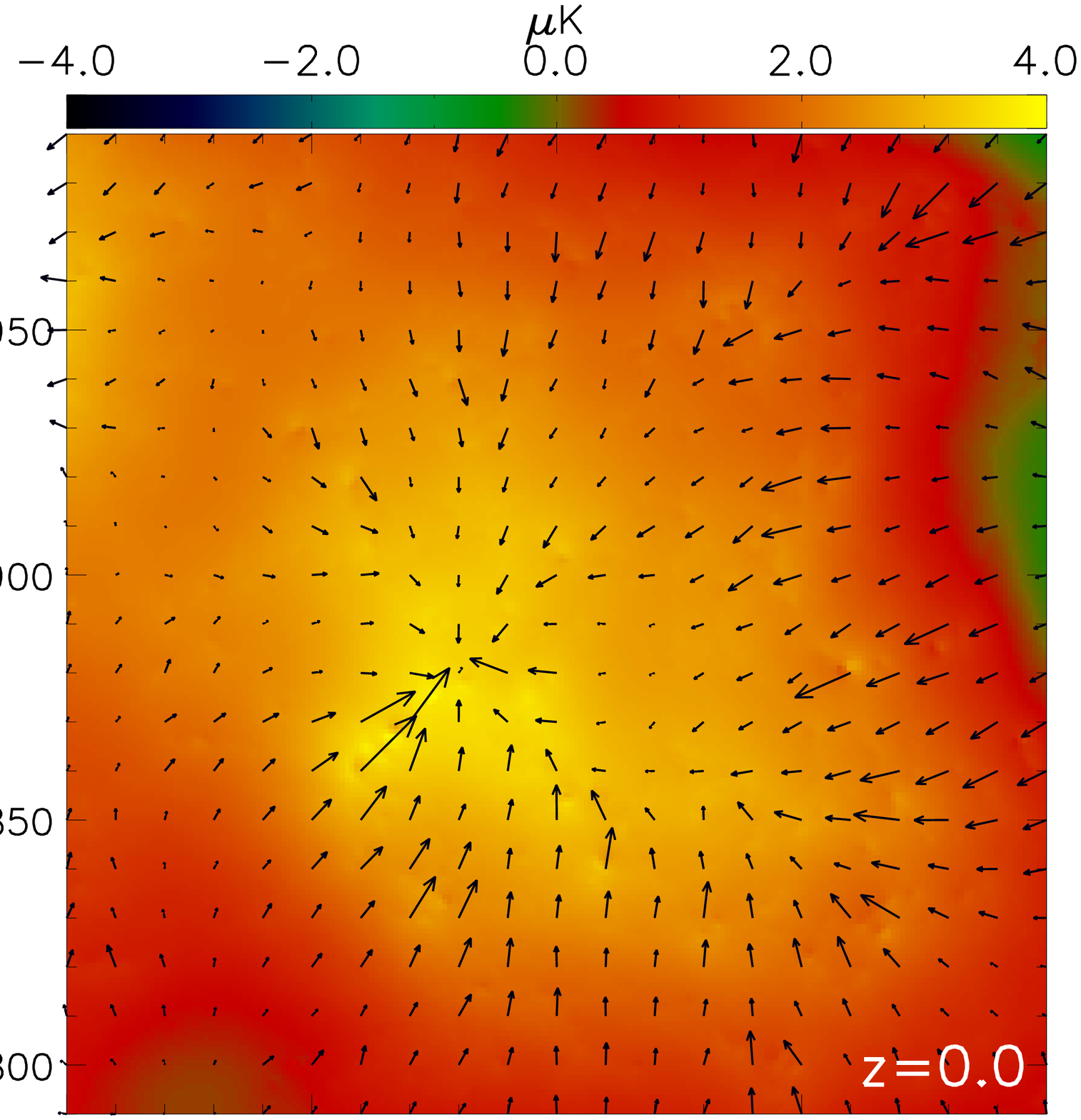}
\includegraphics{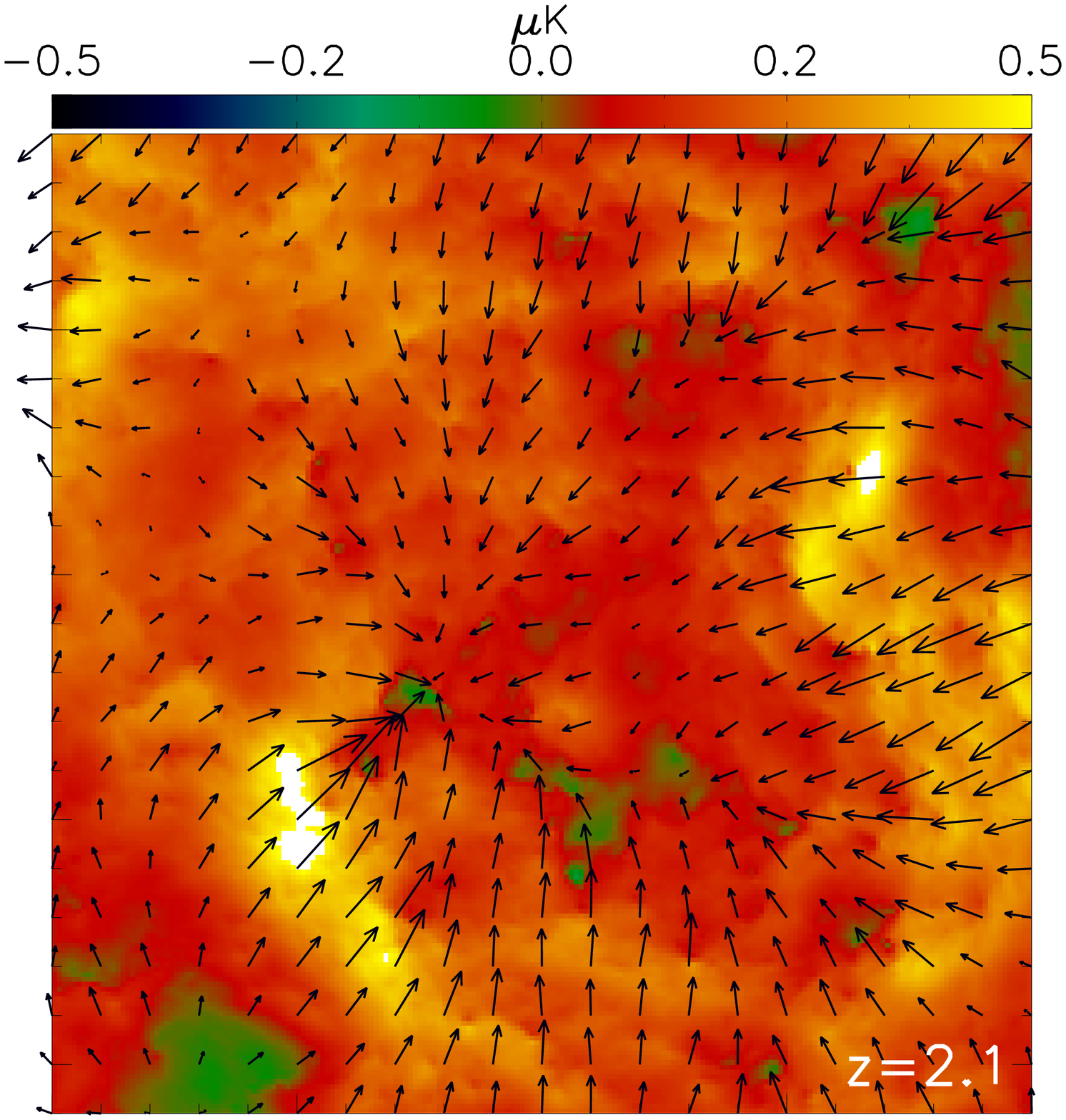}
\includegraphics{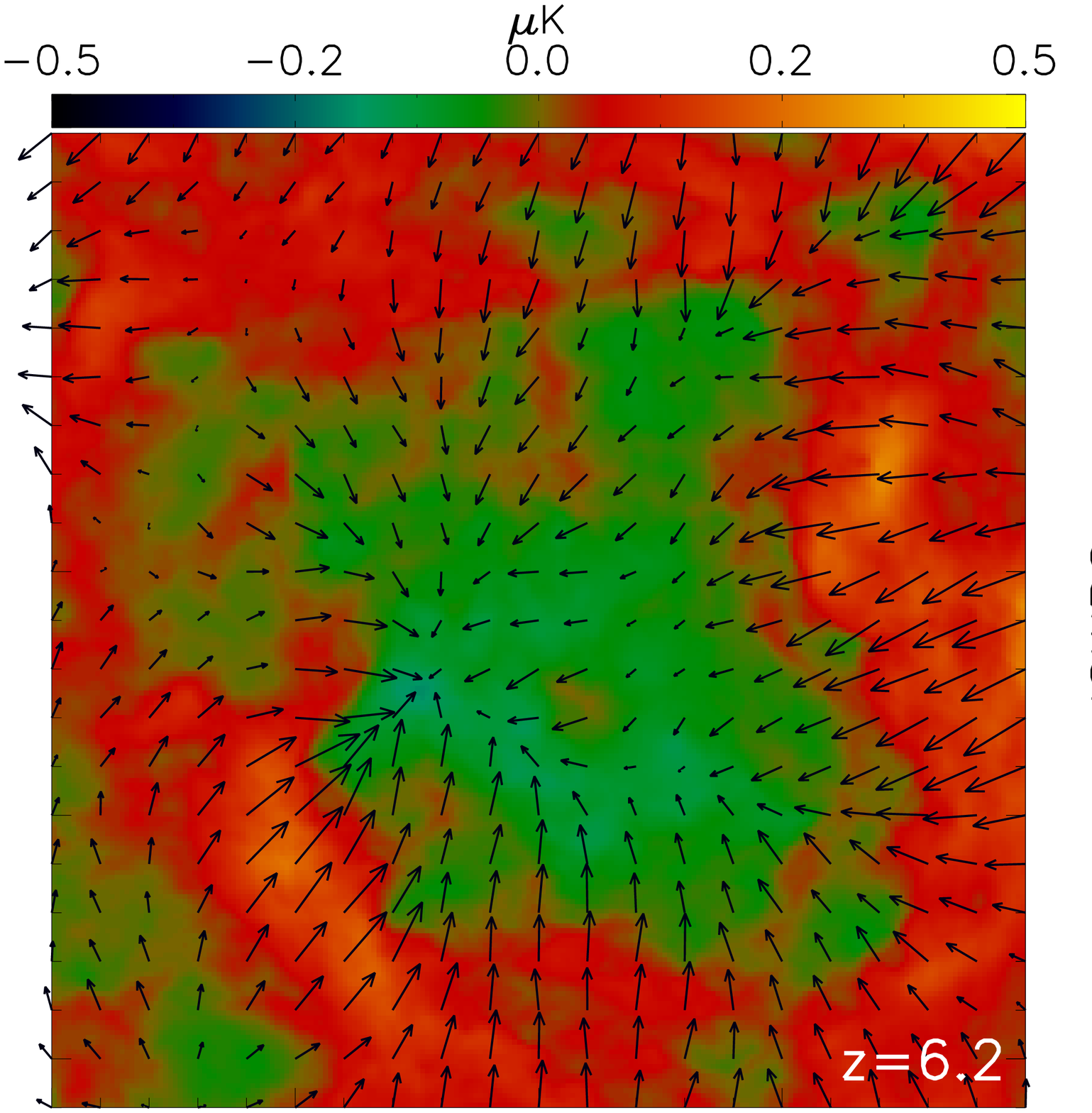}}
\resizebox{\hsize}{!}{
\includegraphics{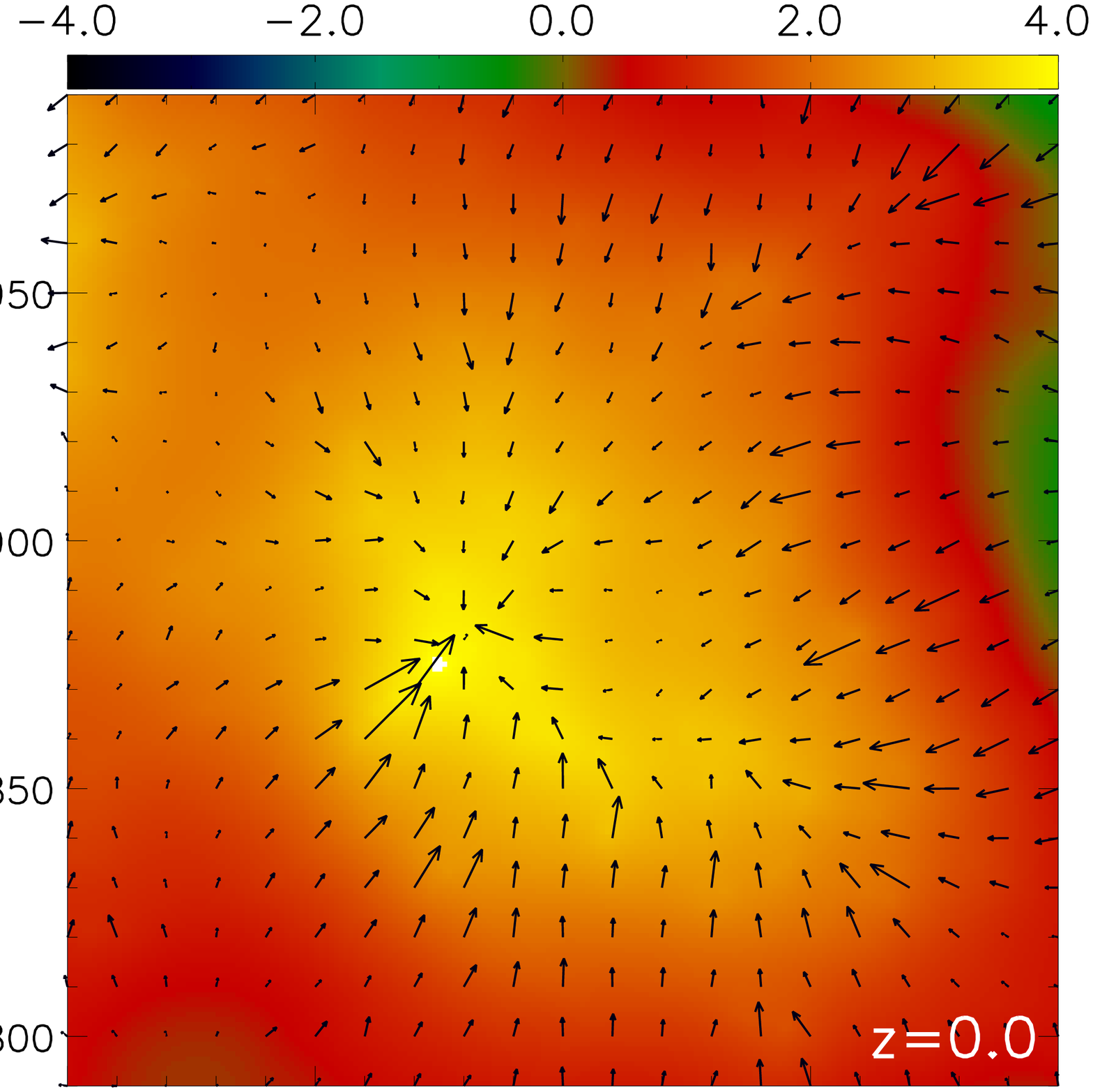}
\includegraphics{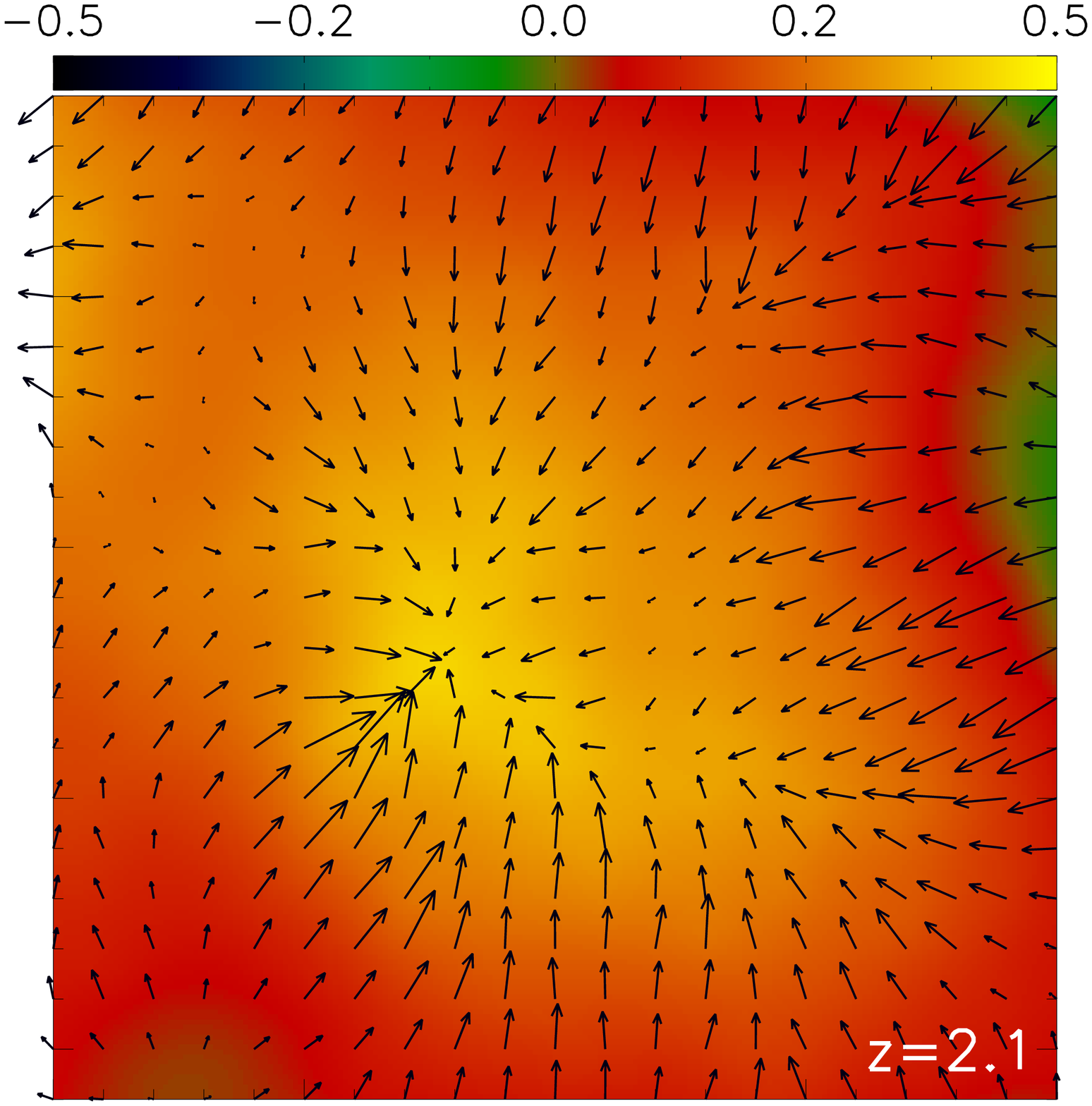}
\includegraphics{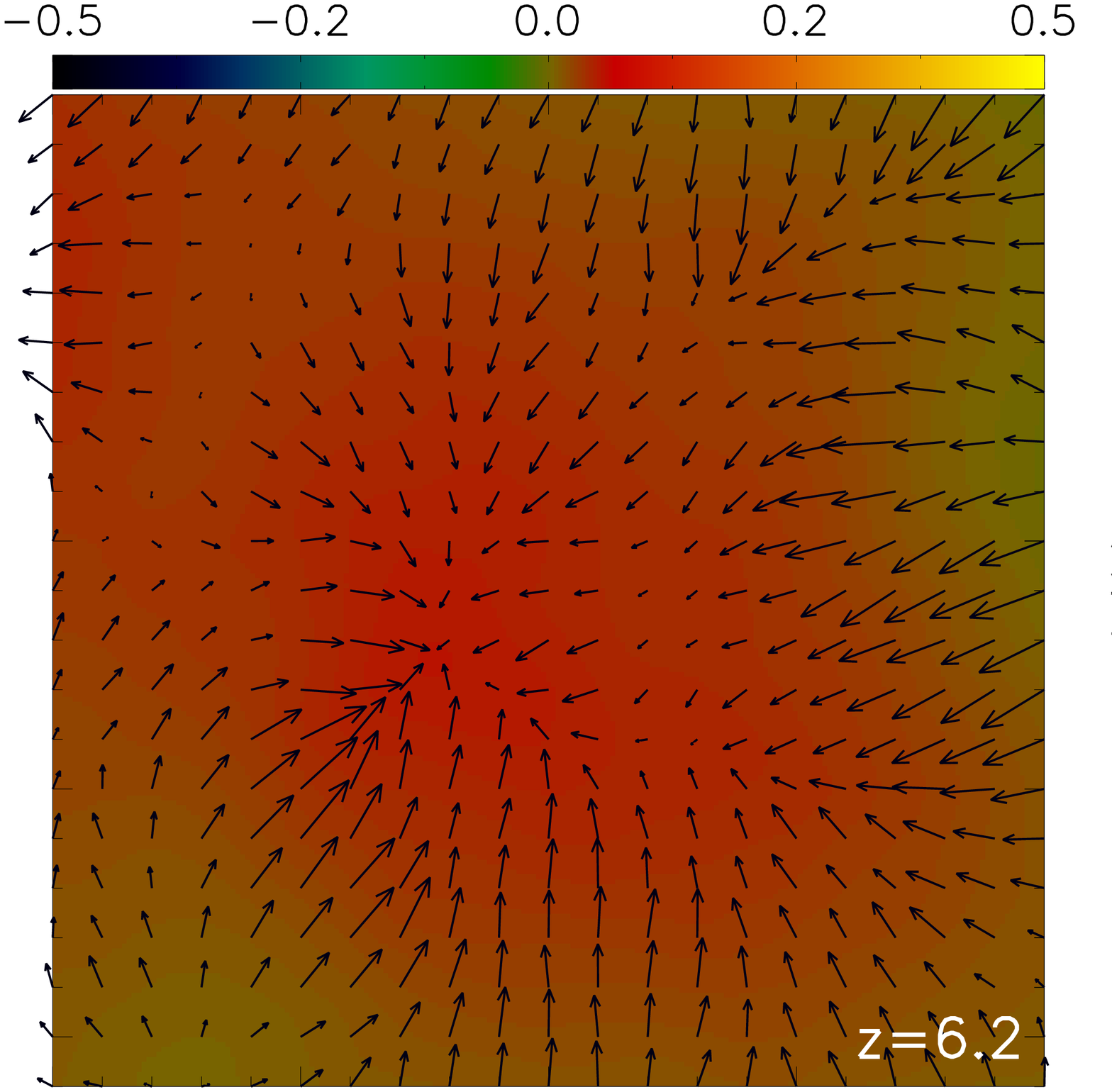}}
\resizebox{\hsize}{!}{
\includegraphics{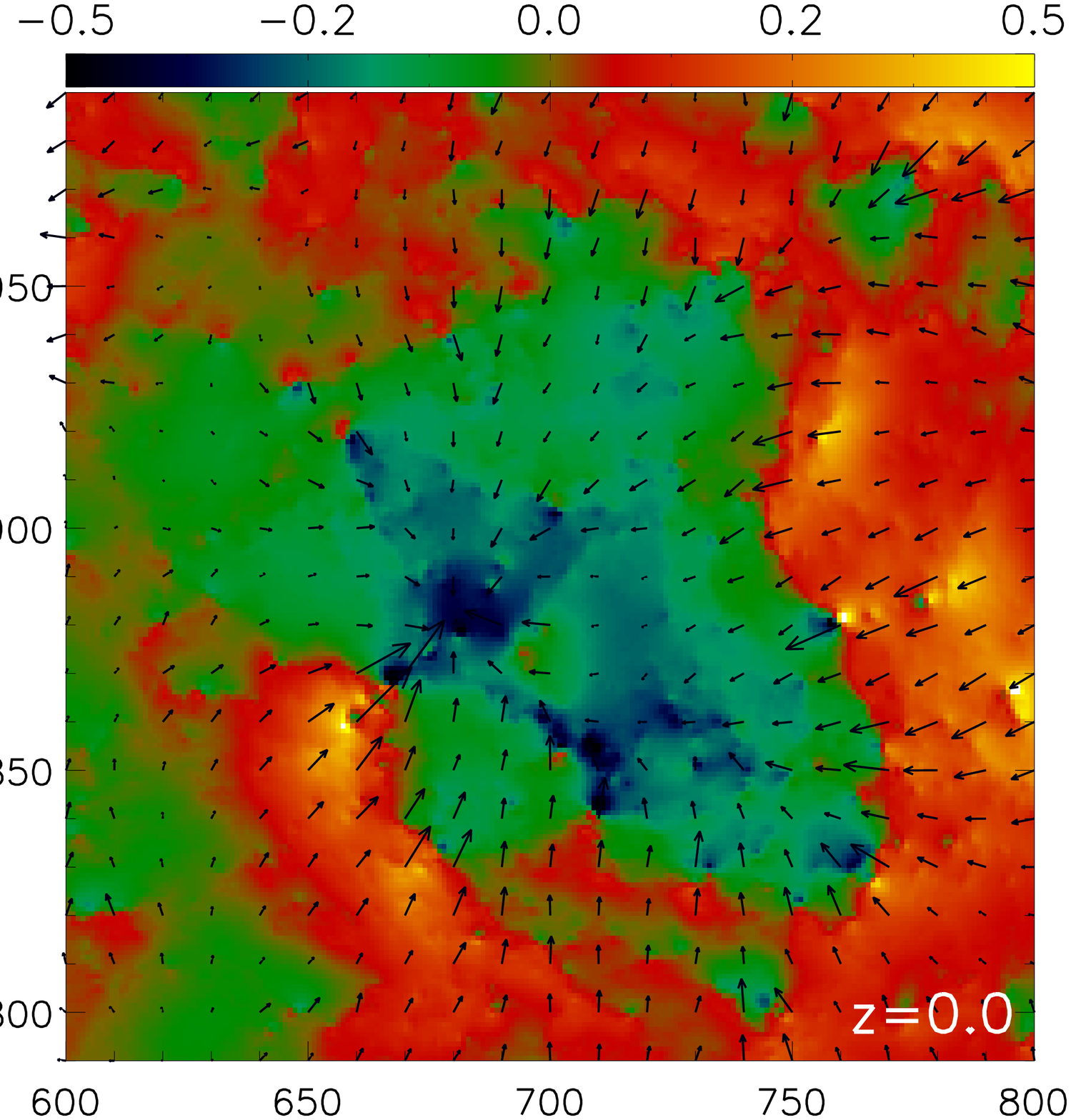}
\includegraphics{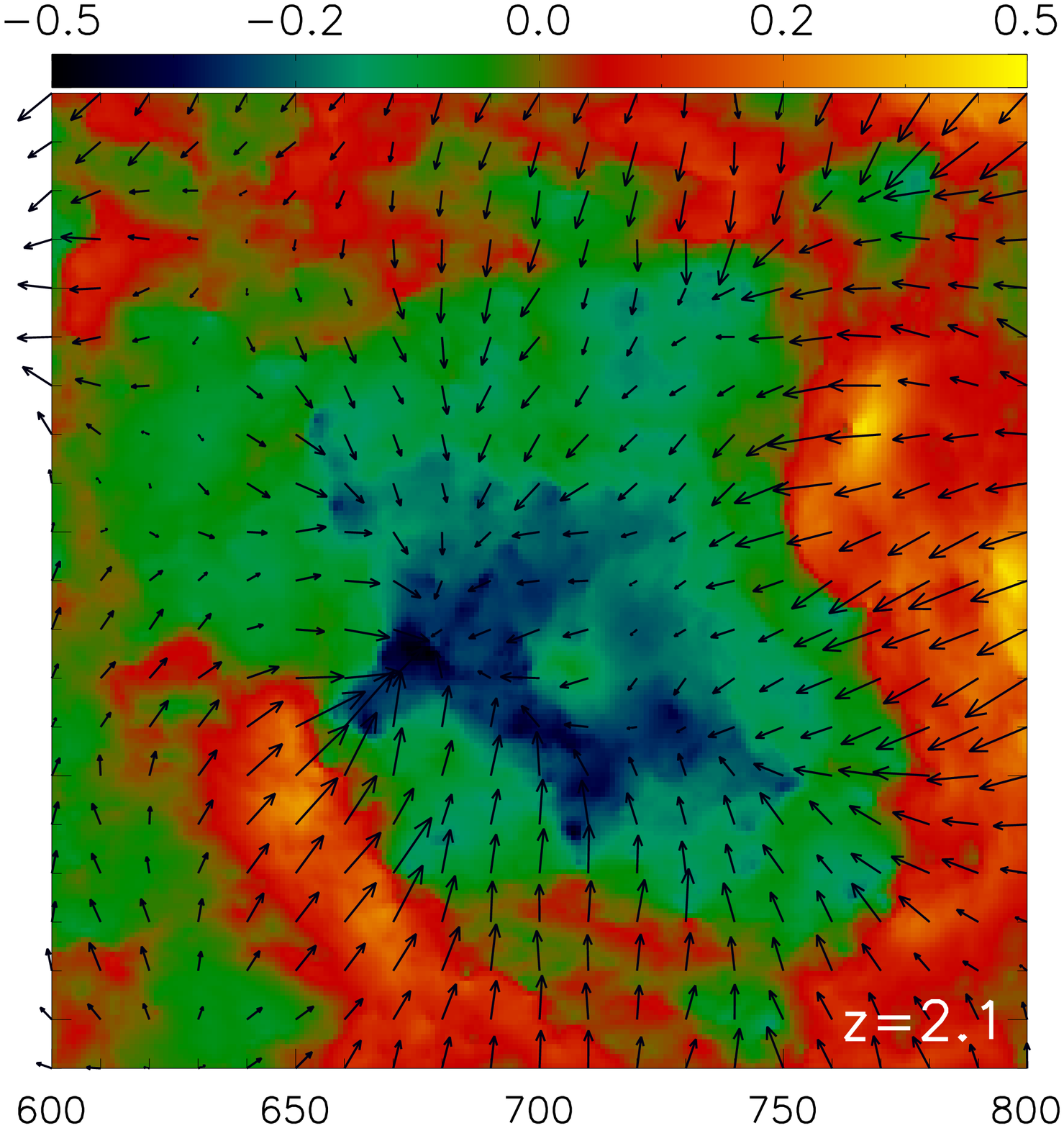}
\includegraphics{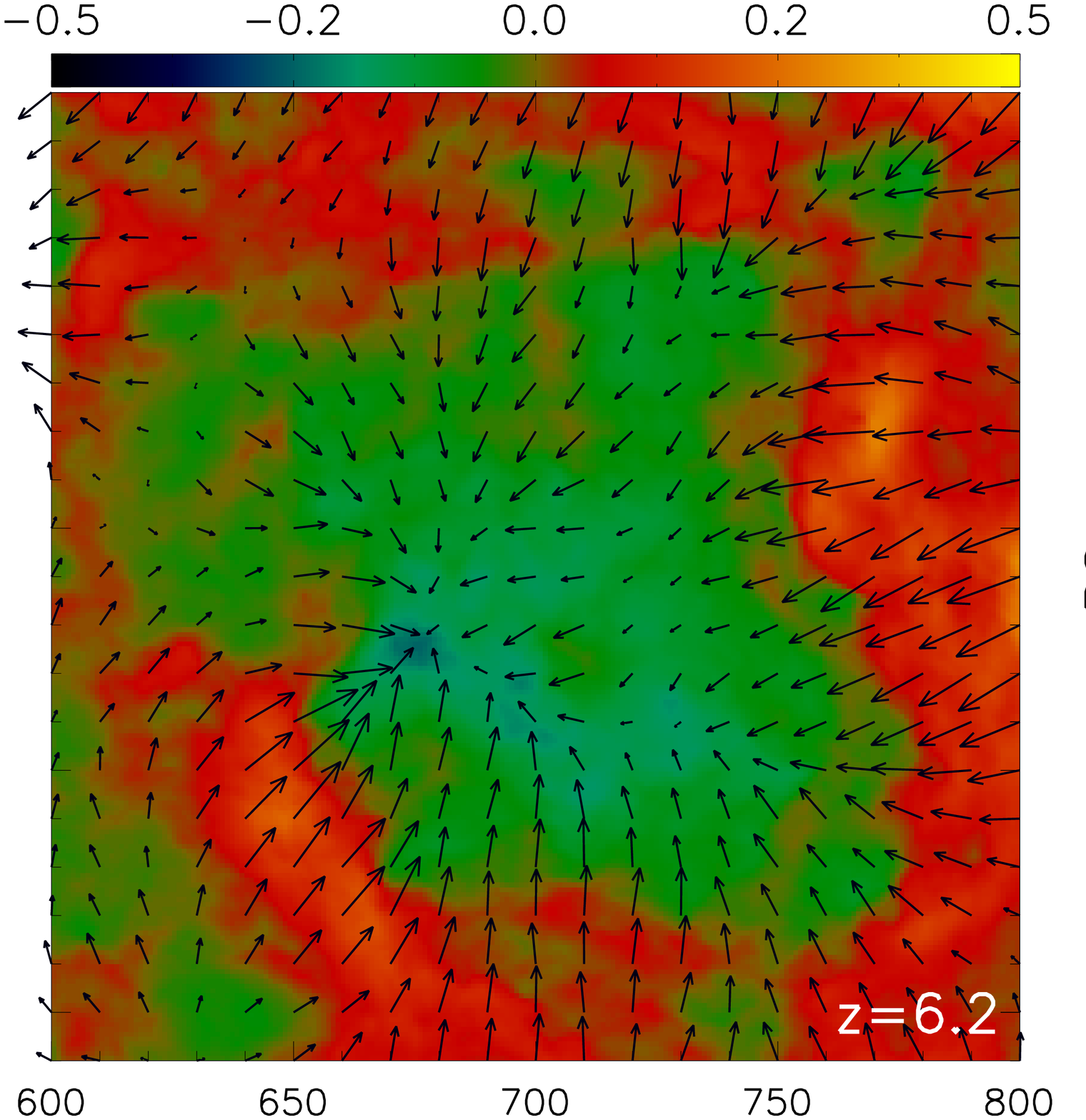}}
\caption{\label{ColdInHot} Like Fig.~\ref{Dipole}, but for slice of
  thickness $\Delta r$=100$h^{-1}$~Mpc, with an area of
  $200\times200~[h^{-1}$~Mpc]$^2$ centred on a convergent flow.  From
  left to right the panels show maps at $z=$ 0.0, 2.1, 6.2
  respectively. Note that the region shown here lies at the boundary
  of the simulation box (see Fig.~\ref{FullBox100}).  To show it
  properly, we shifted the simulation box along the $y$-axis by
  $-100~h^{-1}$~Mpc using the periodic boundary conditions, so that
  the $y$-axis ranges shown in the maps correspond to the combination
  of 890 to 1000$h^{-1}$~Mpc and 0 to 90$h^{-1}$~Mpc in the original
  simulation cube.}
\end{center}
\end{figure*}

%Zoom in of divergent flow
\begin{figure*}
\begin{center}
\resizebox{\hsize}{!}{
\includegraphics{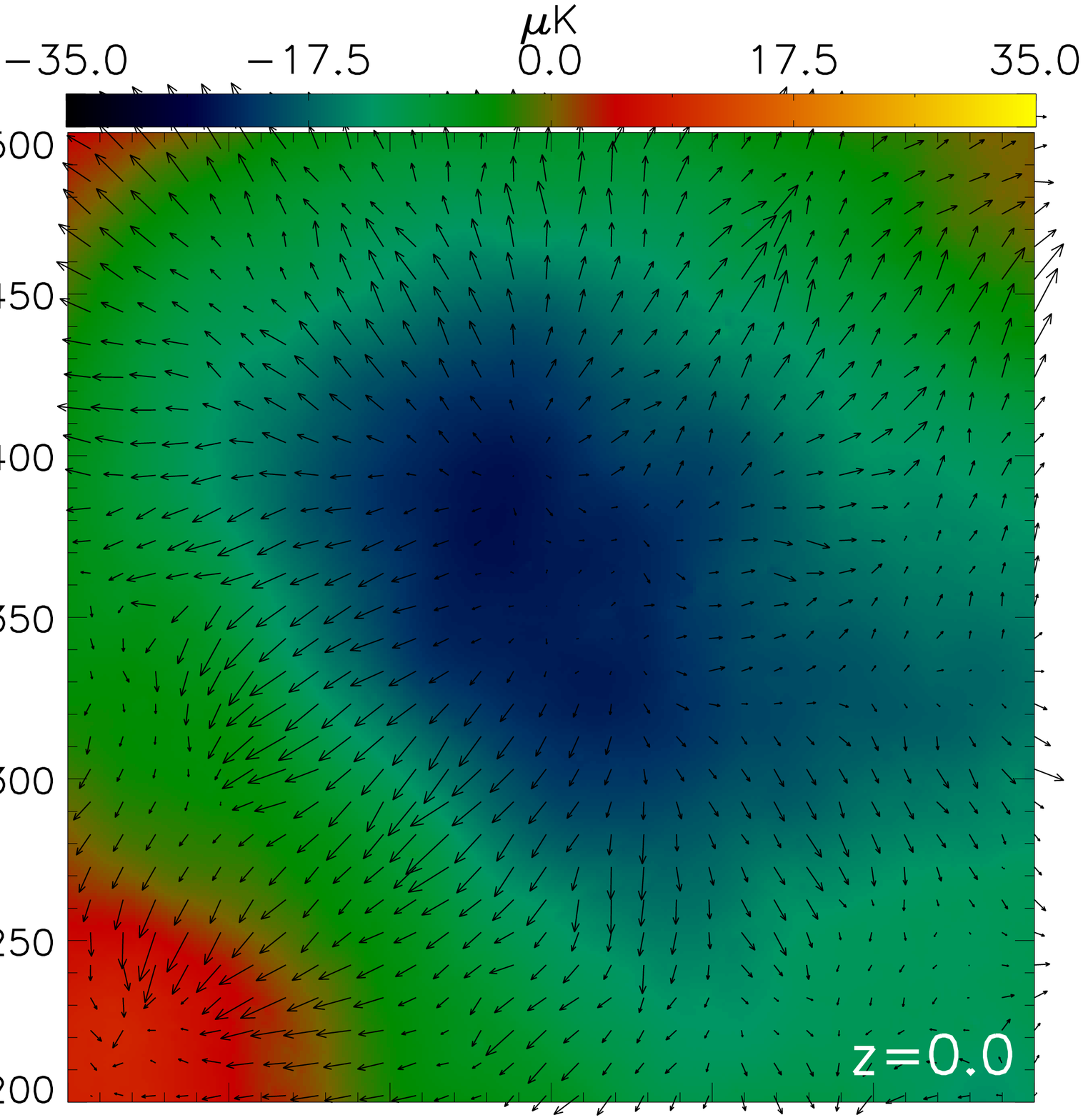}
\includegraphics{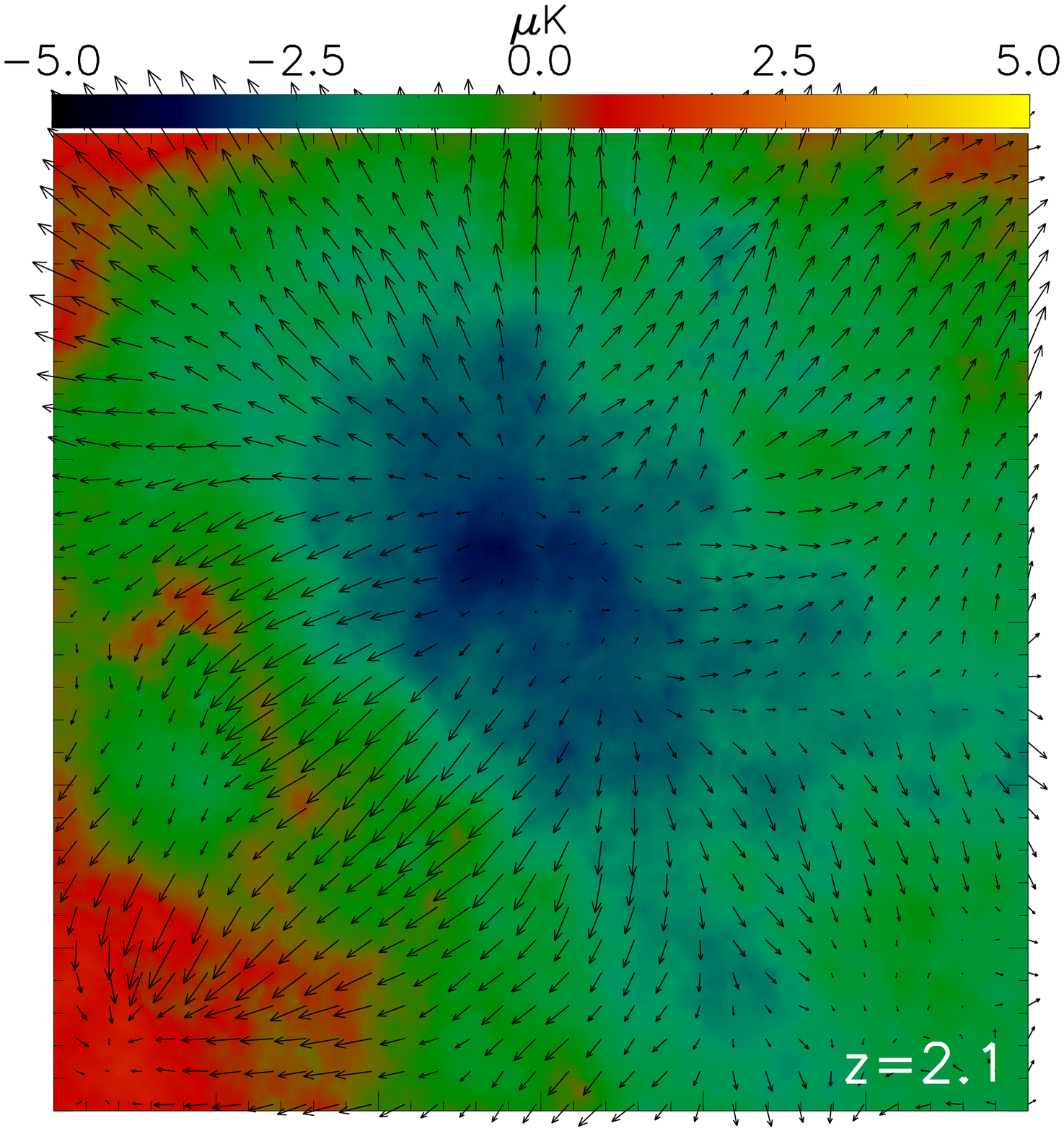}
\includegraphics{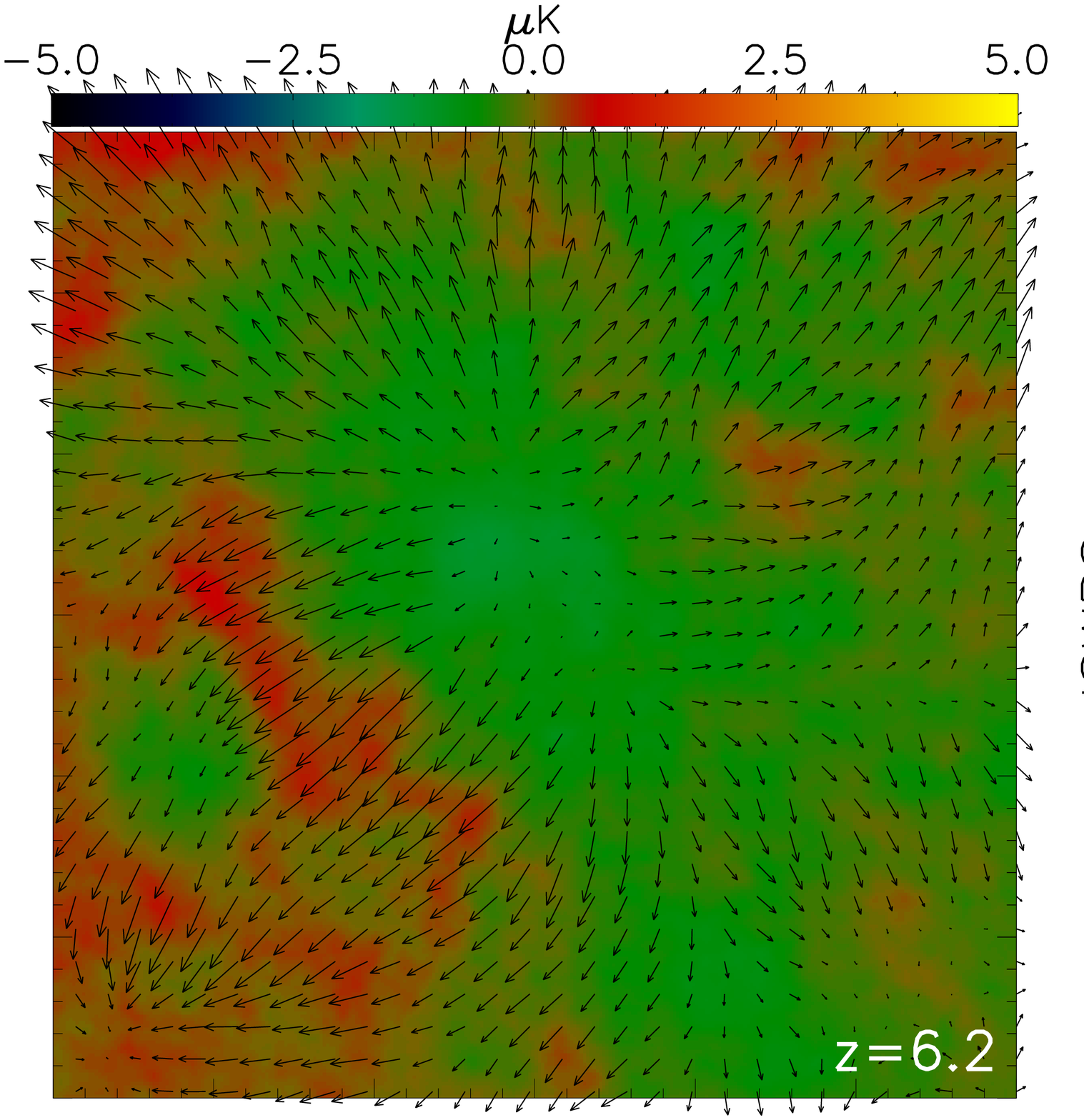}}
\resizebox{\hsize}{!}{
\includegraphics{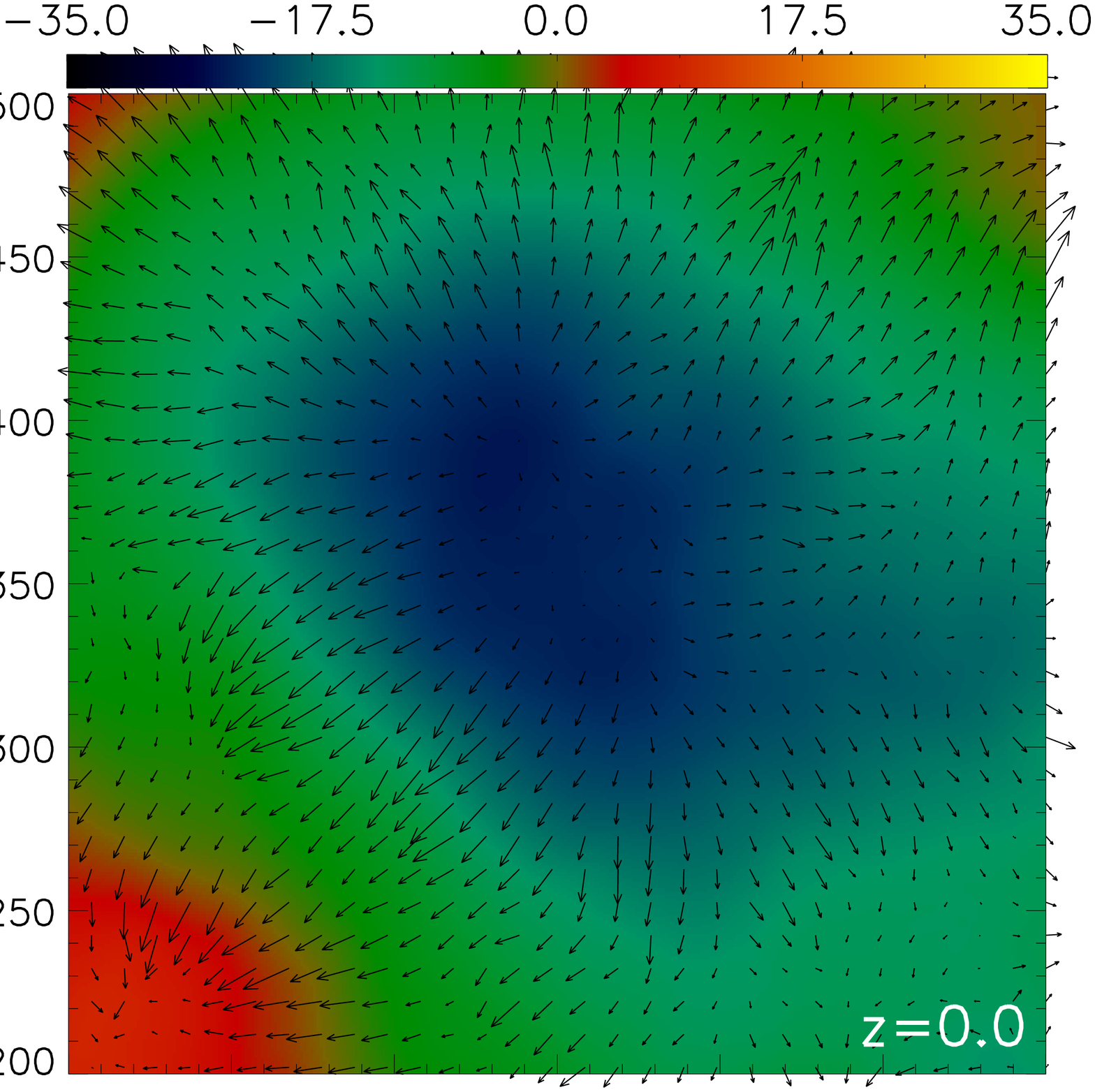}
\includegraphics{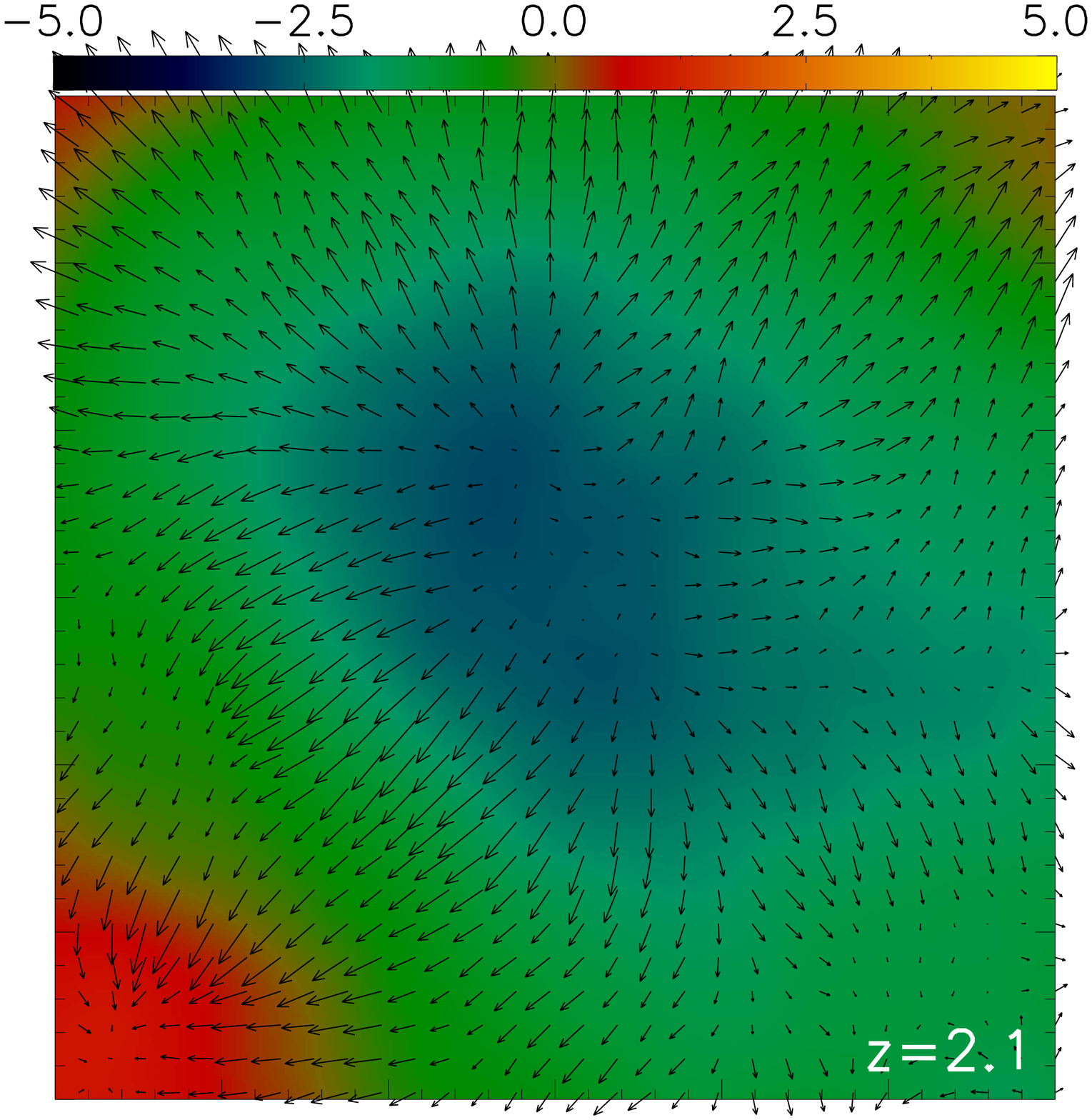}
\includegraphics{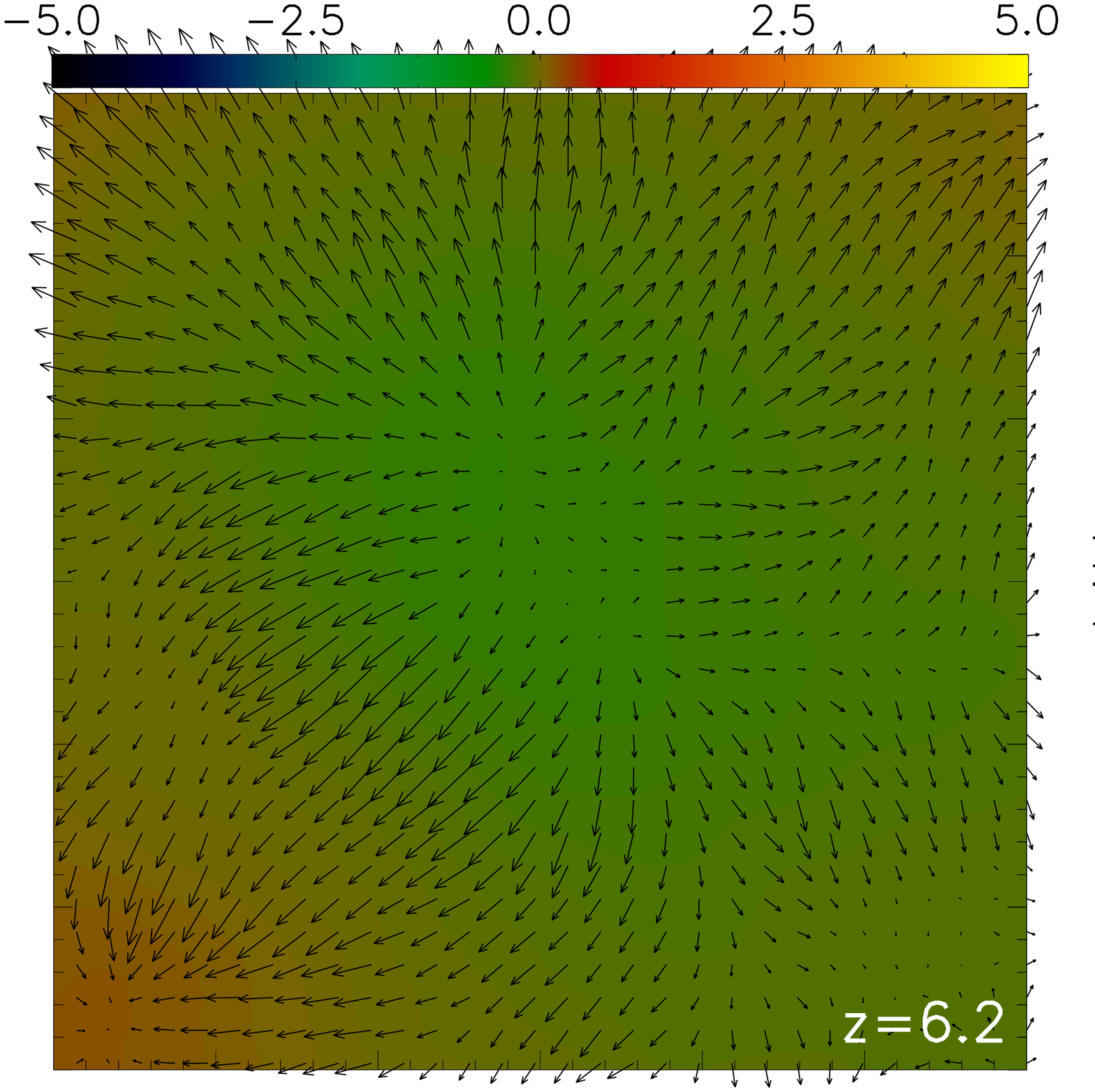}}
\resizebox{\hsize}{!}{
\includegraphics{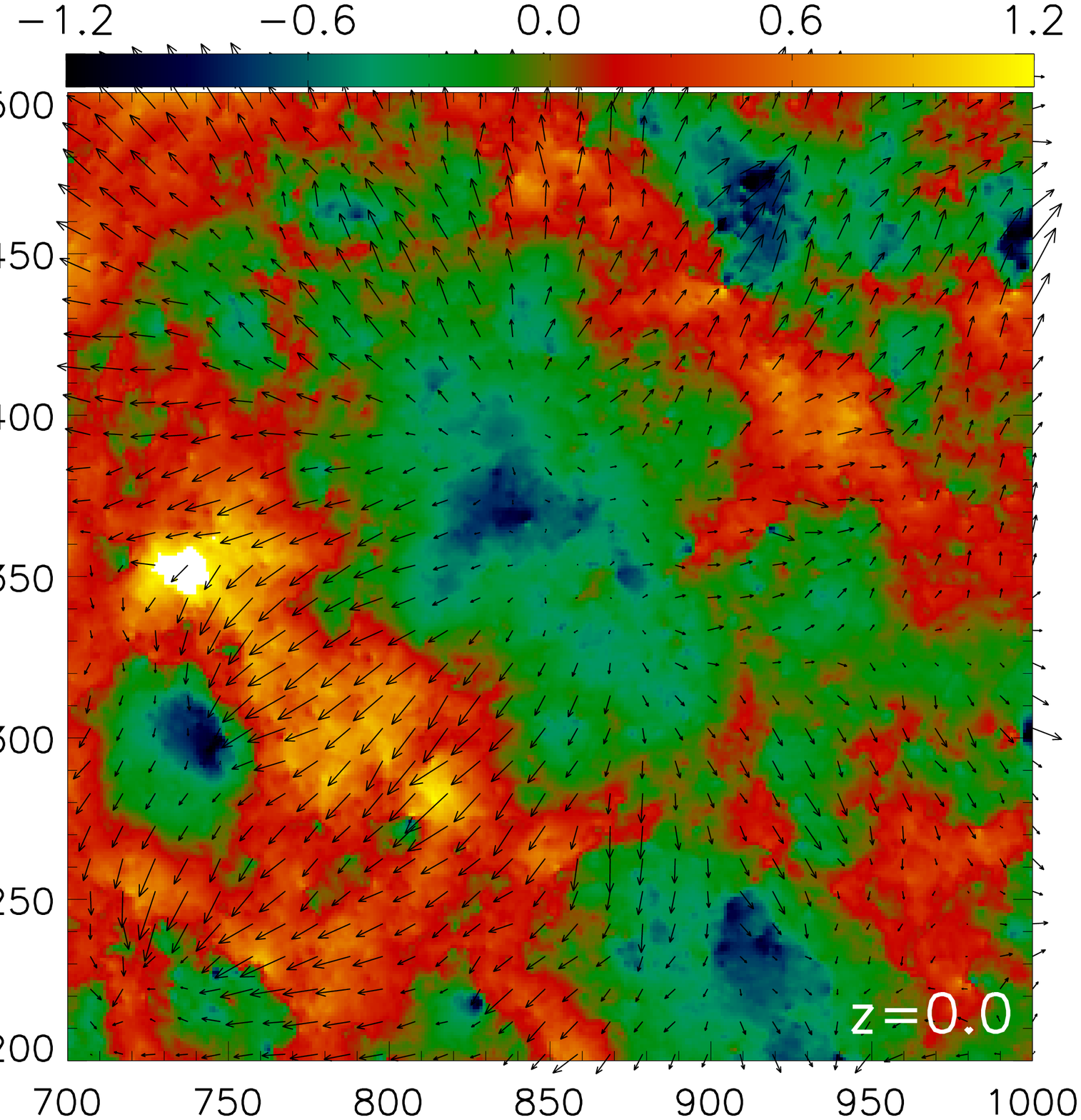}
\includegraphics{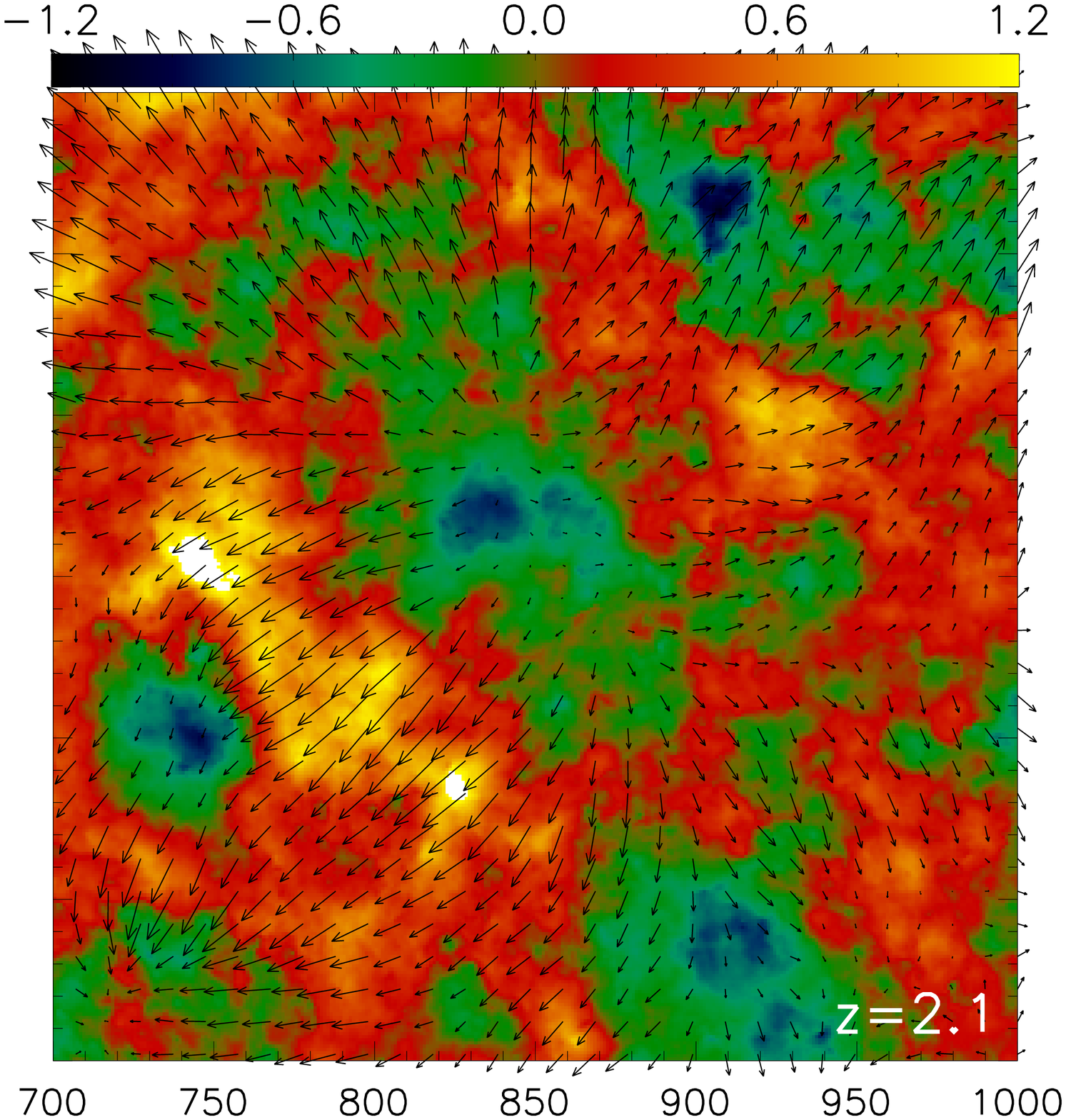}
\includegraphics{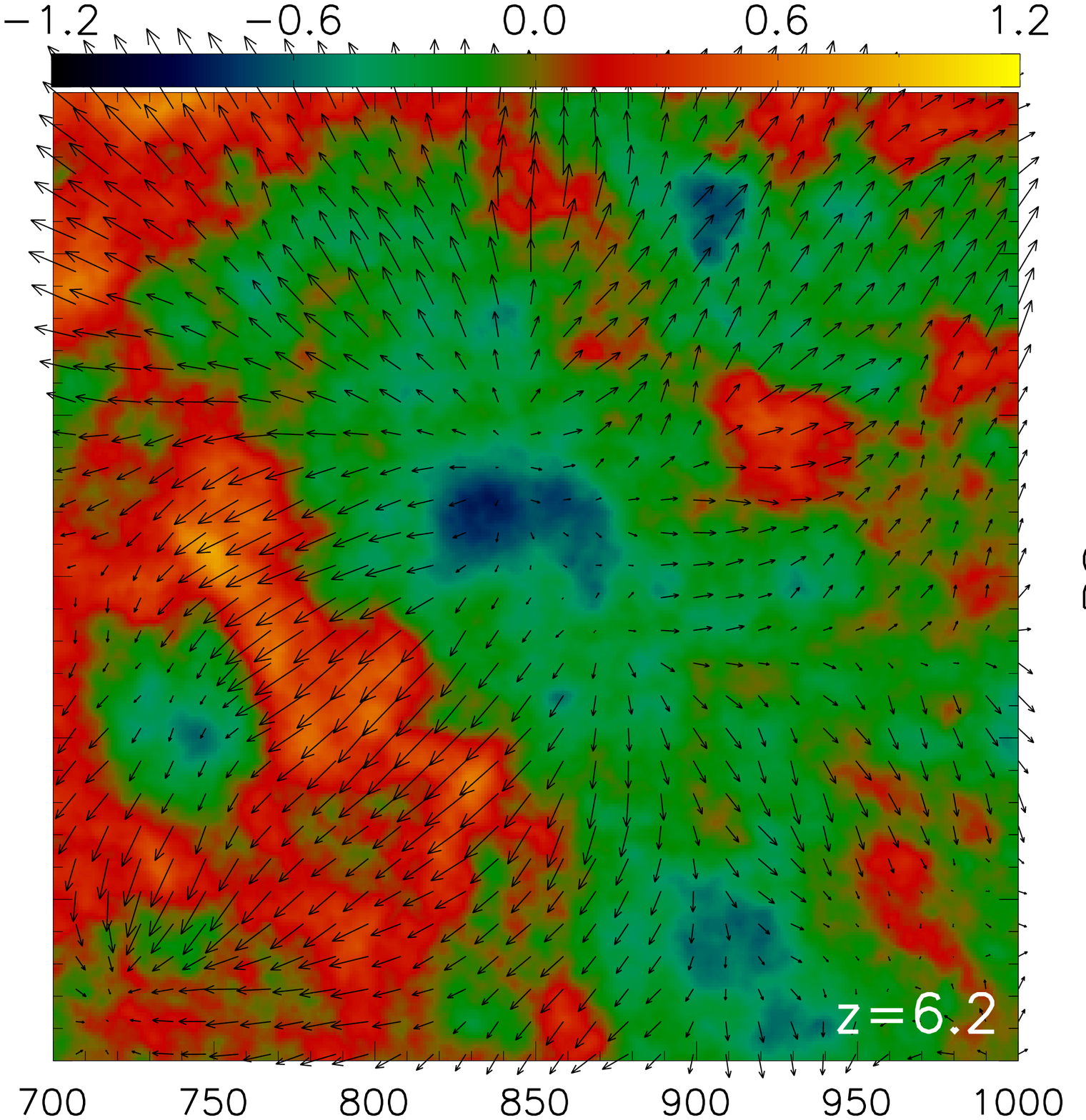}}
\caption{\label{ColdInVoid} As Fig.~\ref{Dipole}, but for
 a slice of thickness
$\Delta r$=1000$h^{-1}$~Mpc, with an area of
$300\times300~[h^{-1}$~Mpc]$^2$ centred on a divergent flow.
From left to right the panels show maps at $z=$ 0.0, 2.1, 6.2 respectively.}
\end{center}
\end{figure*}

Fig.~\ref{RayTraceAll} compares the ISWRS and LAV temperature
perturbations  along three randomly
chosen light rays (avoiding the principal axes of the
simulation box).  In each of the top three panels the solid line shows
the overall ISWRS perturbation and the smoother dashed line the LAV
contribution from $z=0$ to $z=5.8$, corresponding to the comoving
distance from $r_c=0$ to $r_c=6000~h^{-1}$~Mpc.  As in the maps of
Fig.~\ref{FullBox100}, we see that the LAV contribution varies more
smoothly and rapidly damps in amplitude as one progresses along the
rays to higher redshift. The RS contribution produces high frequency
fluctuations around the LAV predictions with the amplitude of these
high frequency fluctuations decreasing with redshift at a much more
modest rate. The lower two panels of Fig.~\ref{RayTraceAll} show that
while a large portion of the high frequency RS signal cancels out when
accumulated along the line-of-sight, persistent residuals of the order
of $1~\mu$K are also accumulated with a large contribution coming
from the redshift range around a radial distance of about
$2000~h^{-1}$~Mpc.

In order to identify the non-linear physical processes that give rise
to the persistent features in the ISWRS maps, we have studied the
momentum field of our simulation. The overlaid white arrows in the
left hand panels of Fig.~\ref{FullBox100} indicate the projected
momentum field in this slice of the simulation.  At higher redshifts
we find that the orientation of the momentum vectors is similar to
that at $z=0$, but
their amplitude increases, roughly in accord with the expected linear
growth rate factor, $\beta$. In the LAV panel (middle row) we see that
on large scales the momentum field is well correlated with the
temperature perturbations and hence with the $\dot \Phi$ field.  Dark
matter particles are moving towards hot lumps (overdense regions,
potential wells) and flowing away from cold lumps (underdense regions,
potential hills).  This is the scenario expected in linear theory.
However, looking at the smaller scale features evident in the
ISWRS (top panel) maps we find that this correlation between
temperature and momentum does not always hold. In fact, at high
redshift, violations of the correspondence expected in linear theory
are very strong. Hence, guided by the
momentum field we have identified three interesting non-linear
phenomena in the ISWRS maps, which are related to dipoles, convergent
flows and divergent flows. Three regions exhibiting these phenomena
are highlighted by the square boxes in Fig.~\ref{FullBox100} (one of
which is
split across the periodic boundary of the simulation). We zoom in and
study each in detail in the following subsections.

\subsection{Dipoles}

In RS maps, at scales of tens of Megaparsecs, large lumps of dark matter
moving perpendicular to the line-of-sight give rise to dipole
features, i.e. a cold spot on the leading part of the lump and a hot
spot on the trailing part.  Fig.~\ref{Dipole} shows zoomed-in maps of
the ISWRS, LAV and RS temperature perturbations in the region around
one such dipole at redshifts $z=0$, $2.1$ and $4.2$ for a
$50~h^{-1}$Mpc.
The overplotted momentum vectors clearly show the bulk motion of a
large lump of dark matter. The length of the arrows indicate
that this velocity is being damped and so decreases with decreasing
redshift.
The LAV maps are very smooth and show no sign of the dipole
feature at all, while the strikingly large dipole with amplitude
of $\sim 1~\mu$K is clearly visible in the ISWRS maps (top row),
particularly at the higher redshifts.

The physical origin of these
dipole features, which are just a special case of the RS effect, can
be understood in terms of the evolution of the gravitational
potential, $\Phi$, on either side of the moving mass. At a fixed
position ahead of the moving mass, $\Phi$ decreases as the mass and its
gravitational potential well approaches. This will create a CMB cold
spot as CMB photons passing through this point will gain less energy
falling into the potential well than they will subsequently loose climbing
out of the then deeper well. Conversely, behind the mass, $\Phi$ is
increasing (becoming less deep) and so CMB photons gain more energy
than they loose and this creates a hot spot.

As the maps in Fig.~\ref{Dipole} are just the
contributions to $\Delta T$ from a thin, $50~h^{-1}$Mpc, slice of our
simulation, one might worry that this artificial truncation gives rise to
artificial edge effects. We have checked that this is not the case by
making corresponding maps projected through a depth of $1~h^{-1}$Gpc,
the box size of the periodic simulation. We find that the dipole
features are still clearly visible, but, of course, they are somewhat
perturbed by superposition of other perturbations along the longer
line-of-sight.  This finding is consistent with the accumulated
difference between the ISWRS and LAV $\Delta T$ shown earlier in
Fig.~\ref{RayTraceAll} (bottom panel), which remain roughly constant 
for $r_{\rm c}>3000~h^{-1}$Mpc in spite of some small-scale variations.

The amplitude of the temperature fluctuations generated by the dipoles
is not large, but their characteristic dipole signature might enable
them to be detected \citep{Rubino-Martin04,Maturi06,Maturi07b}.
Dipoles generated by moving galaxy clusters, which can equally be
thought of as `moving lenses', were predicted by \citet{Birkinshaw83},
and later discussed by \citet{Gurvits86}.  Their detectability in the
CMB and possible uses, for example to measure transverse motions of
dark matter, have been further explored by \citet{Tuluie95,Tuluie96,
Aghanim98,Molnar00, Aso02, Molnar03, Rubino-Martin04,
Cooray05,Maturi07b}. These studies considered only dipoles on the scale of
galaxy clusters while our simulations reveal dipoles of $\sim \mu$K
amplitude on scales ranging from $10~h^{-1}$Mpc to a few tens of
$h^{-1}$Mpc, the larger of which are seeded by bulk motions on scales
far larger than galaxy clusters.

Moving dark matter lumps will perturb the energies of CMB photons even
if they move along, rather than transverse to the line-of-sight.  Examples
of the perturbations such bulk motions cause are evident at several
points along the rays shown in Fig.~\ref{RayTraceAll}.  They appear as a
sharp peak followed very closely by a deep dip.  The left-hand panels
of Fig.~\ref{RayTraceDetail} show a close-up view of one of these
features.  Following the ray on the top panel from the right to the
left (i.e. moving in the direction of a CMB photon), the line dips and
then peaks, indicating that a lump of dark matter is moving in the opposite
direction to the CMB photon.  The local potential on the lump's
leading part is getting deeper (cooling down photons) while on the
trailing edge the potential is becoming shallower (heating up
photons). The accumulated $\Delta T$ (middle panel) is boosted and
then suppressed as the ray passes through and the net effect is extremely
small. This is to be contrasted to the case discussed previously when
the lump moves transverse to the line-of-sight for which there is
no cancellation.

\subsection{Convergent Flows}
\label{sec:cflow}

In the RS and the higher redshift ISWRS maps of Fig.~\ref{FullBox100}
we can see several examples of cold regions surrounded by hot rings or
filaments which are centred on converging velocity flows. These
features are larger in scale than the dipoles, ranging from
tens to hundreds of Megaparsecs.  Fig.~\ref{ColdInHot} shows zoomed-in maps of
the ISWRS, LAV and RS temperature perturbations and overlaid velocity
vectors at redshifts $z=0$, $2.1$ and~$6.2$ for a $200~h^{-1}$Mpc box
centred on one such feature selected from Fig.~\ref{FullBox100}.  In
the LAV maps (middle row), we see only a smooth hot lump, centred on
the convergent flow, whose amplitude increases towards low redshift as
$\Omega_\Lambda$ becomes increasingly important.  In contrast, the RS
contribution to the temperature perturbations (bottom row) is negative
at the centre of the convergent flow and is surrounded by a positive
shell. Its amplitude evolves with redshift much more weakly than the
amplitude of the LAV perturbation. It is strongest in the central
panel, decreasing at the lowest redshift due to the damping of
non-linear growth by the late time acceleration of the
universe. At all but the lowest redshift, this RS perturbation is a
significant contribution to the overall ISWRS maps (top row). At high
redshift, the RS completely dominates over the LAV contribution turning
the hot spot predicted by linear theory into a cold spot, almost
$200~h^{-1}$Mpc in extent, with amplitude of order a $\mu$K, surrounded
by a hot shell.  Even in the redshift $z=2.1$ slice, the RS
contribution drastically changes the morphology of the LAV hotspot,
producing a cold region in the very centre of the flow that is
surrounded a hot filamentary shell.

The explanation of this counter-intuitive conclusion that overdense
regions can become cold spots surrounded by hot-rings, rather than the
hotspots predicted by the ISW, involves principally the same physics as
is responsible for the dipoles.  In the $\Lambda$CDM model, overdense
regions grow as the result of the inflow of lumpy material, often
along filaments. Each of these inflowing lumps will give rise to a
dipole feature. On the leading edge of the lump, the potential is
decreasing and CMB photons loose energy, while on the trailing edge
the potential is increasing and CMB photons are heated.  The only
difference is that dipoles seen on the sky are indicators of lumps of
dark matter with large transverse momentum, while these larger scale
cold regions surrounded by hot rings are produced by larger scale
coherent convergent flows.  One can imagine splitting the convergence
flow into many small lumps of dark matter moving towards the same
centre: the cold region in the centre is just the result of stacking
many leading parts of those lumps, while the hot ring consists
of their trailing parts. Our findings regarding the morphology of the
RS effect in converging flows are in general agreement with analytic
models of forming clusters that have been discussed by other authors
\citep[e.g.][]{Martinez-Gonzalez90b, Lasenby99, Dabrowski99, Inoue06,
Inoue07}.

The maps shown in Fig.~\ref{ColdInHot} are the contribution to the
$\Delta T$ perturbation of just a $100~h^{-1}$Mpc slice of our
simulation. Again, one should consider whether this truncation gives
rise to artificial edge effects that would qualitatively affect the
appearance of the cold spots and hot rings.  In particular, if the slice
artificially removes a foreground or background section of the hot
shell this will enhance the visibility of the cold central region. We
have directly tested whether this is a strong effect by making
untruncated maps whose depth is the full $1000~h^{-1}$Mpc size of the
periodic simulation.  We find that while there are some cases where
the heating and cooling cancel each other out, normally the visibility
and contrast of the cold spot and hot ring features is not strongly
affected. This is perhaps to be expected as the hot shell is diluted
because it is spread over a much larger area than is subtended by the cold
spot. Moreover, the dominance of the cold spots can be seen directly
in the plots of the $\Delta T$ contributions along particular lines of
sight. An example of a ray passing through a cold spot is shown by the
middle column of Fig.~\ref{RayTraceDetail}.  In the top panel the
broad peak predicted by the LAV (dashed line) corresponds to an
overdense region in the simulation. There is a convergent flow around
this overdense region that gives rise to both the sharp central dip
and surrounding upward fluctuations seen in the ISWRS result (solid
line). The lower panels show that the difference between ISWRS and LAV
perturbations is strongly dominated by the central cold spot.

\subsection{Divergent Flows}
\label{sec:dflow}

Divergent flows surrounding voids or underdense regions also produce
characteristic features in the ISWRS and RS maps. However their effect
is not simply the reverse of that of the convergent flows. Instead,
the non-linear behaviour in these void regions always acts to enhance
the LAV perturbation producing stronger cold spots.
Fig.~\ref{ColdInVoid} shows zoomed-in ISWRS, LAV and RS temperature maps
of a $300~h^{-1}$Mpc region centred on such a cold spot.
This region is underdense and the LAV maps (middle row) show the
expected linear ISW behaviour of a cold spot which grows in amplitude
with decreasing redshift as $\Omega_\Lambda$ increases and the
potential hill associated with the underdensity decays.
The RS contribution (bottom row) also produces a cold spot
at the centre of the region, but surrounded by a hot filamentary
shell. Thus, perhaps counter-intuitively, the pattern of the RS contribution
for these divergent flows is the same, and has the same sign, as for
the convergent flows discussed above.

The reason why the effect of non-linearity in an underdense region
is not simply the reverse of its effect in an overdense region
is because the two situations are not symmetrical.
In an overdense region the overdensity
is unbounded while
in a underdense region the density cannot drop below zero, so
$\delta>-1$. It is this saturation of the density contrast in voids
that prevents the perturbation from growing in the non-linear regime even
as fast as linear theory would
predict and so enhances the rate of decay of $\Phi$.

Again, the amplitude of the
RS contribution evolves with redshift much more weakly than the
LAV contribution. At $z=0$, the LAV contribution to the ISWRS is
overwhelmingly dominant, but at higher redshift, the RS effect
contributes by reinforcing the linear effect in the centre of
the cold spot and suppressing it in the outer regions.

% Cartoons depicting LAV and RS in various phenomena
\begin{figure*}
\begin{center}
\resizebox{\hsize}{!}{
\includegraphics[angle=270]{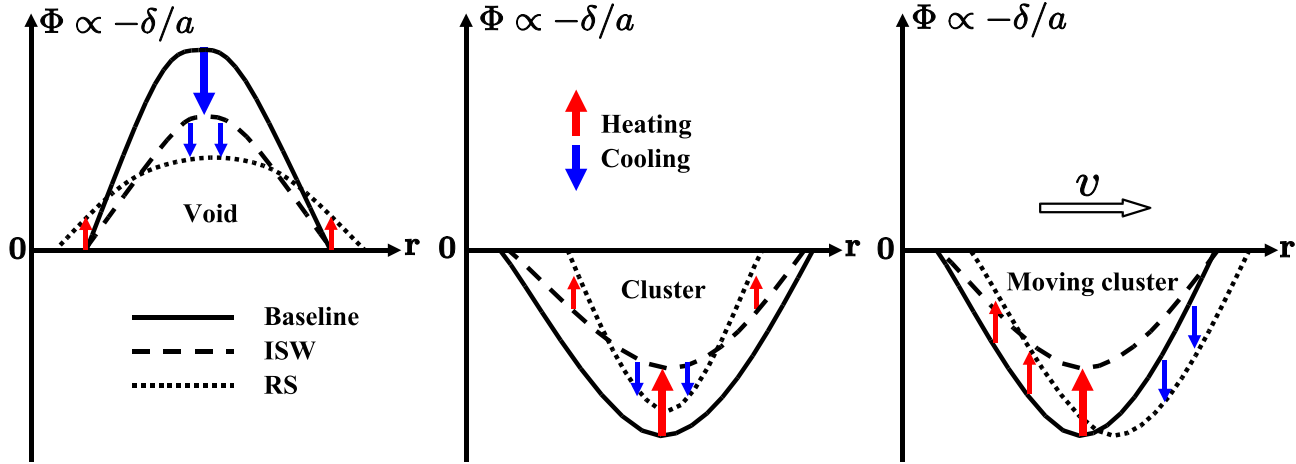}}
\caption{\label{Cartoon} Schematic diagrams of the evolution of the
gravitational potentials that produce the three characteristic
features of the non-linear RS effect discussed in Section~\ref{slices}. The
solid lines depict for a void (left), cluster (middle) and moving
cluster (right) the gravitational potential at a reference epoch, say
when a CMB photon enters the potential. The dashed and dotted lines
show predictions for the evolved potential at a later time, say
when the CMB photon exits the region.  The dashed lines are the linear
prediction of the LAV approximation and the dotted line represents the
fully non-linear result.  The heavy up and down arrows indicate the
heating or cooling of the CMB that is predicted by the LAV
approximation (essentially the linear ISW effect). The lighter arrows
indicate the RS contribution to the temperature perturbations,
i.e. the difference between the fully non-linear and LAV
contributions. }
\end{center}
\end{figure*}

\subsection{Overview of non-linearity}

A useful summary of the three physical effects we have discussed above
is given by the schematic diagrams in Fig.~\ref{Cartoon}. They
depict the gravitational potential, $\Phi \propto -\delta/a$, at a
baseline reference epoch, say when the CMB photon enters the region,
and both the LAV and fully non-linear predictions for the evolved
potential at a later time, say when the CMB photon exits the
region. These schematics yield  straightforward interpretations of
each of the phenomena we have identified in the ISWRS and RS maps.

In void regions, the ISW effect is easily understood as the result
of the linear decay of the gravitational potential. CMB photons are
cooled as they loose more energy climbing the initial potential hill
of the void region than they subsequently regain on departing the
now shallower potential hill. This cooling is depicted by the heavy
arrows in the left-hand panel of Fig.~\ref{Cartoon}.
One would expect non-linearity to accelerate the growth
of $\delta$ and hence increase the rate of growth of
the potential hill. This is what we observe to occur at the edge of
the voids and leads to a component of heating of the CMB photons
in this outer shell, as depicted by the light upward arrows.
However, once the centre of a void region becomes almost empty, i.e. $\delta
\approx -1$, the local density contrast stops growing, as it cannot
become emptier. In this case, the expansion of the
universe will reduce the height of the potential hill, just as it does in the
linear regime in voids. Thus, in the centre of voids the RS effect has
the same sign as the linear ISW effect and the two reinforce each
other to produce cold spots at the centre of voids.

 In overdense regions such as galaxy clusters, the linear decay of the
gravitational potential well results in the heating of CMB photons as
depicted by the heavy arrows in the central panel of
Fig.~\ref{Cartoon}.  The effect of non-linear infall and growth of the
density perturbation is to shrink the scale of the potential well while
deepening its central value. In the centre of the cluster this
non-linear growth acts in opposition to the linear decay of the
potential and cools the CMB photons, depicted by the downward light
arrow. We have seen that at high redshift this effect can be stronger
than the linear ISW effect, resulting in cold spots centred on
overdense regions.  At larger scales, the inflowing region shrinks the
scale of the potential well and results in a further reduction in the depth
of the potential well, reinforcing the linear effect. Hence, in an
outer shell around the cluster, the non-linear effect heats CMB
photons.  This latter effect can also be considered as the
superposition of a set of dipoles arising from lumps in a surrounding
inflowing region.

The right-hand panel of Fig.~\ref{Cartoon} depicts the situation
that gives rise to a characteristic dipole perturbation of the CMB.
A moving cluster, or other overdense region, gives rise to a moving
potential well. Linear theory does not model the movement of the
potential well and would instead predict a simple static decaying
potential well, resulting in a hot spot as indicated by the heavy
upward arrows. In contrast, the movement of the potential well leads
to a rapid deepening of the potential well on the leading edge of the
cluster and a corresponding rapid increase of the potential on the
trailing edge. This gives rise to cooling of the CMB photons on the
leading edge and heating on the trailing edge as indicated by the
light arrows.

\section{Sky maps}
\label{SkyMap}

\begin{figure*}
\begin{center}
\resizebox{\hsize}{!}{
\includegraphics[angle=90]{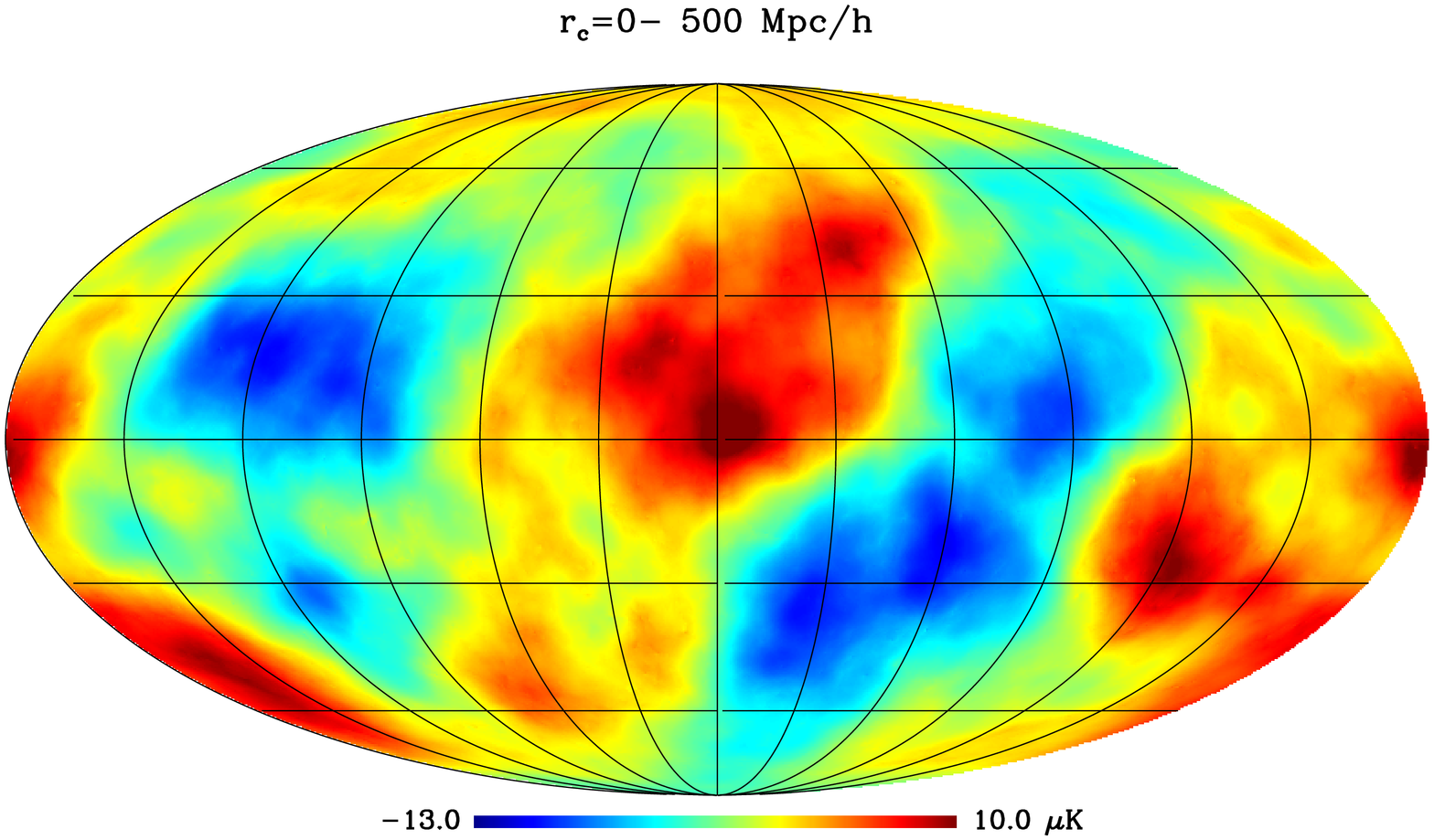}}
\resizebox{\hsize}{!}{
\includegraphics[angle=90]{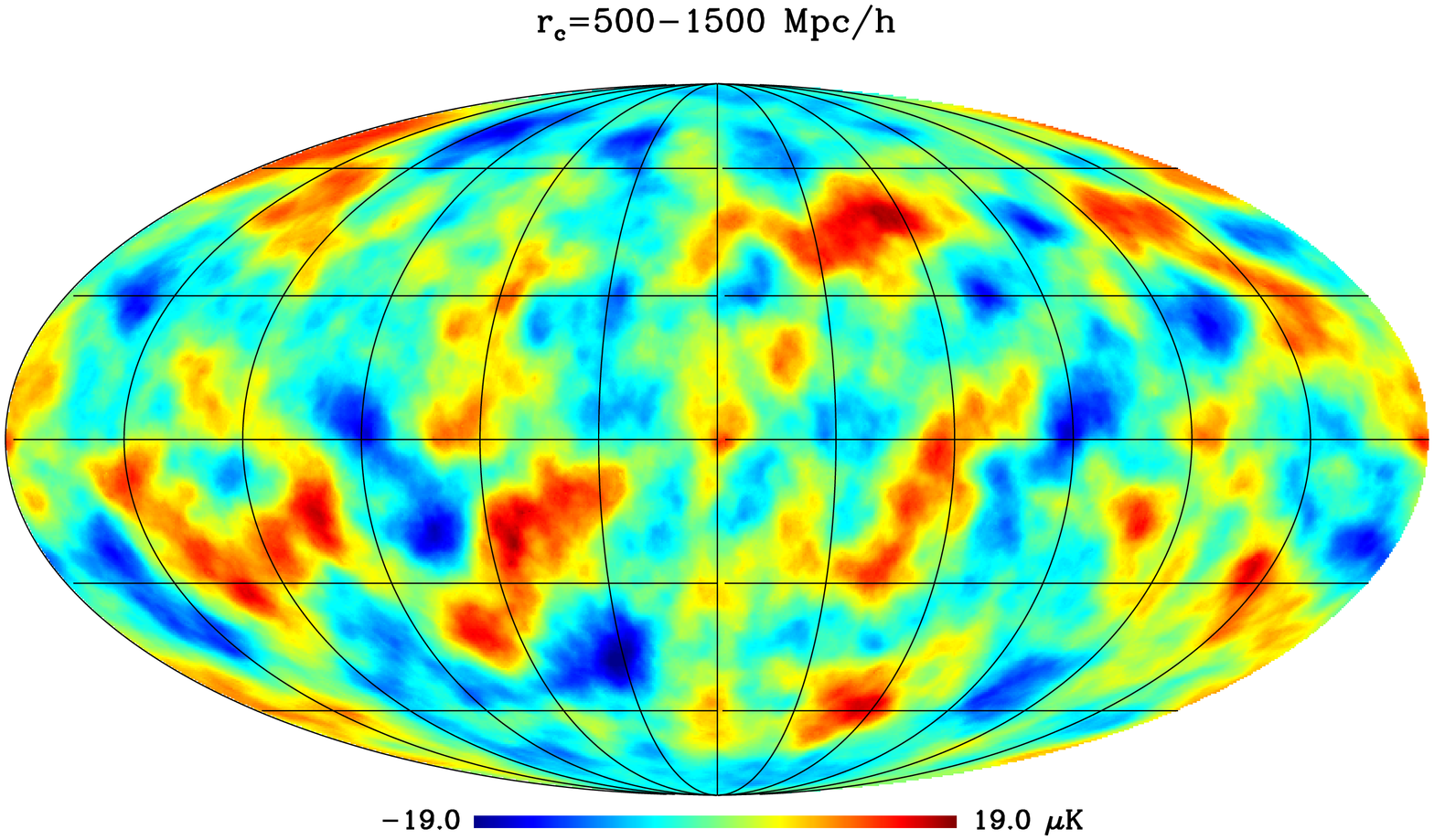}}
\caption{\label{SkyMap_ISWRS}
Full sky maps of the predicted CMB
$\Delta T$ due to the ISW and Rees-Sciama effects (ISWRS) made from
our simulation.  The upper plot is obtained by ray-tracing through the
simulation over the redshift interval $0<z<0.17$, which corresponds
to a range of comoving distance from the observer of 0 -
500 $h^{-1}$~Mpc.  In the lower plot the projection is for the range
$0.17 <z <0.57$ (corresponding to the range 500 - 1500~$h^{-1}$~Mpc in
comoving distance).  These maps, and all subsequent sky maps, use the
Mollweide projection to represent the sky on a plane, with each
pixel having an area of $(6.87')^2$.  The grid spacing is 30$^{\circ}$
in both longitude and latitude. For reference, the box-size of our
simulation is 1000~$h^{-1}$~Mpc.  }
\end{center}
\end{figure*}

\begin{figure*}
\begin{center}
\resizebox{\hsize}{!}{
\includegraphics[angle=90]{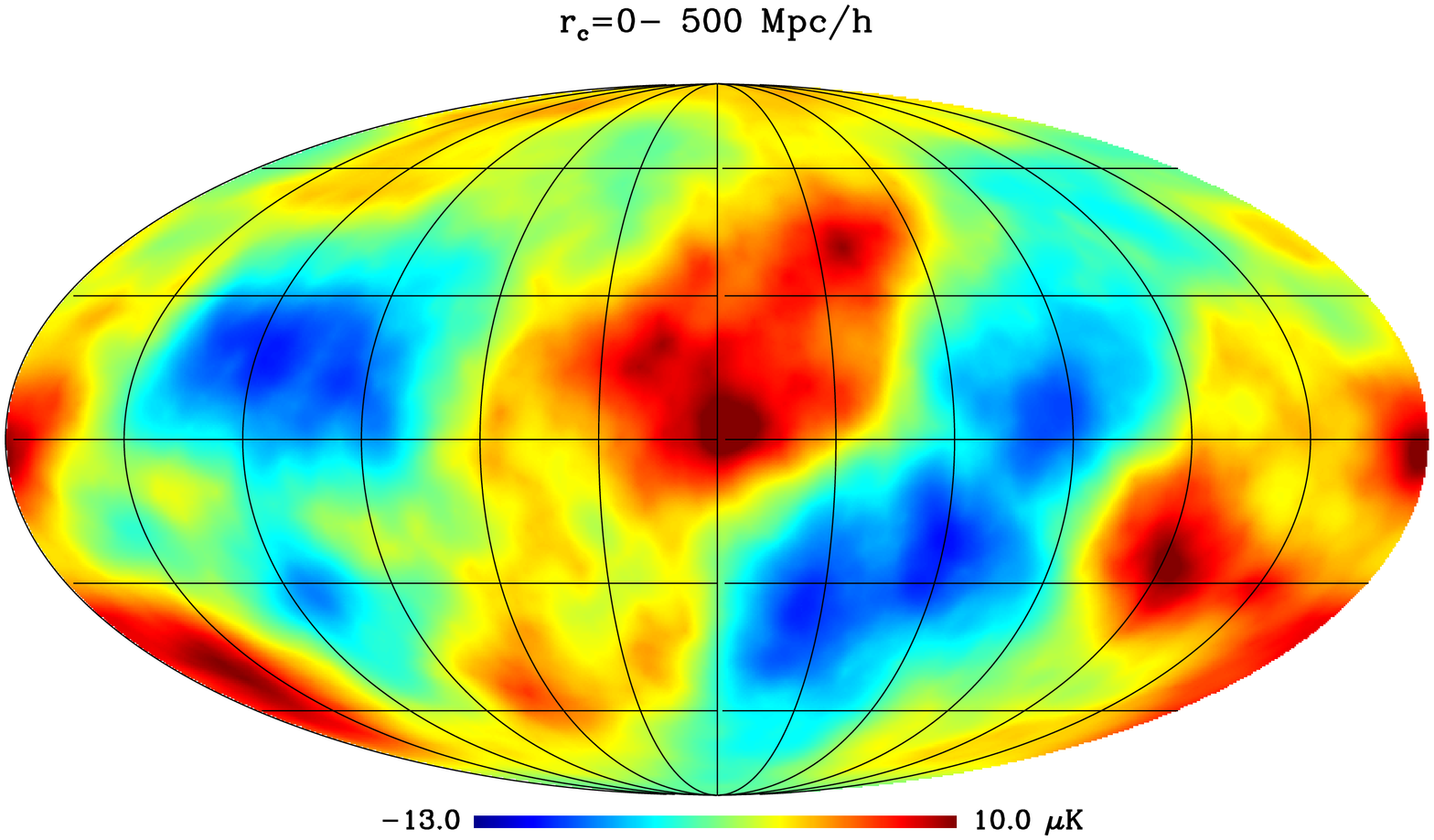}}
\resizebox{\hsize}{!}{
\includegraphics[angle=90]{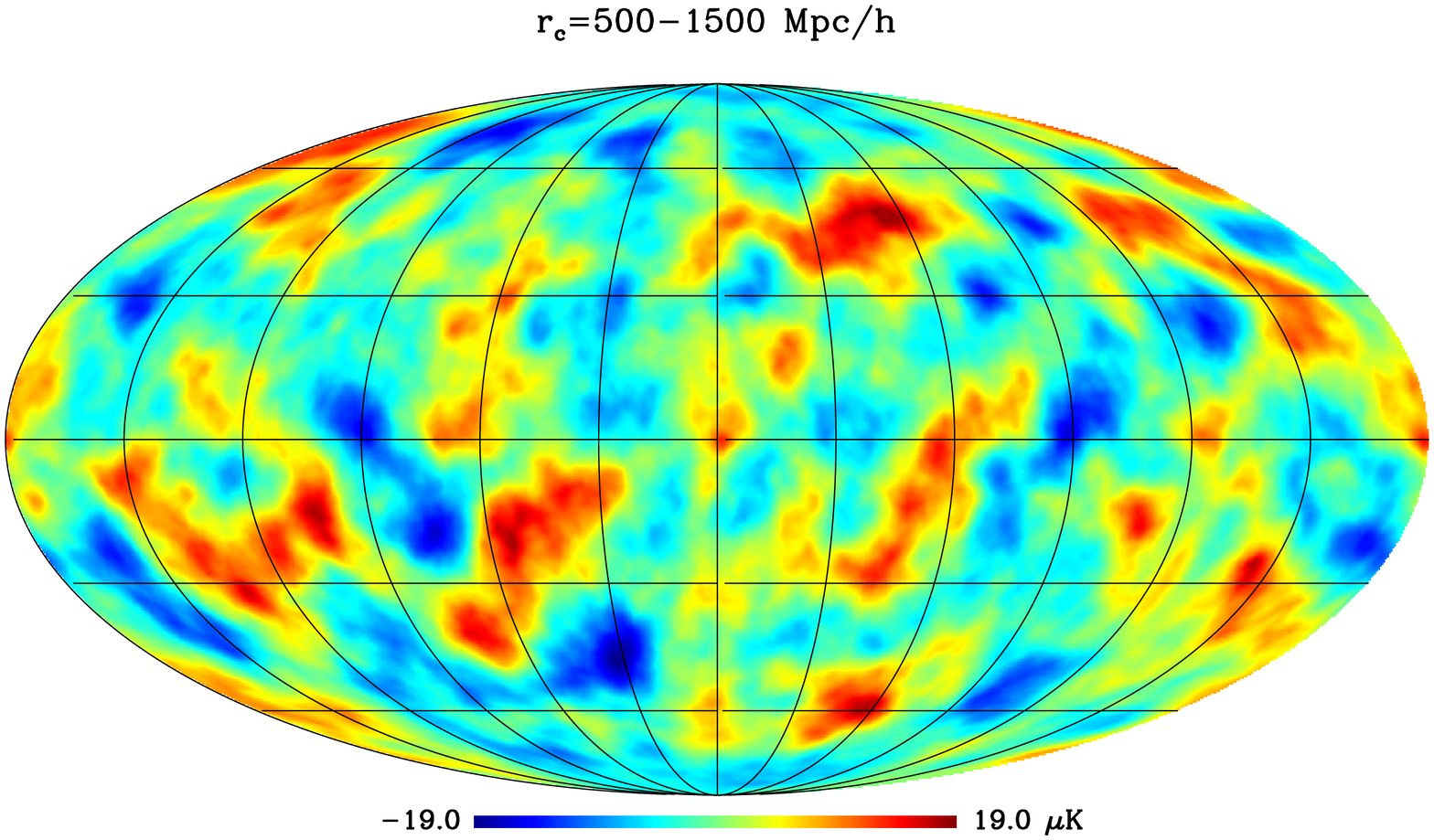}}
\caption{\label{SkyMap_LAV} As Fig.~\ref{SkyMap_ISWRS}, but showing
the full sky maps of the predicted CMB $\Delta T$ due to the ISW
effect constructed using the linear approximation for the velocity
field
(LAV).}
\end{center}
\end{figure*}

\begin{figure*}
\begin{center}
\advance\leftskip 1.7cm
\resizebox{\hsize}{!}{
\includegraphics[angle=90]{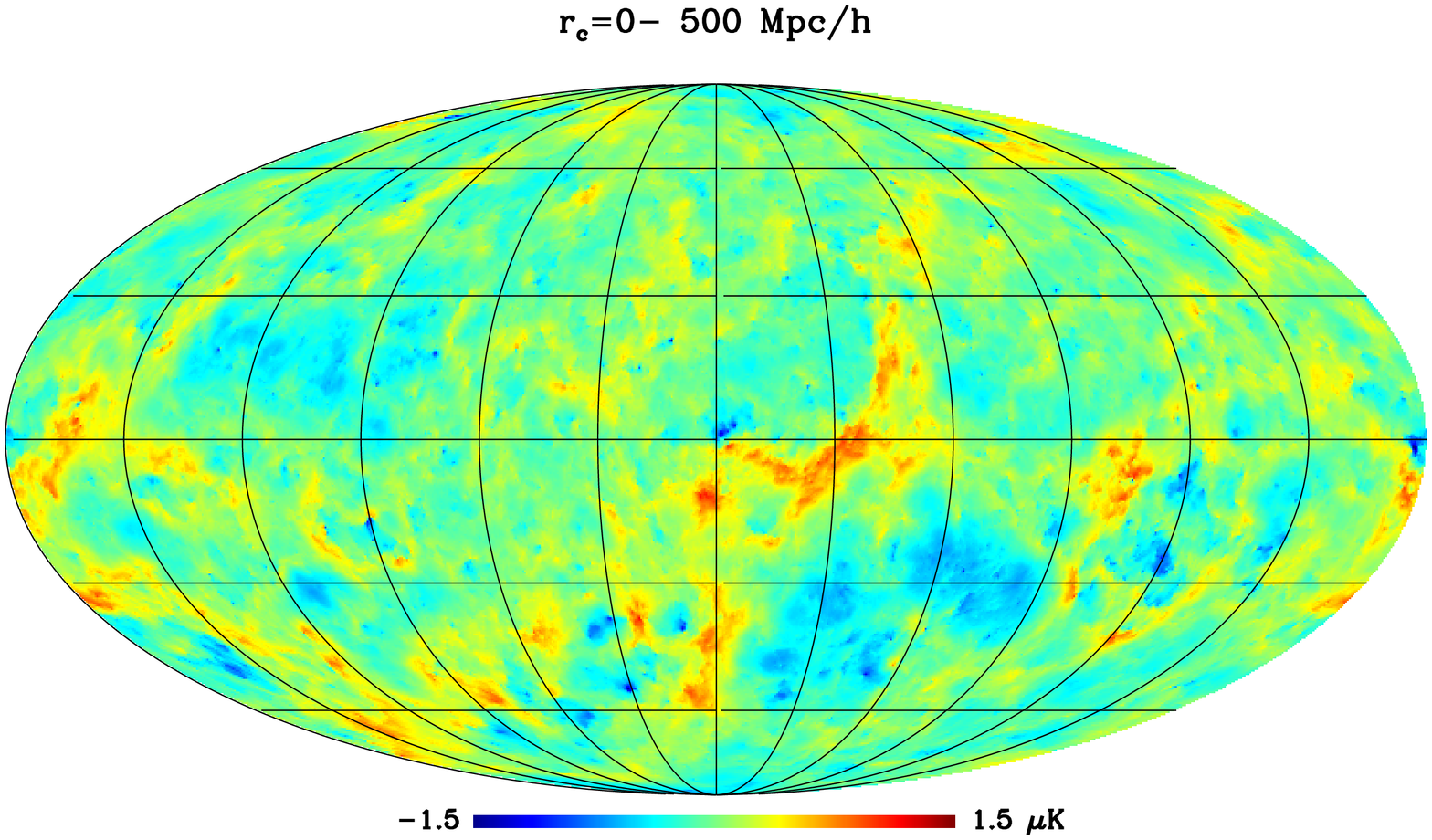}}
\resizebox{\hsize}{!}{
\includegraphics[angle=90]{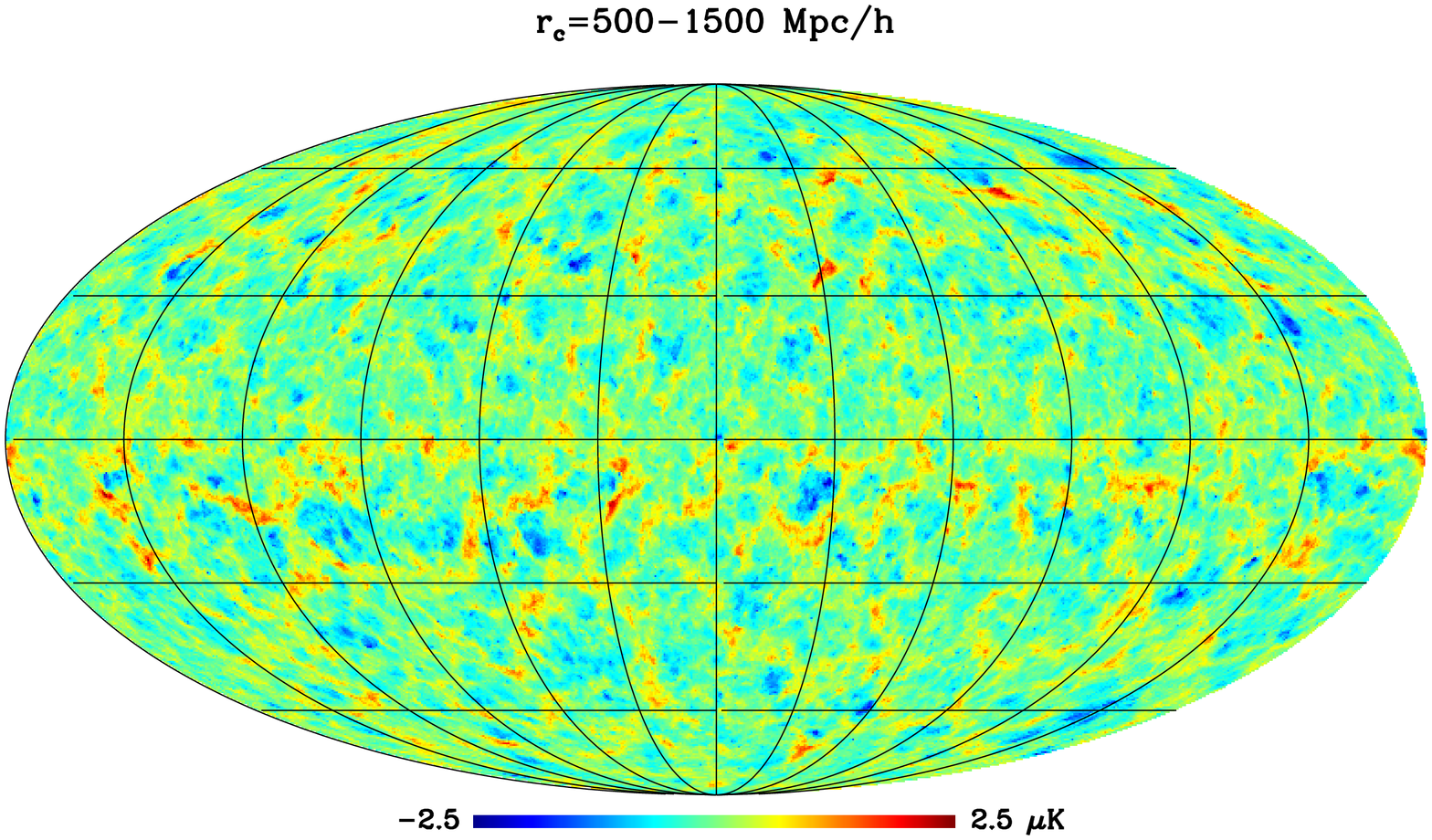}}
\caption{\label{SkyMap_RS} As Fig.~\ref{SkyMap_ISWRS}, but for the
Rees-Sciama (RS) effect, obtained by subtracting
the maps in Fig.~\ref{SkyMap_LAV} from the corresponding maps in Fig.~\ref{SkyMap_ISWRS}.}
\end{center}
\end{figure*}

\begin{figure*}
\begin{center}
\resizebox{\hsize}{!}{
\includegraphics{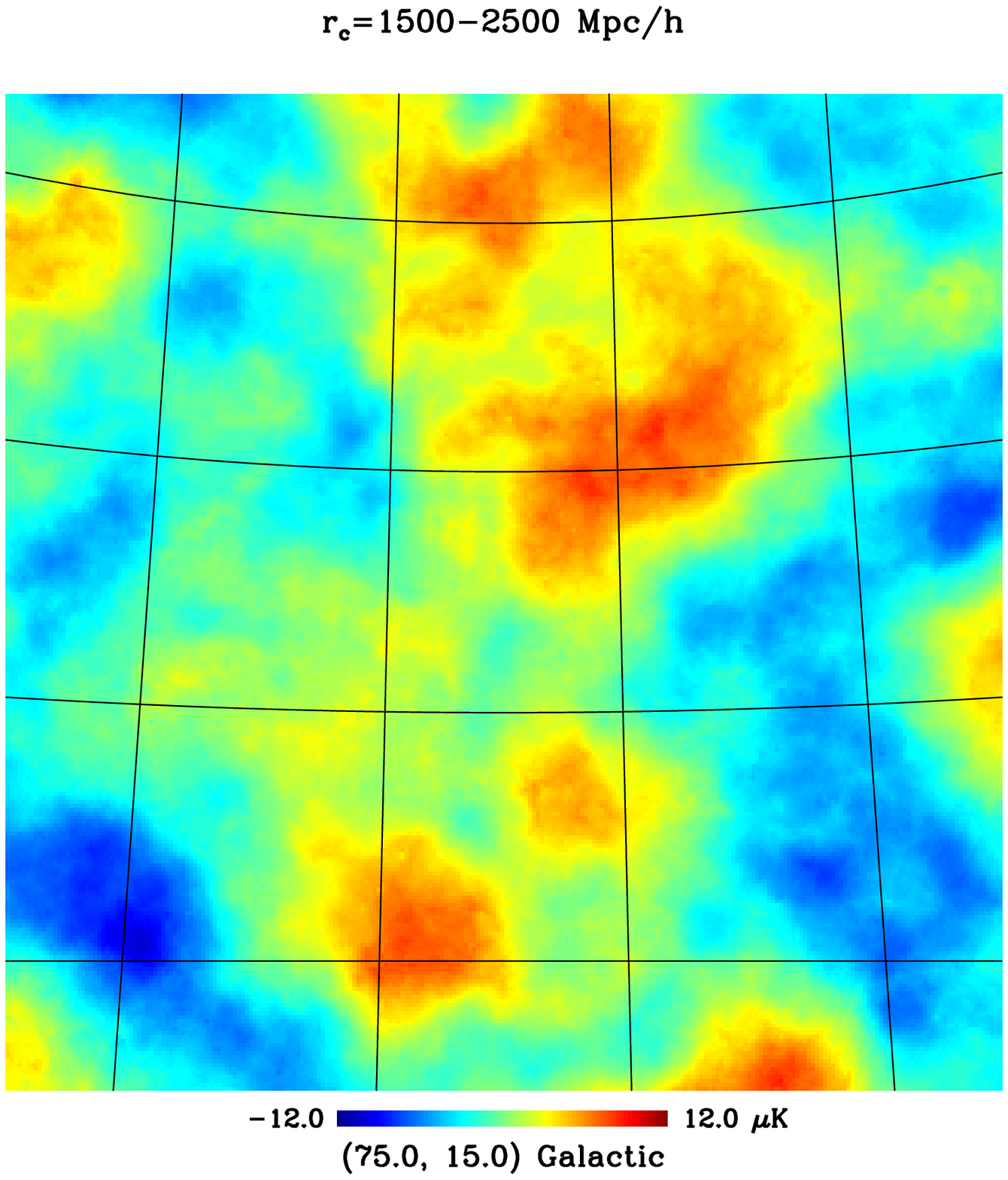}
\includegraphics{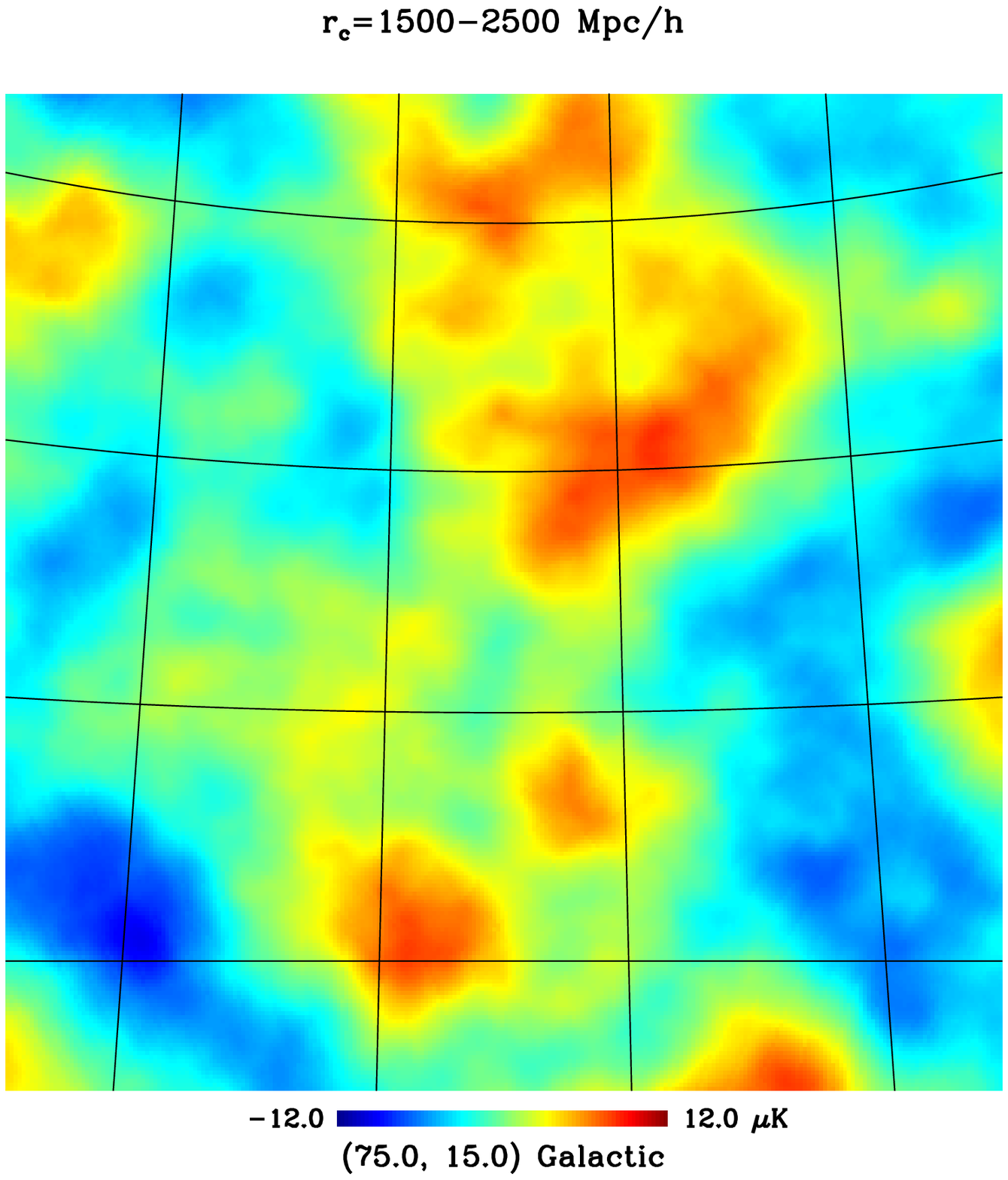}
\includegraphics{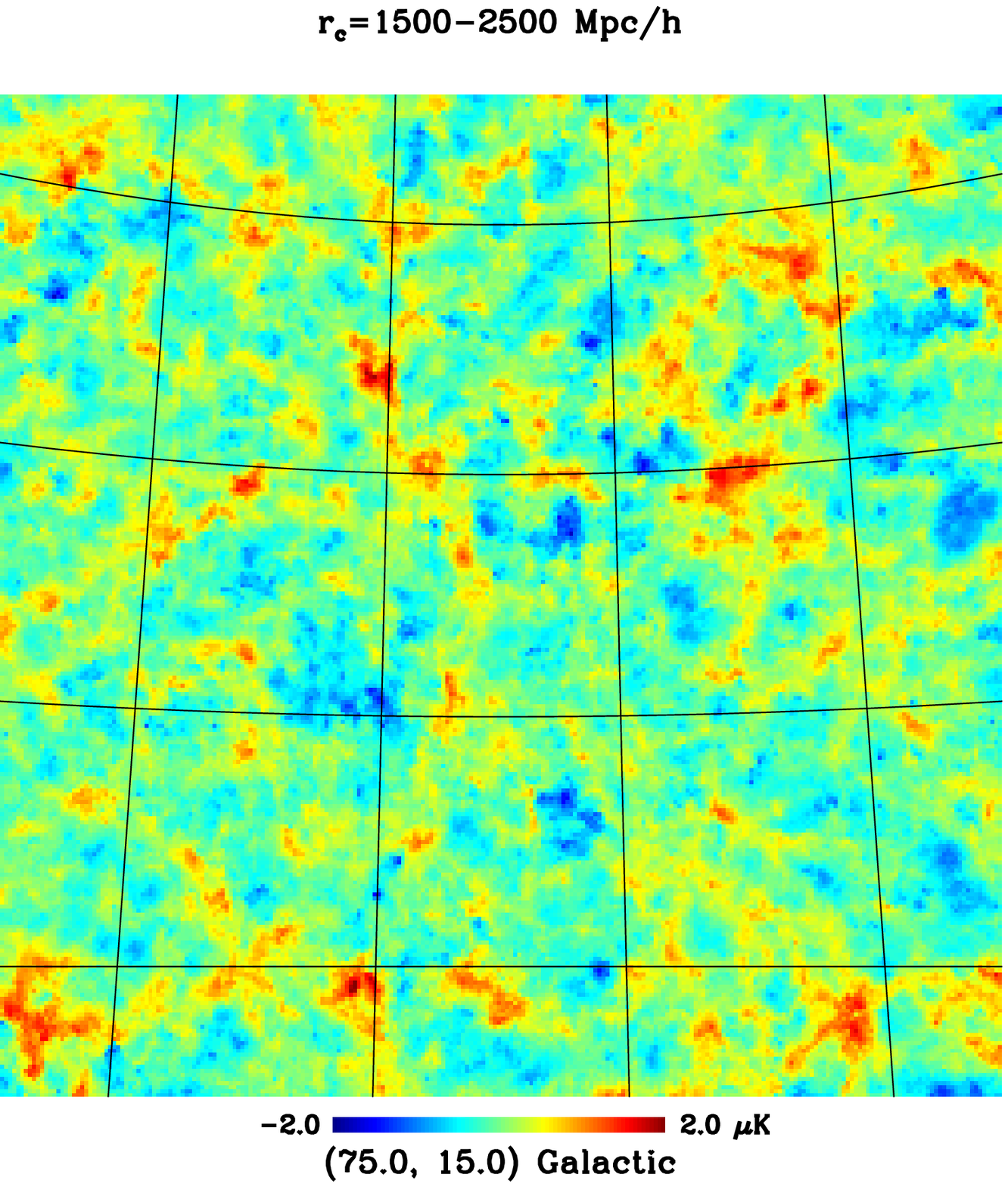}}
\caption{\label{SkyMap1500_2500} Maps of $40\times 40$~degree patches
of the sky of, from left to right: (i)
the ISW and Rees-Sciama effects (ISWRS); (ii) the linear approximation
for the velocity field (LAV); (iii) the Rees-Sciama (RS) effect.  The
projection is over the range $0.57<z<1.07$, corresponding to 1500 -
2500~$h^{-1}$~Mpc comoving distance from the observer.  The grid
spacing is 10$^{\circ}$ in both longitude and latitude.}
\end{center}
\end{figure*}

\begin{figure*}
\begin{center}
\resizebox{\hsize}{!}{
\includegraphics{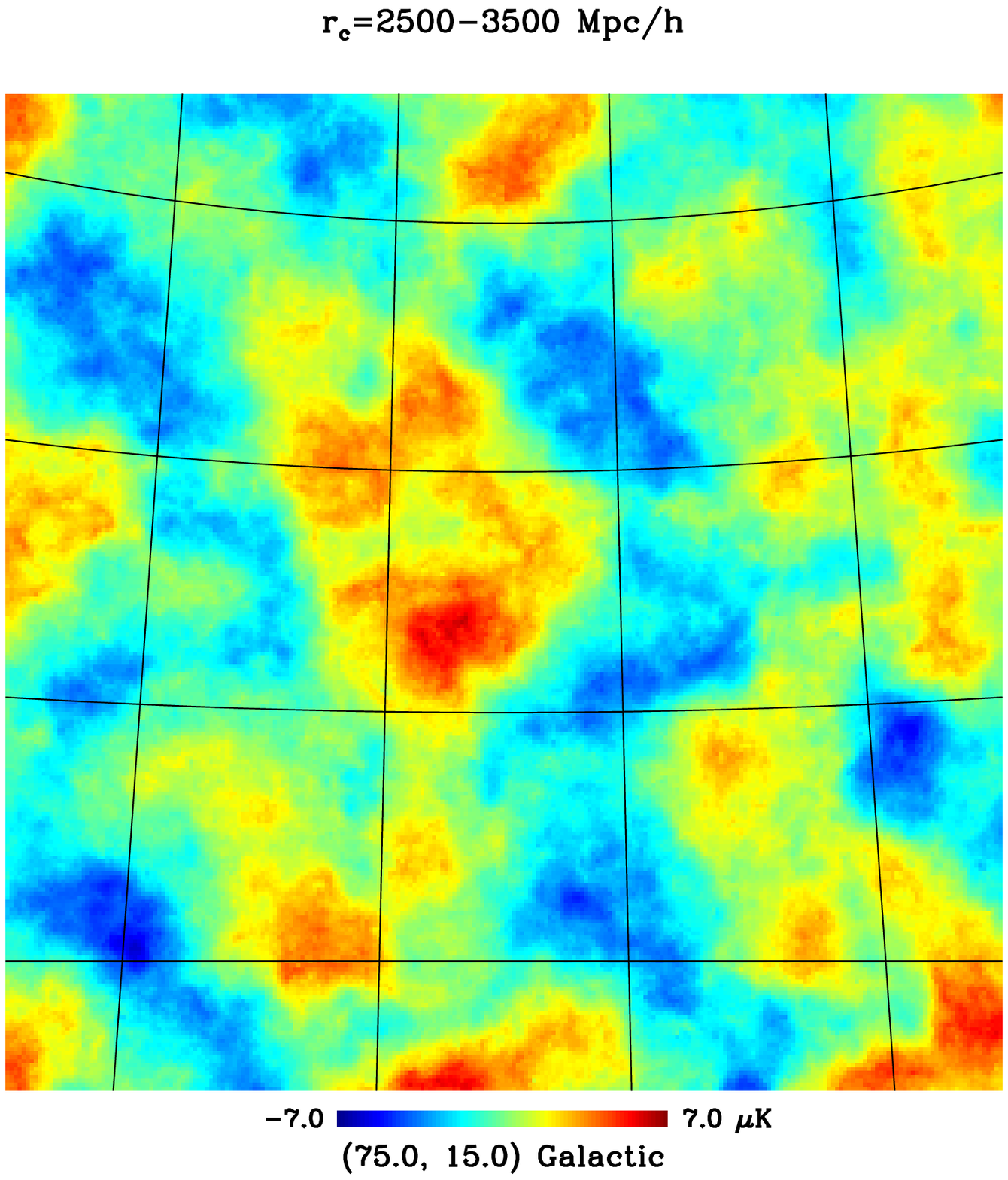}
\includegraphics{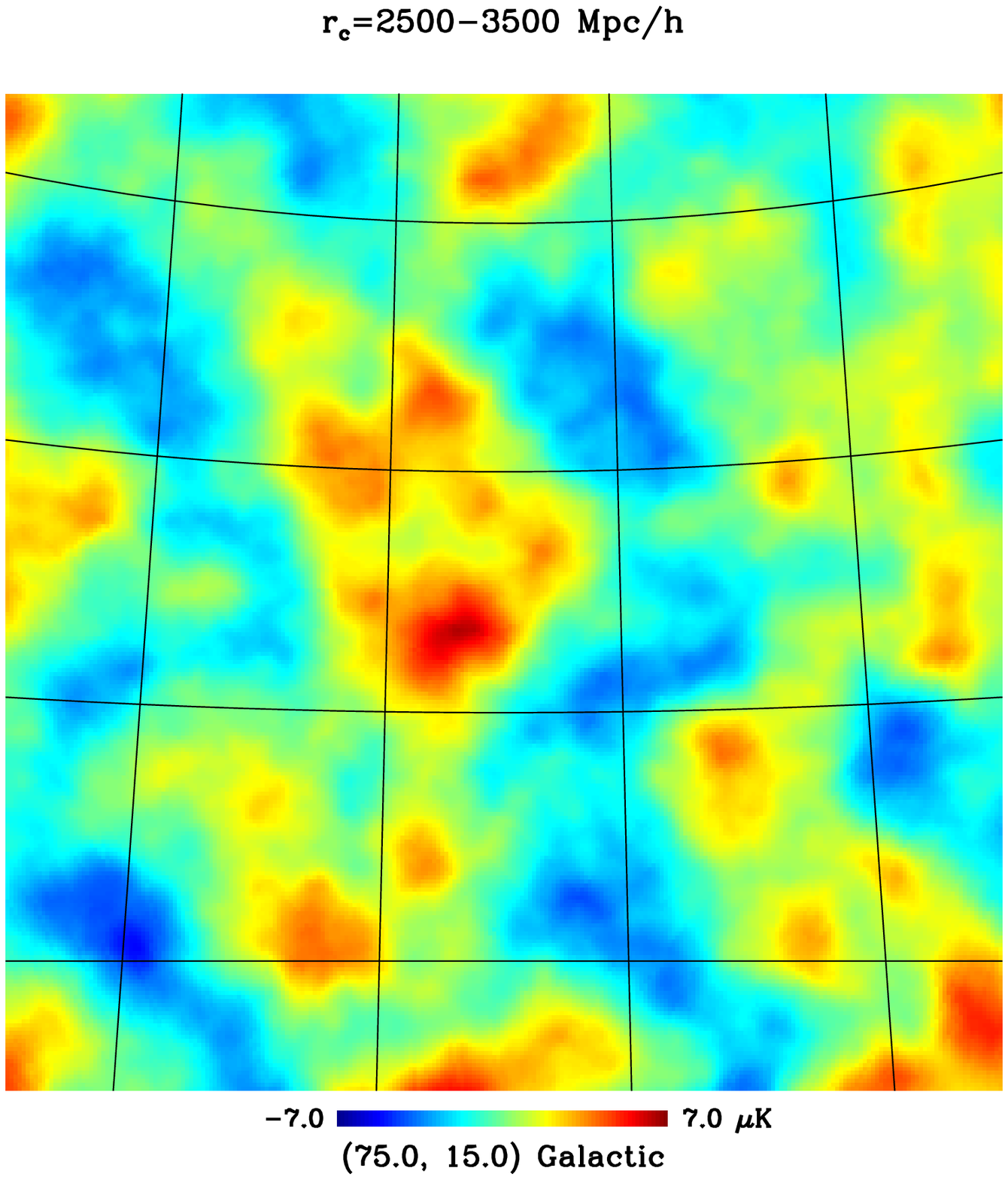}
\includegraphics{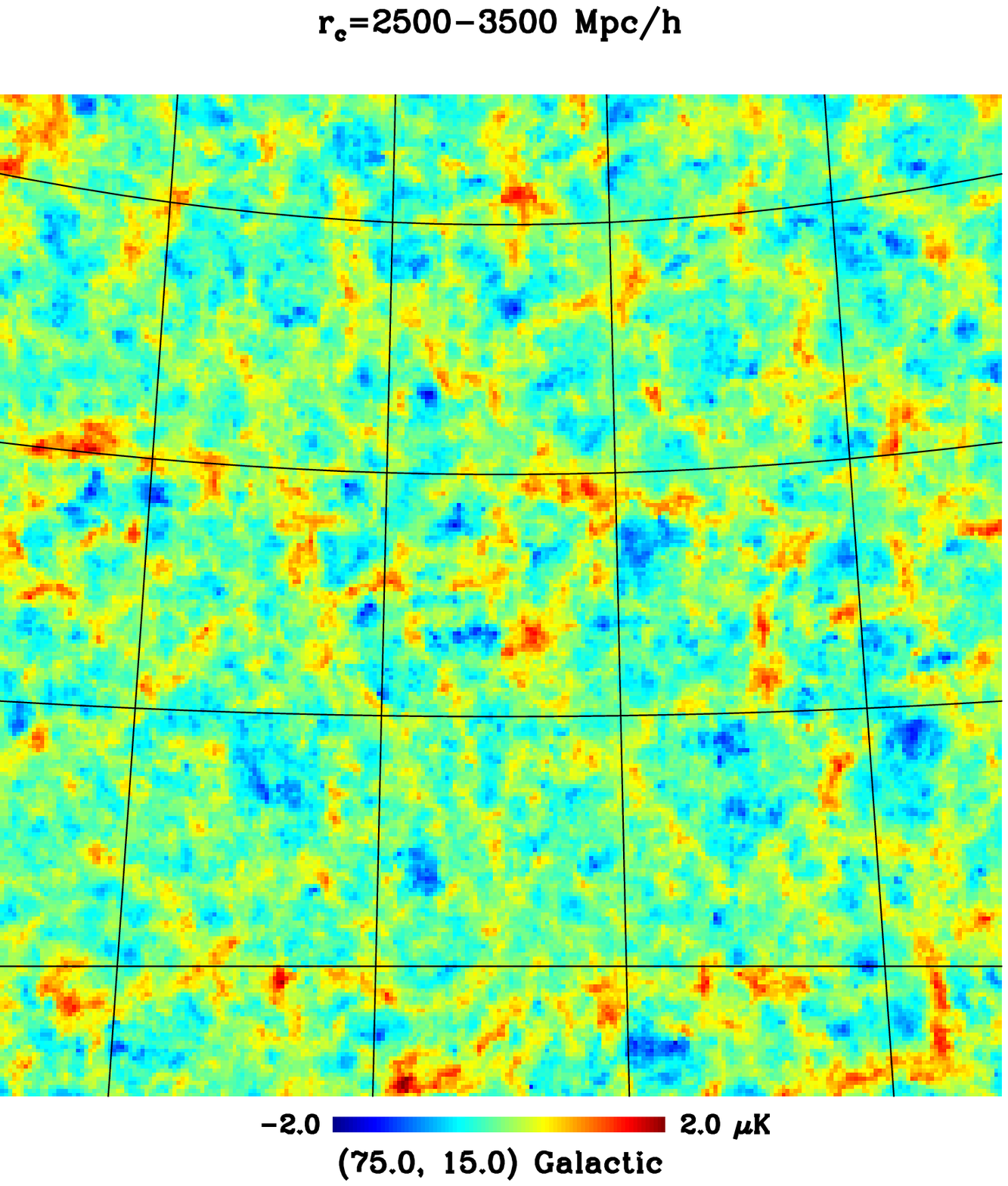}}
\caption{\label{SkyMap2500_3500} As
Fig.~\ref{SkyMap1500_2500}, but projecting over the range
$1.07<z<1.78$, corresponding to a comoving distance range of
2500 - 3500~$h^{-1}$~Mpc. Note that since the simulation volume is only
1000~$h^{-1}$~Mpc on a side, the same features necessarily appear, but at
smaller angular scales than in the previous figure.}
\end{center}
\end{figure*}

\begin{figure*}
\begin{center}
\resizebox{\hsize}{!}{
\includegraphics{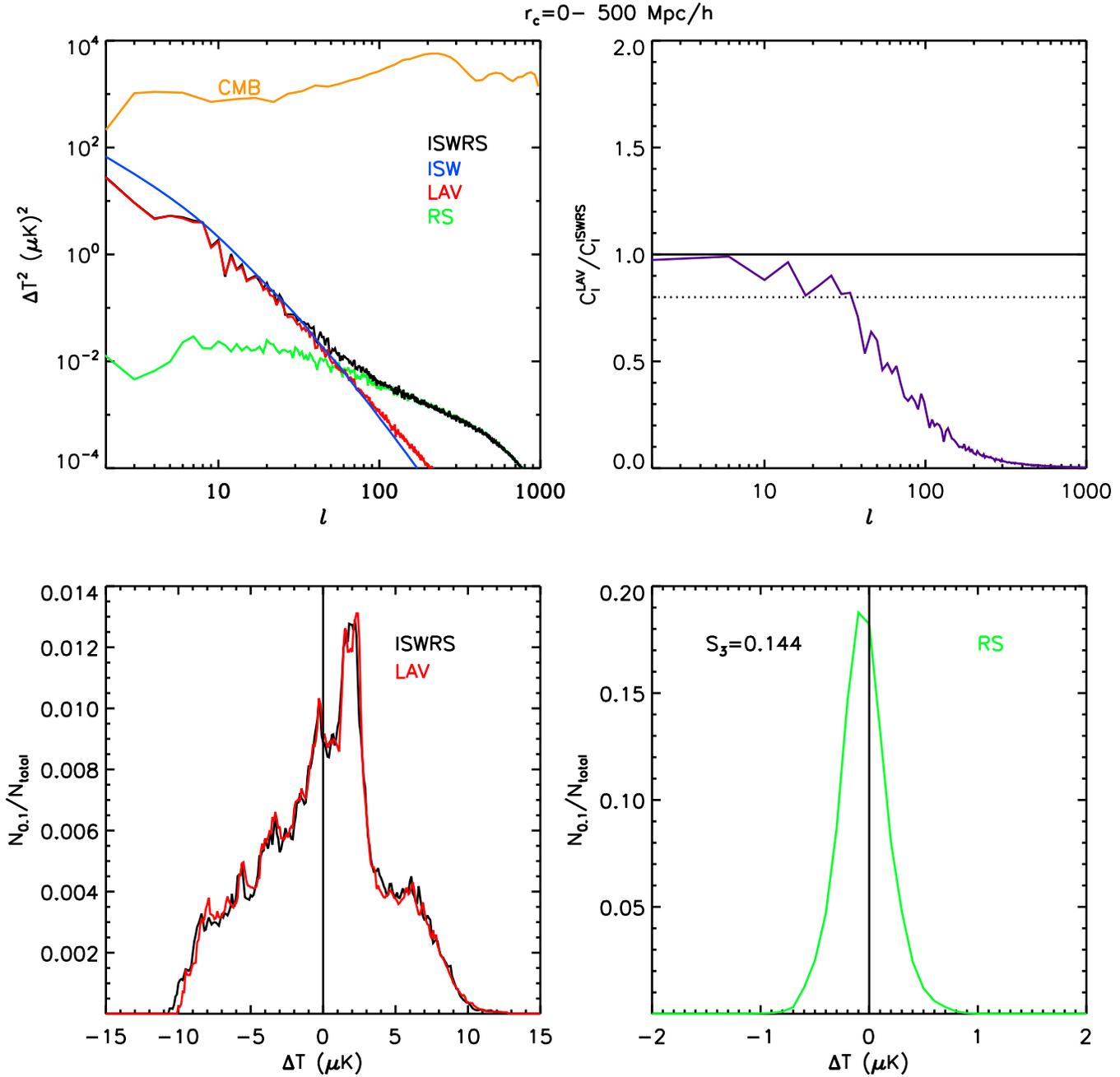}}
\caption{\label{PS0_500} Statistics of the ISWRS, LAV and RS maps
shown at the top panel ($0<z<0.17)$ of
Fig.~\ref{SkyMap_ISWRS}, Fig.~\ref{SkyMap_LAV} and
Fig.~\ref{SkyMap_RS}. Top-left: angular power spectra. The orange
line at the top is the WMAP5 measurement of the CMB temperature
power spectrum.  The ISWRS power
spectrum is shown in black line. The red line is the power spectrum of
the map constructed by applying the linear approximation to the
velocity field (LAV), i.e. equation~(\ref{eq3}). The blue line is the
linear theory prediction for the same redshift interval. Top-right:
the ratio of the LAV and ISWRS in the top-left panel.  Bottom-left:
the histogram of pixel temperatures of the ISWRS (black) and LAV
(red) maps. The number of pixels within each bin of $0.1~\mu$K has
been divided
by the total number of pixels. Bottom-right: the histogram of
pixel temperatures of the residual map (RS). $S_3$ is the skewness
of the map temperature.}
\end{center}
\end{figure*}

\begin{figure*}
\begin{center}
\resizebox{\hsize}{!}{
\includegraphics{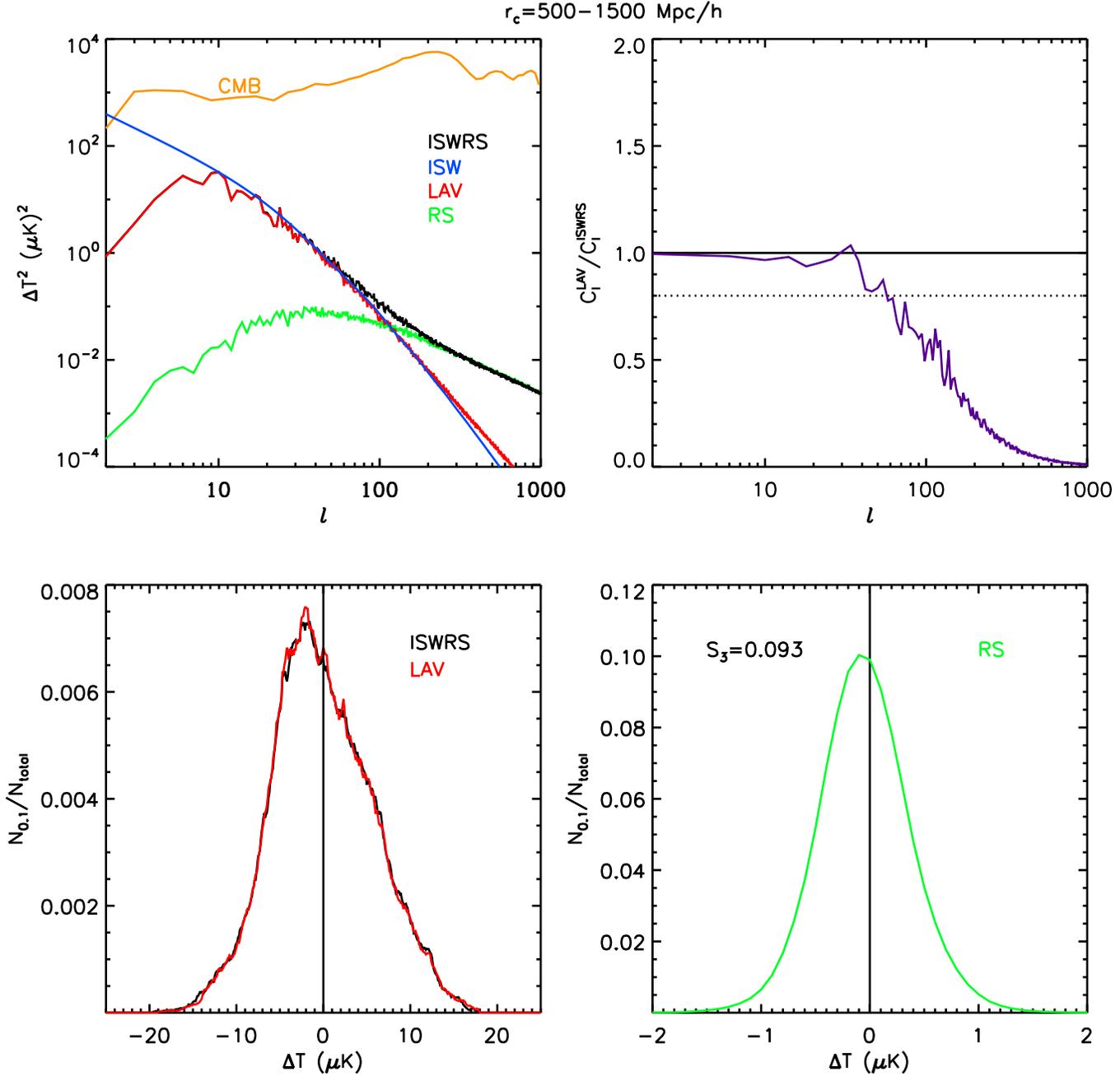}}
\caption{\label{PS500_1500} As Fig.~\ref{PS0_500}, but showing
the statistics for the maps in the bottom panel of
Figs~\ref{SkyMap_ISWRS}, \ref{SkyMap_LAV} and \ref{SkyMap_RS}, projecting
over the range $0.17<z<0.57$.
}
\end{center}
\end{figure*}

\begin{figure*}
\begin{center}
\resizebox{\hsize}{!}{
\includegraphics{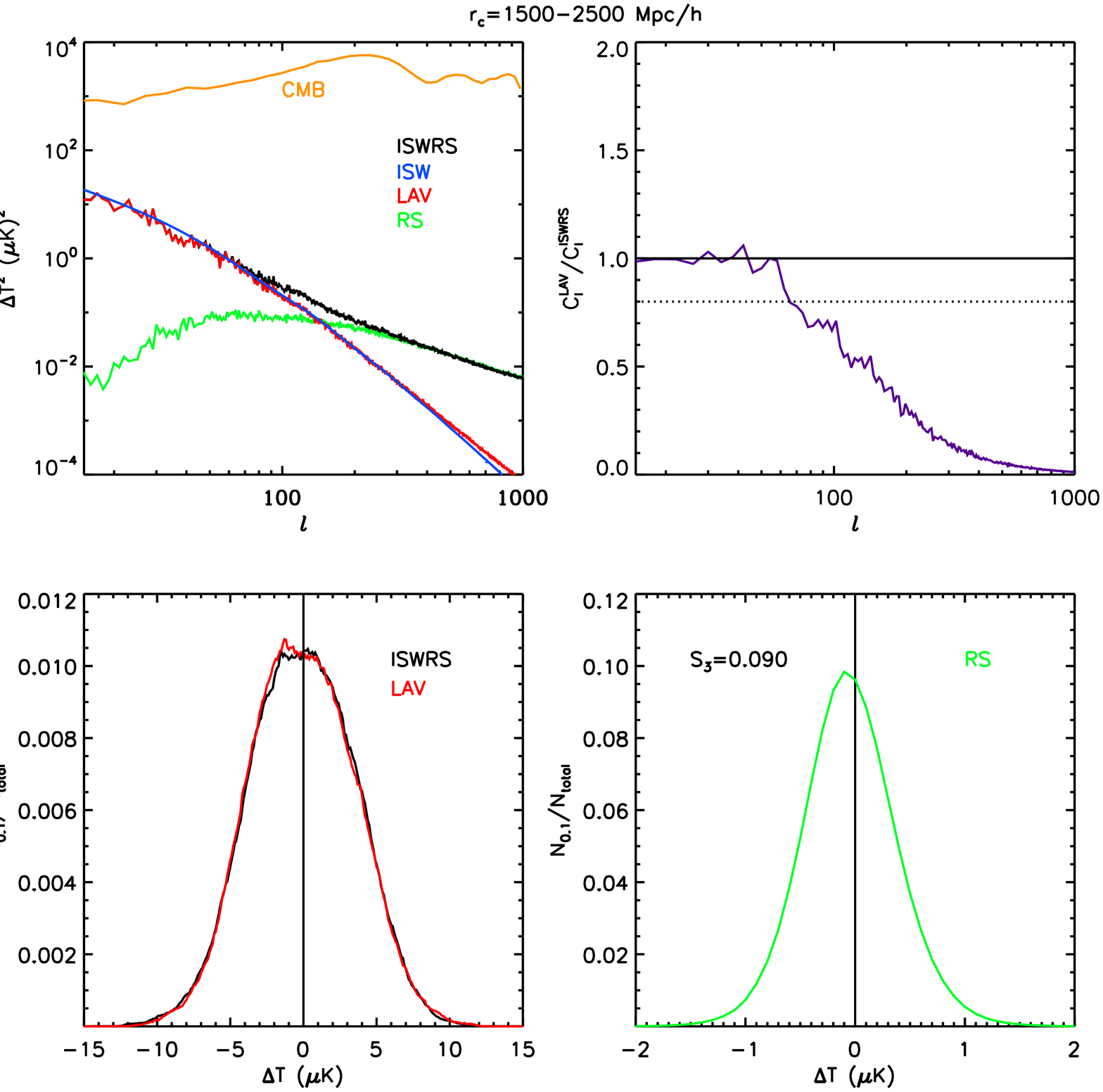}}
\caption{\label{PS1500_2500}
As Fig.~\ref{PS0_500}, but showing
the statistics for the  maps in Fig.~\ref{SkyMap1500_2500},
projecting over the range $0.57<z<1.07$, corresponding to the comoving distance
from 1500 to 2500~$h^{-1}$~Mpc.}
\end{center}
\end{figure*}

\begin{figure*}
\begin{center}
\resizebox{\hsize}{!}{
\includegraphics{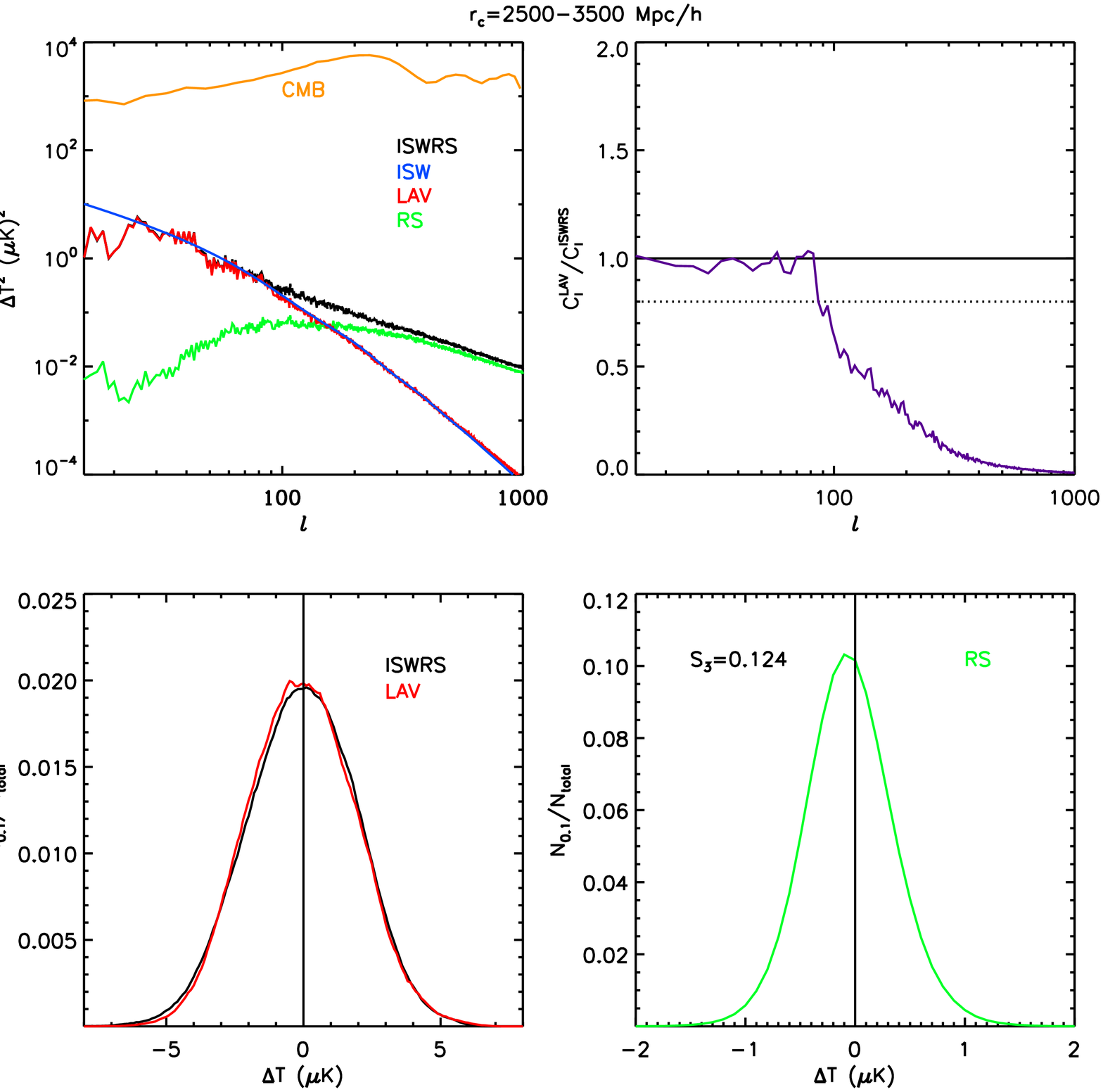}}
\caption{\label{PS2500_3500}As Fig.~\ref{PS0_500} but showing the
statistics for the maps  in Fig.~\ref{SkyMap2500_3500}, projecting
over the range $1.07<z<1.78$, corresponding to the comoving distance
from 2500 to 3500~$h^{-1}$~Mpc. }
\end{center}
\end{figure*}

\begin{figure*}
\begin{center}
\resizebox{\hsize}{!}{
\includegraphics{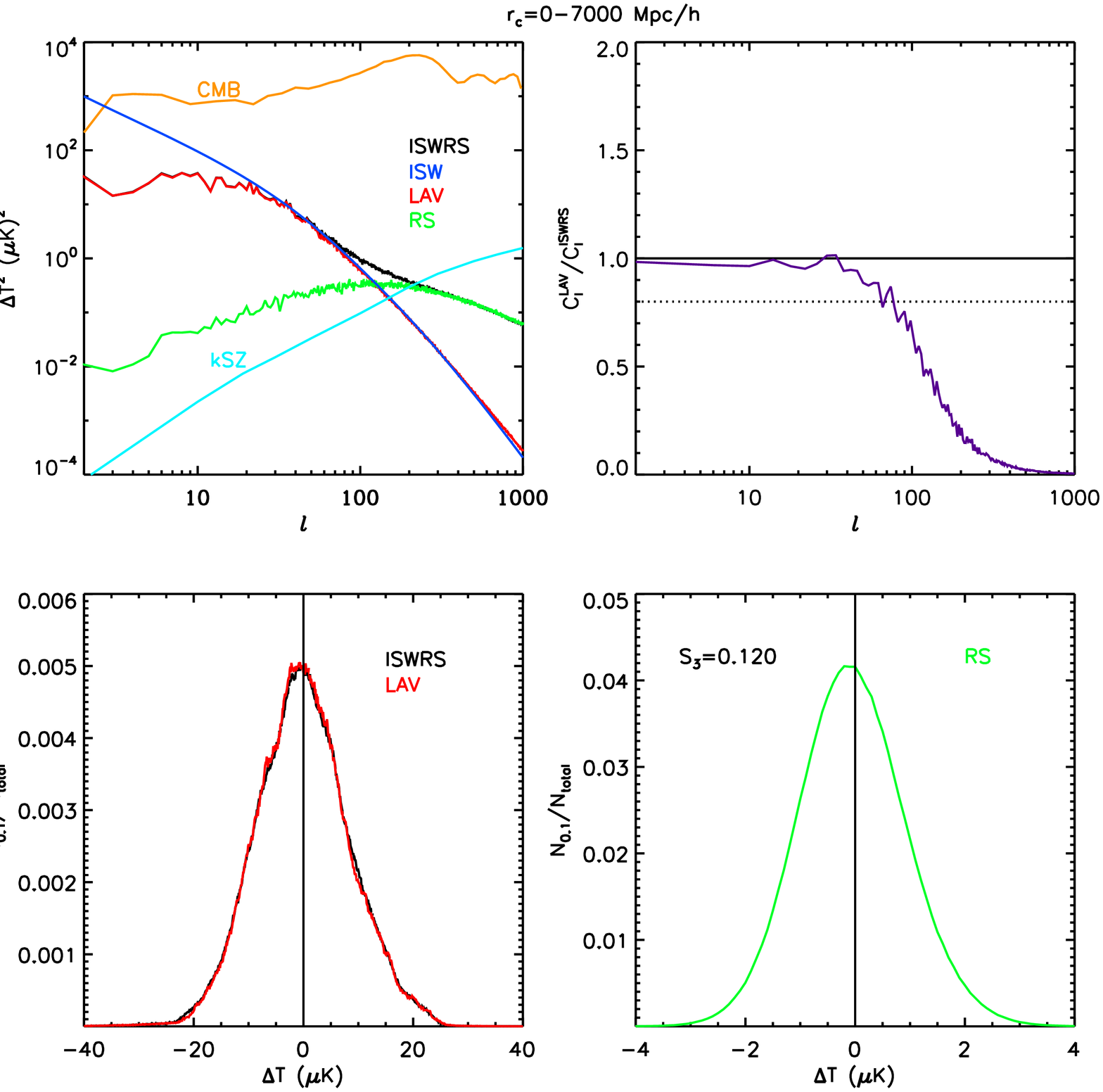}}
\caption{\label{PS0_7000}As Fig.~\ref{PS0_500}, but showing
statistics of maps from $z=0.0$ to $z=10.3$, corresponding to the
comoving distance from 0 to 7000~$h^{-1}$~Mpc.
The cyan line, in the top left panel, indicates the predicted
kinetic SZ effect from \citet{Cooray01}  }
\end{center}
\end{figure*}

We have used the methods described in \S\ref{method} to construct full
sky maps of the ISWRS effect using both the density and velocity
fields, and using the LAV approximation. We have also generated maps
of the RS effect, by subtracting the LAV maps from the corresponding
ISWRS maps. In all of the sky maps we have removed both the monopole
and dipole using the {\sc REMOVE$\_$DIPOLE} subroutine of HEALPix
\citep{Gorski05}.

The top panel of Fig.~\ref{SkyMap_ISWRS} shows the ISWRS map made by
integrating along the line-of-sight of the observer over the range
$0<z<0.17$, corresponding to distances from the observer, $r_c$, in the
range $0<r_c< $500$h^{-1}$~Mpc. The whole sky is dominated
by a few large hot and cold features with amplitudes of a few $\mu$K
to 10 $\mu$K, as expected from linear theory. The bottom panel shows
the ISWRS map integrated from $0.17<z<0.57$, corresponding to $500 <
r_c < 1500$ $h^{-1}$~Mpc.  The typical angular size of the features in
this shell is smaller than in the top panel.  As we will discuss later
in this subsection, the smaller angular size of the features in the
bottom map relative to the top map is due largely to the fact that
there is a cut-off in the power spectrum of the simulation used to
build the sky maps. Unlike in the real universe, there is no power
contributing to our maps on scales larger than the fundamental modes
of the simulation. This defect can easily be remedied: as the missing
modes are essentially in the linear regime, it would be
straightforward
simply to add additional longer wavelength modes by hand
before doing the line-of-sight projection to make the maps.  This
extra power, however, would be the same in maps of the ISWRS effect and
the LAV approximation, and so would vanish identically from the maps
of the RS effect. For this reason we have chosen not to add these
linear modes to the ISWRS and LAV maps.

The corresponding maps, but with the LAV approximation, are shown in
Fig.~\ref{SkyMap_LAV}.  Comparing
Fig.~\ref{SkyMap_ISWRS} with Fig.~\ref{SkyMap_LAV}, we see
that the large-scale distribution is essentially identical, but
differences are apparent on smaller scales.  The difference map made
by subtracting the LAV map from the ISWRS map to reveal the RS effect,
is shown in Fig.~\ref{SkyMap_RS}.  In the top panel, covering the
range $0<z<0.17$, we see some striking large-scale structures with
amplitudes that are about 1$\mu$K, which is about 10$\%$ of the
amplitude of the features seen in the ISWRS and LAV maps themselves. A
few large dipoles, ranging from a few degrees to over 10 degrees are
clearly visible, indicating the large-scale bulk flow of matter in the
local universe of our observer.  The bottom panel shows the RS effect
for a projection covering the range $0.17<z<0.57$.  Many more dipoles
are visible, with typical angular scales of a few degrees.  In addition,
the map has large-scale filamentary structure which is coherent over
lengths of up to tens of degrees. The typical amplitude of these features
is a few $\mu$K, again about 10$\%$ of the amplitude of the features
in the corresponding ISWRS and LAV maps.

In Fig.~\ref{SkyMap1500_2500} and Fig.~\ref{SkyMap2500_3500},
we show a patch of sky of size $40\times40$~degrees, for the redshift
intervals of $0.57<z<1.07$ ($1500 <r_c< 2500 h^{-1}\rm{Mpc}$) and
$1.07<z<1.78$ ($2500 <r_c< 3500~h^{-1}\rm{Mpc}$).  For comparison, at
$r_c=3~h^{-1}$Gpc, the simulation box-size subtends an angle of about $18$
degrees on the sky.  We find, as expected, that the intensity of the ISWRS
and LAV maps drops with increasing redshift. By contrast, the intensity of
the RS maps remains similar at the two epochs so that this becomes
dominant effect at high redshift.

 We now consider the one and two point statistics measured from the
maps. The results are summarised in
Figs.~\ref{PS0_500} - \ref{PS0_7000} which correspond to projections over the
following respective redshift ranges: $0<z<0.17$,
$0.17<z<0.57$, $0.57<z<1.07$, $1.07<z<1.78$, and $0<z<10.3$.  Starting with the
bottom left-hand panels, we see that the histograms of the temperature
fluctuations of the ISWRS and LAV maps are highly non-Gaussian at low
redshift (particularly in Figs.~\ref{PS0_500} and~\ref{PS500_1500}),
but are closer to Gaussian at higher redshifts.  At low redshift
the distributions are significantly skewed and irregular due to the
fact that only a small number of features contribute to these maps.
It is to be expected that potential variations in our local universe
will also generate similar large-scale temperature fluctuations in the
observed CMB and these too will be non-Gaussian.
To quantify this effect on the observed CMB, it is
necessary to investigate the contribution of the local environment on
scales up to at least a few hundred Megaparsecs. (See \cite{Maturi07a}
for discussion of the contribution from within 110~Mpc of Earth.)  At
higher redshifts, the relatively small size of the simulation box
means that the same features will appear more than once in the maps,
which will lead to some artificial residual non-Gaussianity, but only
on large angular scales.

Although the one point distributions are very similar there is a
noticeable systematic difference between the histograms of the ISWRS
and LAV effects at all redshifts. The ISWRS line is higher/lower than
the LAV line in the negative/positive tails of the
distribution. Because the monopoles of the maps have been set to zero,
so the means of the distributions are the same, these features are
indicative of difference in the skewness of the distributions.  The
bottom-right panel shows the one point distribution of RS effect, given
by the difference map formed by subtracting the LAV from the ISWRS
map.  The resulting distribution, which has zero mean, is clearly
skewed.  Measuring the skewness of the RS distribution, $S_3\equiv
\mu_3/\sigma^3$, where $\mu_3$ is the third moment about the mean and
$\sigma$ is the standard deviation of pixel temperatures, we find
$S_3\sim 0.1$ over all the redshift intervals.
The positive skewness indicates a shift of the mode of the distribution
to negative temperature values, consistent with
our findings that the RS effect
produces negative perturbations in both overdense and underdense regions.
This skewness is a further indication, already
visible in the RS sky maps, that the RS effect is significantly
non-Gaussian.

We show the results of the two point statistics, in the form of the
angular power spectrum, in the top-left panels for all five redshift
ranges in Figs~\ref{PS0_500} - \ref{PS0_7000}. The expected (linear)
ISW angular power spectrum is shown as a blue line.  Since the
simulation we use to compute the maps is periodic, and there is
no power on scales larger than the box,  the
angular power in our maps drops below the linear ISW prediction at the largest
angular scales.
The angular scale at which this departure occurs
depends on the redshift range of the projection, and occurs at higher
$l$ with increasing redshift.  Our simulation box-size is
1000$h^{-1}$~Mpc on a side so there are waves in the simulation which
are well described by linear theory. So, as expected, over intermediate
values of $l$, we find excellent agreement between our ISWRS (and LAV)
maps and the ISW prediction. We plot the ratio of the LAV and ISWRS
power in the top-right panel. Using the LAV approximation
significantly underestimates the power at large scales. Comparing
the LAV and ISWRS power spectra, we see, in Fig.~\ref{PS0_500},
that the LAV approximation underestimates the power by 20\% at around $l=35$.
At higher redshifts the
value of $l$ at which the power is underestimated by 20\%
increases gently to around 90
(Fig.\ref{PS2500_3500}). For the full projection (Fig.\ref{PS0_7000}) a
20\% deviation occurs at around $l=80$.

In Fig.~\ref{PS0_7000} we have also plotted the power spectrum of the
kinetic Sunyaev-Zel'dovich (kSZ) effect taken from \citet{Cooray01}
for a similar cosmological model.  The thermal and kinetic SZ effects
both arise from Compton scattering of CMB photons off electrons in the
ionized intracluster medium. The power spectrum of the thermal effect
is about one order of magnitude greater than the kSZ effect at arc
minute and degree scales \citep[e.g.][]{Cooray01, Hu02}. However, the
transfer of the thermal energy of the electrons to the CMB photons
produces a characteristic distortion of the CMB spectrum and so, in
multi-band CMB observations, the thermal SZ signal can be separated
from that of the ISWRS. In contrast, the kSZ effect, which is due to
the bulk motion of the ionized material, does not share this spectral
signature and so cannot be separated from the ISWRS signal. At
$l\approx 80$ where the RS effect begins to dominate over the linear
ISW contribution, the kSZ effect is more than an order of magnitude
weaker than the ISWRS effect. However, the kSZ effect has a much
steeper power spectrum and becomes comparable to the ISWRS power at
$l\sim 225$, corresponding to an angular scale of $\sim $0.5 degree.
Thus, on sub-degree scales, both the ISWRS and kSZ effects must be
considered. We note that their spatial characteristics are quite
distinct. A moving lump composed of both dark matter and ionized gas
moving along the line-of-sight towards the observer will produce a kSZ
hotspot while one moving away from the observer will produce a
coldspot, but neither case will produce an RS distortion.  In
contrast, if the lump is moving in the transverse direction it will
produce an RS dipole, but no kSZ distortion.

\subsection{Cold spots in the CMB?}

In this section, we compare the properties of cold spots produced
by the ISWRS with the notable CMB cold spot found in the southern
Galactic hemisphere \citep{Vielva04,Cruz05,McEwen05, Cruz06,McEwen06,
Cruz07,McEwen08}. In Sections~\ref{sec:cflow} and~\ref{sec:dflow} we have
illustrated how both convergent and divergent flows can generate
negative temperature perturbations via the non-linear RS effect.
For convergent flows into overdense regions this temperature
decrement will be swamped by the increment caused by the linear
ISW effect, except at high redshift where the linear effect diminishes
more rapidly than the non-linear effect. The amplitude of both
the high-$z$ decrements and low-$z$ increments is expected to be
of order 1~$\mu$K or less
(see, also the studies of non-linearity in cluster regions
\citet{Martinez-Gonzalez90a,
Martinez-Gonzalez90b,Panek92, Saez93,Arnau94,
Fullana94,Quilis95,Quilis98,Lasenby99,Dabrowski99} ).
In contrast, in the divergent flows both the linear ISW effect and
the non-linear RS effect reinforce each other to produce decrements
that have amplitudes of around 10~$\mu$K at low redshift (e.g.
Fig.~\ref{SkyMap_ISWRS}) with approximately 10\% contributed by the
non-linear RS effect.

\begin{figure}
\resizebox{\hsize}{!}{
\includegraphics[angle=0]{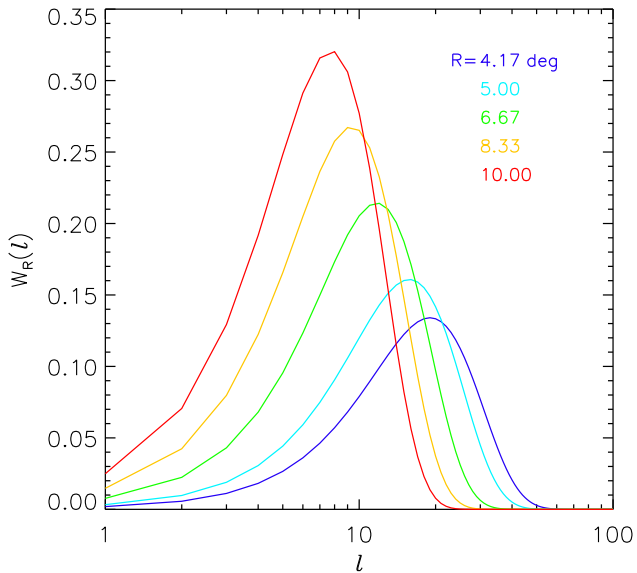}}
\caption{\label{SMHW} Legendre transforms of the Spherical Mexican
Hat Wavelet given by equation \ref{eqSMHW} for different widths of
the filter scale, $R$. The filters
are normalized so that $2\pi\int \Psi^2(y,R)d\cos\theta=1$
and $\Psi$ is related to $W_R(l)$ by equation (9).}
\end{figure}

\begin{figure*}
\begin{center}
\advance\leftskip 0.8cm
\resizebox{\hsize}{!}{
\includegraphics[angle=90]{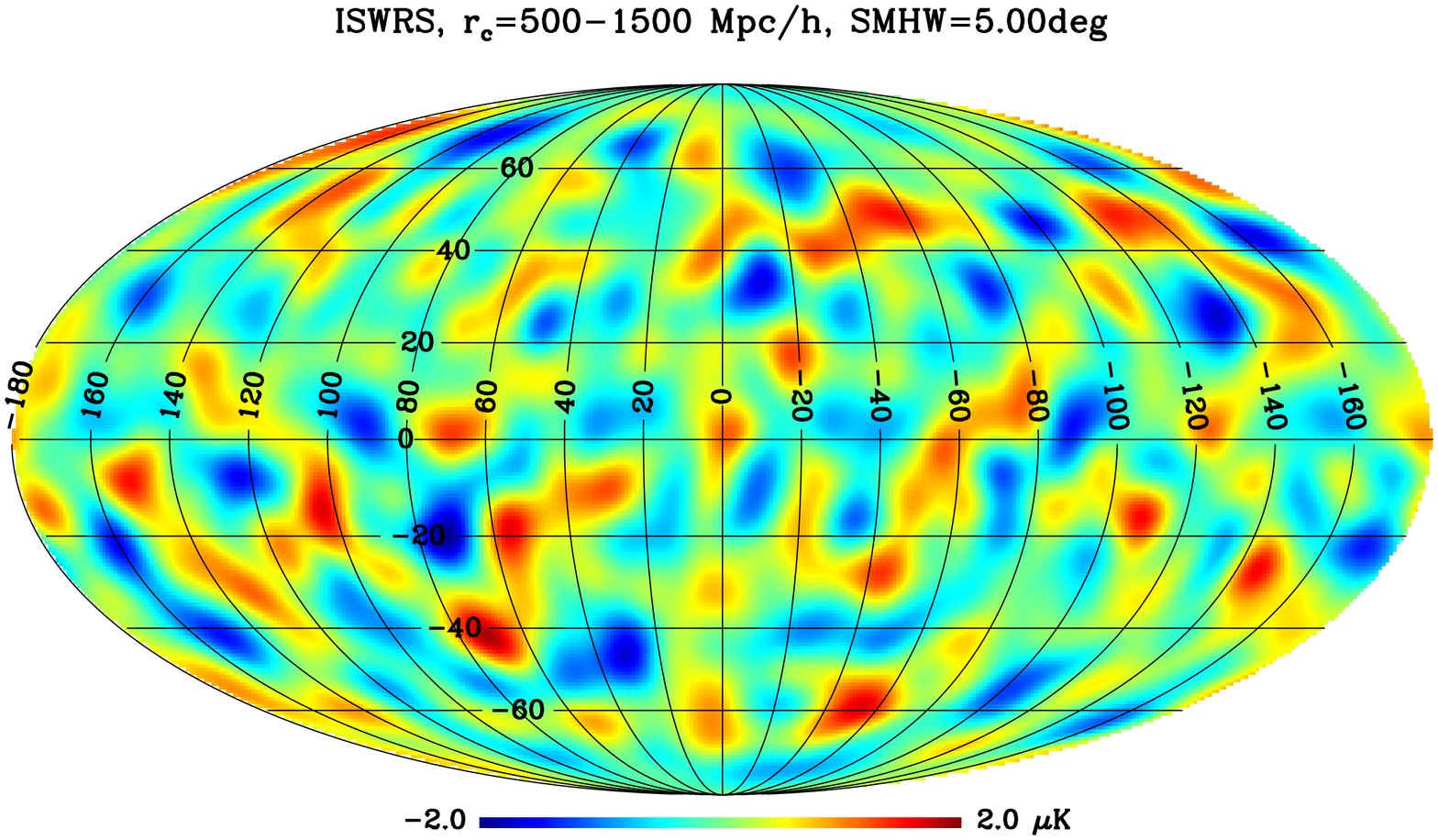}
\includegraphics[angle=90]{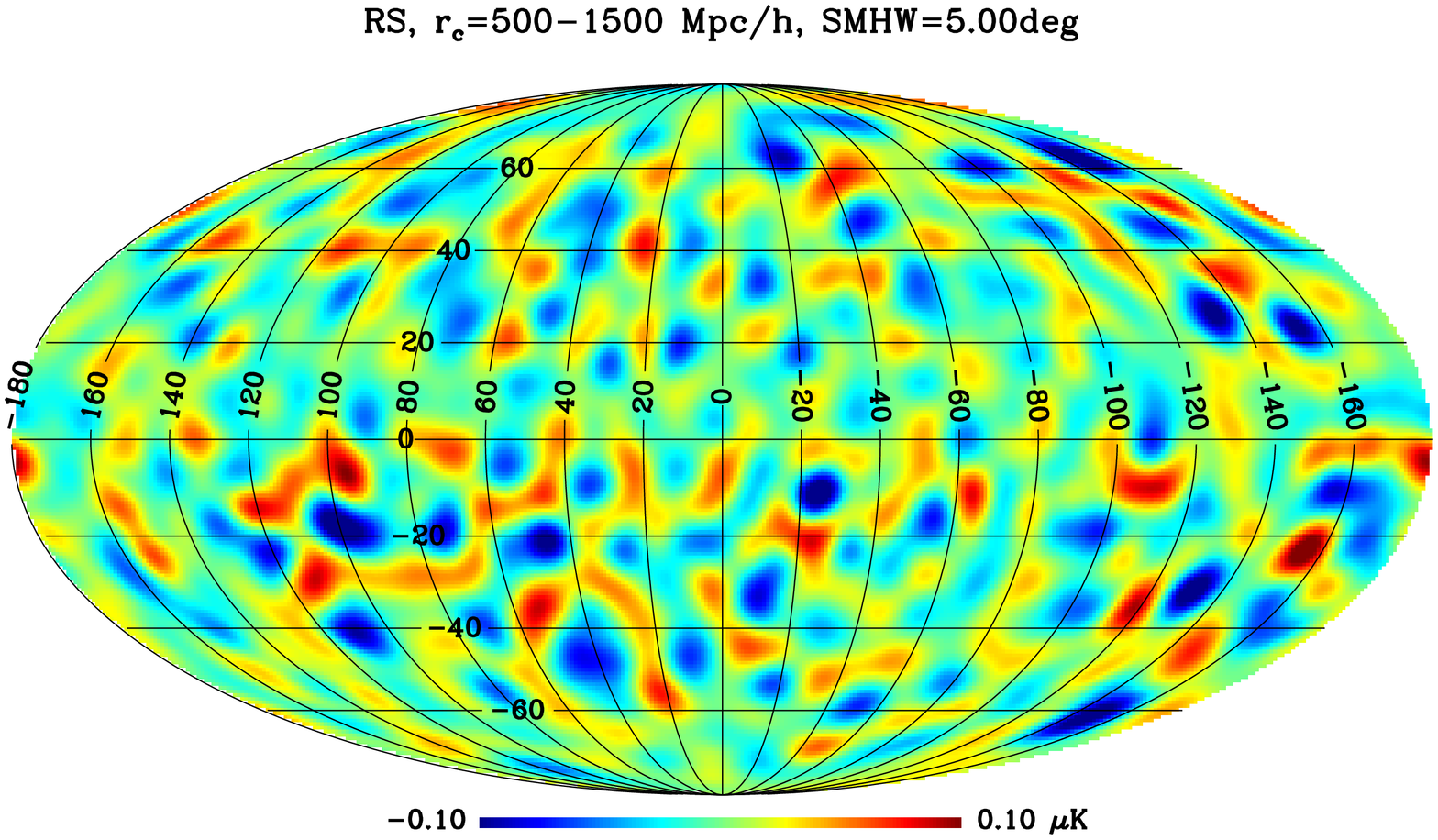}
}
\end{center}
\begin{center}
\scalebox{0.5}{
\includegraphics[angle=0]{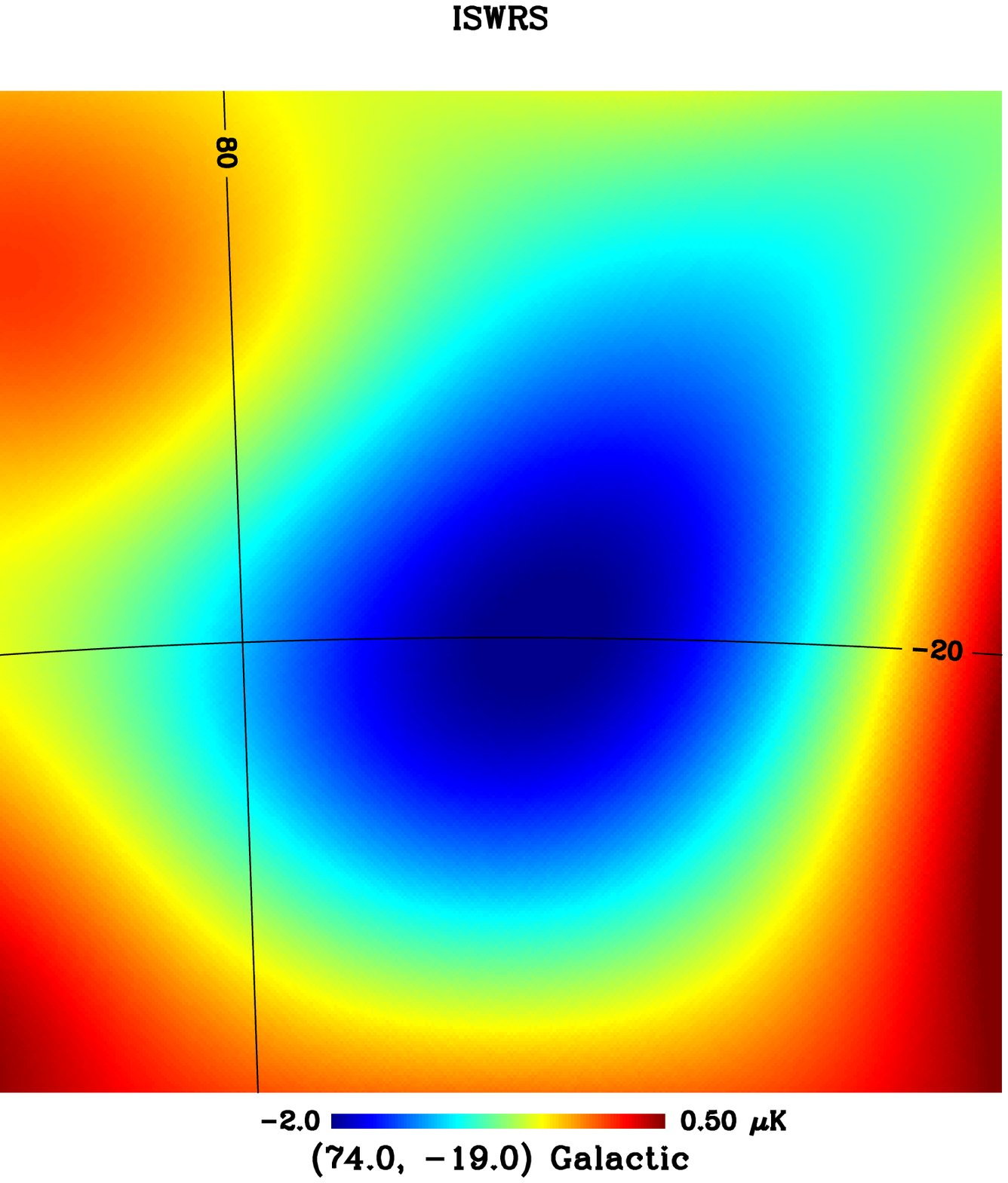}
\includegraphics[angle=0]{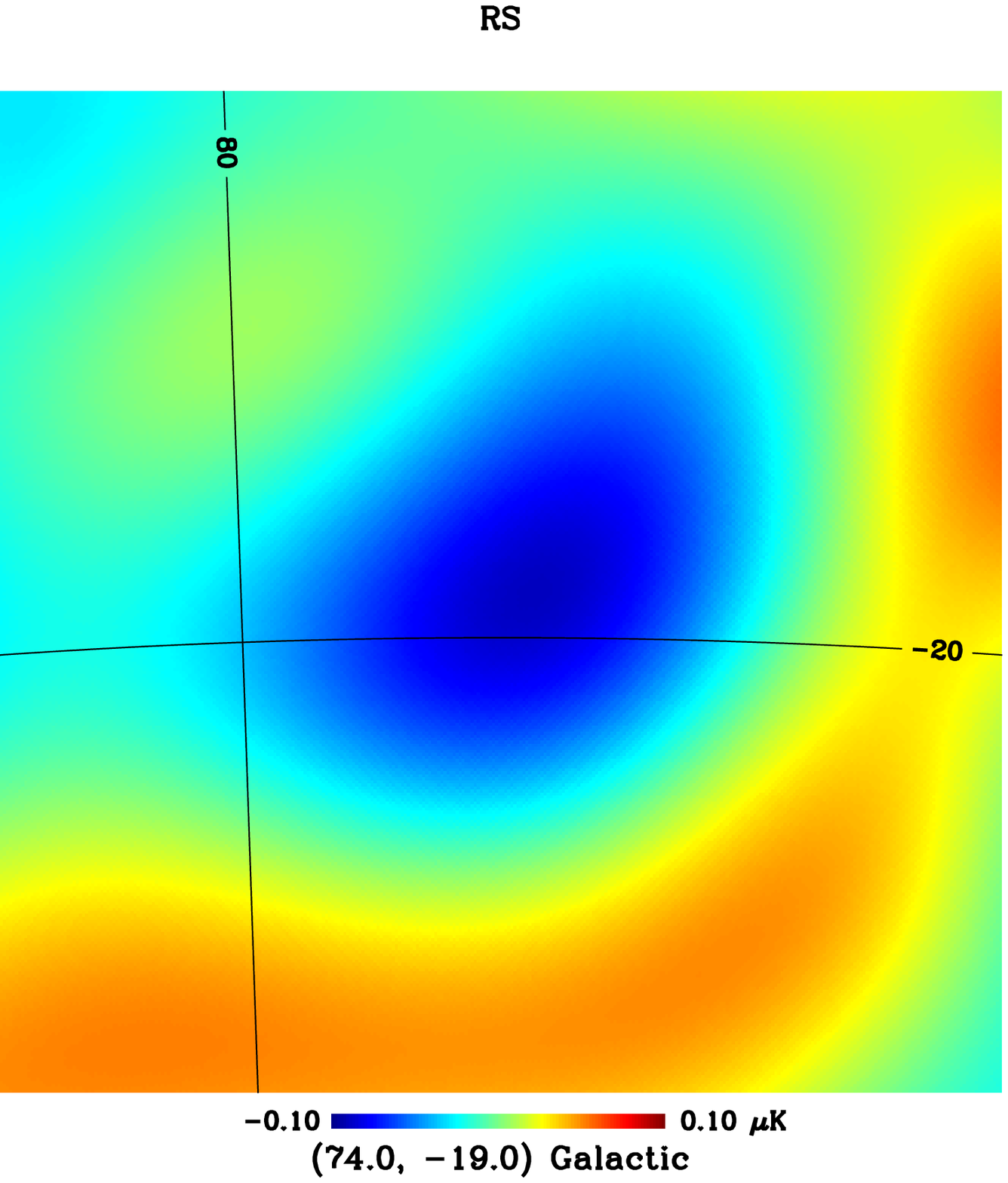}}
\caption{\label{SMHW5.00_500-1500} Full sky maps of the predicted CMB
  $\Delta T$ due to the ISWRS (top-left) and solely the RS, i.e.
  ISWRS-LAV, (top-right) effect made from our simulation smoothed by a
  SMHW with $R=5.00$~degrees, the same as $R_9$ in \citet{Cruz07}.  The grid
  spacing is 20~degrees in both longitude and latitude.  The bottom
  panels show a zoomed in version of a  $22\times 22$~degrees patch
  taken from the top plots. These maps
  are made by ray-tracing through the simulation over the redshift
  interval $0.17<z<0.57$, which corresponds to a range of comoving
  distance from the observer of 500 - 1500~$h^{-1}$~Mpc.}
\end{center}
\end{figure*}

\citet{Cruz05} estimate that the southern cold spot has
a temperature decrement of 73~$\mu$K and an angular extent of
$10^{\circ}$ when smoothed by a Gaussian filter of width $4$~degrees.
To suppress the contributions to this decrement from
primordial perturbations of both smaller and larger angular scales
and hence better characterize the
properties of the cold spot \citet{Cruz05}
\citep[see also][]{Vielva04, Cruz06, Cruz07} filter the sky map
with a spherical Mexican hat wavelet (SMHW). The SMHW filter can
be expressed as
\begin{equation}\label{eqSMHW}
\Psi(y,R)=\frac{1}{\sqrt{2\pi}N(R)}\left[1+\left(\frac{y}{2}\right)^2\right]^2
\left[2-\left(\frac{y}{R}\right)^2\right]e^{-y^2/2R^2},
\end{equation}
where $N(R)=(1+R^2/2+R^4/4)^{0.5}$, $y=2 \tan(\theta/2)$ and $\theta$
is the polar angle \citep{Martinez-Gonzalez02}. The effect of smoothing
with this form of filter can be understood most easily in
spherical harmonic space. In Fig.\ref{SMHW},
we show the Legendre transform of the SMHW,
\begin{equation}\label{eqWl}
W_R(l)= 2\pi \int \Psi(y,R)P_l(\cos \theta)d\cos\theta,
\end{equation}
where $P_l$ is the Legendre polynomial of order $l$ \citep{Page03},
for various smoothing scales, $R$.
Note that these
filters are normalized as in \citet{Vielva04,Cruz05,Cruz06,Cruz07}
such that $2\pi\int \Psi^2(y,R)d\cos\theta=1$. We see that this set of
filters retain sensitivity to modes in the range $5\le l\le40$,
while effectively suppressing the primordial CMB perturbations that peak
at $l=200$.
\citet{Cruz05} find that when filtered in this way with an SMHW of
$R=4.17$~degrees the temperature decrement of the cold spot in the WMAP1
data is $-16~\mu$K\footnote{Often spectral filters
are instead normalized such that their peak value $W_R(l_{\rm
  max})=1$. If this were done here, the temperature decrement would be
$\sim120~\mu$K. }, with similar results for $R=4$
and $5$~degrees \citep{Cruz06} and with WMAP3 \citep{Cruz07}.
The probability of
finding such a high deviation in Gaussian simulations of the CMB is
 $1.85\%$ \citep{Cruz07}, but see \citet{Zhang09}. There is no
evidence that the decrement is  frequency dependent \citep{Cruz06, Cruz07}
and hence any contribution from the thermal SZ effect must be small.

To compare our simulated sky maps with the properties of this observed
cold spot we first smoothed the ISWRS maps of Figs.~8,~11, and~12
using a Gaussian filter with the FWHM of $4$~degrees. The resulting
maps contain cold spots of $\sim 10$~degrees extent, but only with
the amplitudes of up to $-20$ $\mu$K, approximately a
factor of three lower than the reported amplitude, $-73~\mu$K, though,
of course, the observed cold spot may contain contributions from
primordial CMB perturbations, which are missing from out maps.
 We find that the majority of the deepest
cold spots come from the redshift range $0.17<z<0.57$, which corresponds 
to a range of comoving  distance from the observer of 500 - 1500~$h^{-1}$~Mpc.
The optimum smoothing scale is  $R=6$-$10$~degrees.

A more stringent comparison with the observed cold spot can be made by
applying the SMHW filtering to our ISWRS maps. An example with an
$R=5$~degrees SMHW is shown in Fig.~\ref{SMHW5.00_500-1500} which shows
the map resulting from the redshift range $0.17<z<0.57$.  It should be
noted that the amplitude of the fluctuations in this map are much less
than those in the Gaussian filtered maps reported above simply because
of the \citep{Cruz06} choice of normalization of the SMHW filter.
We, again, find the
typical size of coherent structures is comparable with that found in
the observations. However, the coldest region we find from the map
after SMHW filtering is $\sim -2~\mu$K, accounting for 12.5\% of the reported
CMB cold spot. The left-hand panels of Fig.~\ref{SMHW5.00_500-1500}
show the difference between the full ISWRS prediction and that of only
ISW (LAV). Careful comparison shows that this non-linear RS
contribution acts to enhance the depth of cold spots (an example of
which is shown in the lower panels of  Fig.~\ref{SMHW5.00_500-1500})
and to diminish the height of the hot spots. However it is a modest
effect, contributing only $0.1~\mu$K (5\%) to the depth of the
illustrated cold spot.

Previous studies have mainly focused on the ISW(+RS) effect
arising from a single void, i.e, a region that is significantly
underdense relative to the mean. They have concluded that a void of
size $100$~Mpc at $z<1$ would be needed to produce a cold spot
with depth of a few tens of micro-Kelvin
via gravitational effects \citep{Rudnick07,Inoue06,Inoue07,
Tomita08, Masina09b, Masina09a, Granett09,Francis09, Granett09b} and
that such a huge void is very unlikely to occur in the standard $\Lambda$CDM
cosmology. Here we have generalised this comparison by simulating the
correct 3D $\Lambda$CDM distribution of evolving density perturbations
and mimicked the observations by producing projected maps. However
the conclusion remains the same that the ISW+RS effect in the standard
$\Lambda$CDM cosmology does not produce cold spots as deep as the one
discussed.

\section{conclusion and discussion}

Our aim in this paper is to investigate fully the gravitational effect that CMB
photons suffer when passing through the evolving non-linear
gravitational potential, $\Phi$, of intervening large-scale structure.
We have developed a method of using large cosmological N-body
simulations to compute the time derivative of the potential,
$\dot{\Phi}$, along the past light-cone of an observer.  By integrating
along light rays we have created full sky maps, both using the full
non-linear calculation of $\dot{\Phi}$ and an alternative in which the
dark matter velocity field is assumed to be related to its density
field by the normal linear theory relation. By comparing the results
of the two calculations we were able to assess fully the linear
Integrated Sachs-Wolfe (ISW) and non-linear Rees-Sciama (RS)
contributions to the induced CMB temperature fluctuations.

In general, in a $\Lambda$CDM universe, the linear ISW effect is
dominant at low redshift where the accelerating expansion, driven by
dark energy, causes the decay of perturbations in the gravitational
potential. The propagation of CMB photons through these decaying
potential wells and hills produces hot and cold regions respectively
with $\Delta T$ of the order of
$10\mu$K on scales of hundreds of Megaparsecs.  At low redshift the non-linear
(RS) effect produces only small scale perturbations to this large scale
pattern. However, with increasing redshift
the RS contribution decreases in amplitude much more slowly than does
the contribution from the ISW effect, which vanishes as $\Omega_{\rm
m}(z)$ approaches unity.  Hence, the importance and scale of the RS
effect becomes larger at higher redshift, confirming the conclusions of
\citet{Cai09} and \citet{Smith09}.
We have shown that the origin
of the RS effect is primarily the non-linear relation between the
velocity and density field rather than the non-linearity of the
density field itself.\footnote{The fact that to model
$\dot \Phi$ and the growth of large-scale structure requires more
accurate modelling of the large-scale velocity field than is provided
by linear theory may suggest that using redshift space distortions to
measure the growth rate of density perturbations, $\beta$, may also
not be sufficiently accurate.} Our investigation of the RS effect
has revealed three distinct non-linear phenomena that give rise to
corresponding characteristic features in the temperature perturbation
maps.
\begin{itemize}
\item Dipoles are produced by the
transverse motion of large lumps of dark matter, with
typical sizes of tens of Megaparsecs, much larger than the scale
of individual galaxy clusters.
\item Convergent flows, on the scale of up to 100$h^{-1}$Mpc,
around non-linear overdense regions give rise to cold spots
of order $\mu$K surrounded by hot filamentary shells. At high
redshift these can be strong enough to dominate over the linear
ISW effect and change the sign of the temperature perturbation
centred on overdense regions.
\item Divergent flows around void regions are characterised by RS
contributions consisting of hot rings around
cold central regions, where the density
contrast of the void is nearly saturated ($\delta \approx -1$).
This is a small effect at low redshift, but
acts to strengthen the cold spots produced by the linear ISW effect.
\end{itemize}
Unfortunately, none of these effects can be easily detected individually. At low
redshift these phenomena make only 10\% changes to the temperature
perturbations predicted by the linear ISW effect. At very high
redshift they are completely dominant, but their amplitudes are very
low. We find that they contribute to cold spots of comparable physical
scale to those reported in the literature, but their amplitudes are
many times smaller.

It may be possible to detect these large scale features induced
by the non-linear velocity field
by employing stacking techniques. The detectability of RS kinematic
features produced by merging clusters has been discussed by
\citet{Rubino-Martin04} and \citet{Maturi07b}.
They conclude that around a thousand clusters would be needed
to detect the RS signal above the contaminating effects
of the primordial CMB temperature fluctuations and instrument noise.

However, we have found imprints of the RS perturbations on scales of
a few tens of Megaparsecs, much larger than the merging cluster scale, and
with slightly larger amplitudes. Thus, these large-scale features
might be more easily detected, requiring the co-addition of fewer objects.

From our all sky maps we find that the RS contributions to the overall
power spectrum of temperature perturbations become important for $l>
80$ (a few degrees) and completely dominate for $l > 200$. At still
smaller scales the kinetic SZ effect is expected to dominate and at
such scales a full treatment would have to incorporate additional
modelling of this contribution.  The RS contribution to the
temperature maps is strongly non-Gaussian with a skewed one-point
distribution. In future work it will be interesting to investigate the
RS contribution to higher order statistics as its non-Gaussian
characteristics might limit the ability to detect primordial
non-Gaussianity in the underlying primary CMB fluctuations. Combining
our full-sky maps with mock galaxy catalogues built from the same
N-body simulations will be a powerful tool for developing
cross-correlation techniques aimed at extracting the ISWRS signal
from redshift surveys.

\section*{ACKNOWLEDGEMENT}
YC was supported by the Marie Curie Early Stage Training Host
Fellowship ICCIPPP, which is funded by the European Commission. 
YC acknowledges the support of  grant DE-FG02-95ER40893 from the 
US Department of Energy. We thank Carlton Baugh, Elise Jennings and 
Raul Angulo for providing
the Gpc simulation, which was carried out on the Cosmology Machine
at Durham. We also thank Istvan Szapudi and Ben Granett for useful
discussion. YC thanks Andrew Cooper
for useful discussion on technical details and Lydia Heck for
computing support. CSF acknowledges a Royal-Society Wolfson Research
Merit Award. This work was supported in part by the STFC
rolling grant ST/F002289/1.
\bibliography{ISW}

\begin{thebibliography}{}

\bibitem[\protect\citeauthoryear{{Aghanim}, {Prunet}, {Forni} \&
  {Bouchet}}{{Aghanim} et~al.}{1998}]{Aghanim98}
{Aghanim} N.,  {Prunet} S.,  {Forni} O.,    {Bouchet} F.~R.,  1998, \aap, 334,
  409

\bibitem[\protect\citeauthoryear{{Arnau}, {Fullana} \& {Saez}}{{Arnau}
  et~al.}{1994}]{Arnau94}
{Arnau} J.~V.,  {Fullana} M.~J.,    {Saez} D.,  1994, \mnras, 268, L17+

\bibitem[\protect\citeauthoryear{{Aso}, {Hattori} \& {Futamase}}{{Aso}
  et~al.}{2002}]{Aso02}
{Aso} O.,  {Hattori} M.,    {Futamase} T.,  2002, \apjl, 576, L5

\bibitem[\protect\citeauthoryear{{Barreiro}, {Vielva}, {Hernandez-Monteagudo}
  \& {Martinez-Gonzalez}}{{Barreiro} et~al.}{2008}]{Barreiro08}
{Barreiro} R.~B.,  {Vielva} P.,  {Hernandez-Monteagudo} C.,
  {Martinez-Gonzalez} E.,  2008, IEEE Journal of Selected Topics in Signal
  Processing, vol.~2, issue 5, pp.~747-754, 2, 747

\bibitem[\protect\citeauthoryear{{Bielby}, {Shanks}, {Sawangwit}, {Croom},
  {Ross} \& {Wake}}{{Bielby} et~al.}{2009}]{Bielby09}
{Bielby} R.,  {Shanks} T.,  {Sawangwit} U.,  {Croom} S.~M.,  {Ross} N.~P.,
  {Wake} D.~A.,  2009, ArXiv e-prints

\bibitem[\protect\citeauthoryear{{Birkinshaw} \& {Gull}}{{Birkinshaw} \&
  {Gull}}{1983}]{Birkinshaw83}
{Birkinshaw} M.,  {Gull} S.~F.,  1983, \nat, 302, 315

\bibitem[\protect\citeauthoryear{{Boubekeur}, {Creminelli}, {D'Amico},
  {Nore{\~n}a} \& {Vernizzi}}{{Boubekeur} et~al.}{2009}]{Boubekeur09}
{Boubekeur} L.,  {Creminelli} P.,  {D'Amico} G.,  {Nore{\~n}a} J.,
  {Vernizzi} F.,  2009, ArXiv e-prints

\bibitem[\protect\citeauthoryear{{Cai}, {Cole}, {Jenkins} \& {Frenk}}{{Cai}
  et~al.}{2009}]{Cai09}
{Cai} Y.-C.,  {Cole} S.,  {Jenkins} A.,    {Frenk} C.,  2009, \mnras, 396, 772

\bibitem[\protect\citeauthoryear{{Cooray}}{{Cooray}}{2001}]{Cooray01}
{Cooray} A.,  2001, \prd, 64, 063514

\bibitem[\protect\citeauthoryear{{Cooray}}{{Cooray}}{2002}]{Cooray02a}
{Cooray} A.,  2002, \prd, 65, 103510

\bibitem[\protect\citeauthoryear{{Cooray} \& {Seto}}{{Cooray} \&
  {Seto}}{2005}]{Cooray05}
{Cooray} A.,  {Seto} N.,  2005, Journal of Cosmology and Astro-Particle
  Physics, 12, 4

\bibitem[\protect\citeauthoryear{{Cruz}, {Cay{\'o}n},
  {Mart{\'{\i}}nez-Gonz{\'a}lez}, {Vielva} \& {Jin}}{{Cruz}
  et~al.}{2007}]{Cruz07}
{Cruz} M.,  {Cay{\'o}n} L.,  {Mart{\'{\i}}nez-Gonz{\'a}lez} E.,  {Vielva} P.,
   {Jin} J.,  2007, \apj, 655, 11

\bibitem[\protect\citeauthoryear{{Cruz}, {Mart{\'{\i}}nez-Gonz{\'a}lez},
  {Vielva} \& {Cay{\'o}n}}{{Cruz} et~al.}{2005}]{Cruz05}
{Cruz} M.,  {Mart{\'{\i}}nez-Gonz{\'a}lez} E.,  {Vielva} P.,    {Cay{\'o}n} L.,
   2005, \mnras, 356, 29

\bibitem[\protect\citeauthoryear{{Cruz}, {Tucci},
  {Mart{\'{\i}}nez-Gonz{\'a}lez} \& {Vielva}}{{Cruz} et~al.}{2006}]{Cruz06}
{Cruz} M.,  {Tucci} M.,  {Mart{\'{\i}}nez-Gonz{\'a}lez} E.,    {Vielva} P.,
  2006, \mnras, 369, 57

\bibitem[\protect\citeauthoryear{{Dabrowski}, {Hobson}, {Lasenby} \&
  {Doran}}{{Dabrowski} et~al.}{1999}]{Dabrowski99}
{Dabrowski} Y.,  {Hobson} M.~P.,  {Lasenby} A.~N.,    {Doran} C.,  1999,
  \mnras, 302, 757

\bibitem[\protect\citeauthoryear{{Fosalba}, {Gazta{\~n}aga} \&
  {Castander}}{{Fosalba} et~al.}{2003}]{Fosalba03}
{Fosalba} P.,  {Gazta{\~n}aga} E.,    {Castander} F.~J.,  2003, \apjl, 597, L89

\bibitem[\protect\citeauthoryear{{Francis} \& {Peacock}}{{Francis} \&
  {Peacock}}{2009}]{Francis09}
{Francis} C.~L.,  {Peacock} J.~A.,  2009, ArXiv e-prints

\bibitem[\protect\citeauthoryear{{Fullana}, {Saez} \& {Arnau}}{{Fullana}
  et~al.}{1994}]{Fullana94}
{Fullana} M.~J.,  {Saez} D.,    {Arnau} J.~V.,  1994, \apjs, 94, 1

\bibitem[\protect\citeauthoryear{{Giovi}, {Baccigalupi} \& {Perrotta}}{{Giovi}
  et~al.}{2003}]{Giovi03}
{Giovi} F.,  {Baccigalupi} C.,    {Perrotta} F.,  2003, \prd, 68, 123002

\bibitem[\protect\citeauthoryear{{G{\'o}rski}, {Hivon}, {Banday}, {Wandelt},
  {Hansen}, {Reinecke} \& {Bartelmann}}{{G{\'o}rski} et~al.}{2005}]{Gorski05}
{G{\'o}rski} K.~M.,  {Hivon} E.,  {Banday} A.~J.,  {Wandelt} B.~D.,  {Hansen}
  F.~K.,  {Reinecke} M.,    {Bartelmann} M.,  2005, \apj, 622, 759

\bibitem[\protect\citeauthoryear{{Granett}, {Neyrinck} \& {Szapudi}}{{Granett}
  et~al.}{2009}]{Granett09}
{Granett} B.~R.,  {Neyrinck} M.~C.,    {Szapudi} I.,  2009, \apj, 701, 414

\bibitem[\protect\citeauthoryear{{Granett}, {Szapudi} \& {Neyrinck}}{{Granett}
  et~al.}{2009}]{Granett09b}
{Granett} B.~R.,  {Szapudi} I.,    {Neyrinck} M.~C.,  2009, ArXiv e-prints

\bibitem[\protect\citeauthoryear{{Gurvits} \& {Mitrofanov}}{{Gurvits} \&
  {Mitrofanov}}{1986}]{Gurvits86}
{Gurvits} L.~I.,  {Mitrofanov} I.~G.,  1986, \nat, 324, 349

\bibitem[\protect\citeauthoryear{{Hockney} \& {Eastwood}}{{Hockney} \&
  {Eastwood}}{1981}]{Hockney81}
{Hockney} R.~W.,  {Eastwood} J.~W.,  1981, {Computer Simulation Using
  Particles}

\bibitem[\protect\citeauthoryear{{Hu} \& {Dodelson}}{{Hu} \&
  {Dodelson}}{2002}]{Hu02}
{Hu} W.,  {Dodelson} S.,  2002, \araa, 40, 171

\bibitem[\protect\citeauthoryear{{Inoue} \& {Silk}}{{Inoue} \&
  {Silk}}{2006}]{Inoue06}
{Inoue} K.~T.,  {Silk} J.,  2006, \apj, 648, 23

\bibitem[\protect\citeauthoryear{{Inoue} \& {Silk}}{{Inoue} \&
  {Silk}}{2007}]{Inoue07}
{Inoue} K.~T.,  {Silk} J.,  2007, \apj, 664, 650

\bibitem[\protect\citeauthoryear{{Lasenby}, {Doran}, {Hobson}, {Dabrowski} \&
  {Challinor}}{{Lasenby} et~al.}{1999}]{Lasenby99}
{Lasenby} A.~N.,  {Doran} C.~J.~L.,  {Hobson} M.~P.,  {Dabrowski} Y.,
  {Challinor} A.~D.,  1999, \mnras, 302, 748

\bibitem[\protect\citeauthoryear{{Mangilli} \& {Verde}}{{Mangilli} \&
  {Verde}}{2009}]{Mangilli09}
{Mangilli} A.,  {Verde} L.,  2009, ArXiv e-prints

\bibitem[\protect\citeauthoryear{{Mart{\'{\i}}nez-Gonz{\'a}lez}, {Gallegos},
  {Arg{\"u}eso}, {Cay{\'o}n} \& {Sanz}}{{Mart{\'{\i}}nez-Gonz{\'a}lez}
  et~al.}{2002}]{Martinez-Gonzalez02}
{Mart{\'{\i}}nez-Gonz{\'a}lez} E.,  {Gallegos} J.~E.,  {Arg{\"u}eso} F.,
  {Cay{\'o}n} L.,    {Sanz} J.~L.,  2002, \mnras, 336, 22

\bibitem[\protect\citeauthoryear{{Martinez-Gonzalez} \&
  {Sanz}}{{Martinez-Gonzalez} \& {Sanz}}{1990}]{Martinez-Gonzalez90b}
{Martinez-Gonzalez} E.,  {Sanz} J.~L.,  1990, \mnras, 247, 473

\bibitem[\protect\citeauthoryear{{Martinez-Gonzalez}, {Sanz} \&
  {Silk}}{{Martinez-Gonzalez} et~al.}{1990}]{Martinez-Gonzalez90a}
{Martinez-Gonzalez} E.,  {Sanz} J.~L.,    {Silk} J.,  1990, \apjl, 355, L5

\bibitem[\protect\citeauthoryear{{Masina} \& {Notari}}{{Masina} \&
  {Notari}}{2009a}]{Masina09b}
{Masina} I.,  {Notari} A.,  2009a, Journal of Cosmology and Astro-Particle
  Physics, 7, 35

\bibitem[\protect\citeauthoryear{{Masina} \& {Notari}}{{Masina} \&
  {Notari}}{2009b}]{Masina09a}
{Masina} I.,  {Notari} A.,  2009b, Journal of Cosmology and Astro-Particle
  Physics, 2, 19

\bibitem[\protect\citeauthoryear{{Maturi}, {Dolag}, {Waelkens}, {Springel} \&
  {En{\ss}lin}}{{Maturi} et~al.}{2007}]{Maturi07a}
{Maturi} M.,  {Dolag} K.,  {Waelkens} A.,  {Springel} V.,    {En{\ss}lin} T.,
  2007, \aap, 476, 83

\bibitem[\protect\citeauthoryear{{Maturi}, {En{\ss}lin},
  {Hern{\'a}ndez-Monteagudo} \& {Rubi{\~n}o-Mart{\'{\i}}n}}{{Maturi}
  et~al.}{2007}]{Maturi07b}
{Maturi} M.,  {En{\ss}lin} T.,  {Hern{\'a}ndez-Monteagudo} C.,
  {Rubi{\~n}o-Mart{\'{\i}}n} J.~A.,  2007, \aap, 467, 411

\bibitem[\protect\citeauthoryear{{Maturi}, {Ensslin}, {Hernandez-Monteagudo} \&
  {Rubino-Martin}}{{Maturi} et~al.}{2006}]{Maturi06}
{Maturi} M.,  {Ensslin} T.,  {Hernandez-Monteagudo} C.,    {Rubino-Martin}
  J.~A.,  2006, ArXiv Astrophysics e-prints

\bibitem[\protect\citeauthoryear{{McEwen}, {Hobson}, {Lasenby} \&
  {Mortlock}}{{McEwen} et~al.}{2005}]{McEwen05}
{McEwen} J.~D.,  {Hobson} M.~P.,  {Lasenby} A.~N.,    {Mortlock} D.~J.,  2005,
  \mnras, 359, 1583

\bibitem[\protect\citeauthoryear{{McEwen}, {Hobson}, {Lasenby} \&
  {Mortlock}}{{McEwen} et~al.}{2006}]{McEwen06}
{McEwen} J.~D.,  {Hobson} M.~P.,  {Lasenby} A.~N.,    {Mortlock} D.~J.,  2006,
  \mnras, 371, L50

\bibitem[\protect\citeauthoryear{{McEwen}, {Hobson}, {Lasenby} \&
  {Mortlock}}{{McEwen} et~al.}{2008}]{McEwen08}
{McEwen} J.~D.,  {Hobson} M.~P.,  {Lasenby} A.~N.,    {Mortlock} D.~J.,  2008,
  \mnras, 388, 659

\bibitem[\protect\citeauthoryear{{Molnar} \& {Birkinshaw}}{{Molnar} \&
  {Birkinshaw}}{2000}]{Molnar00}
{Molnar} S.~M.,  {Birkinshaw} M.,  2000, \apj, 537, 542

\bibitem[\protect\citeauthoryear{{Molnar} \& {Birkinshaw}}{{Molnar} \&
  {Birkinshaw}}{2003}]{Molnar03}
{Molnar} S.~M.,  {Birkinshaw} M.,  2003, \apj, 586, 731

\bibitem[\protect\citeauthoryear{{Nishizawa}, {Komatsu}, {Yoshida}, {Takahashi}
  \& {Sugiyama}}{{Nishizawa} et~al.}{2008}]{Nishizawa08}
{Nishizawa} A.~J.,  {Komatsu} E.,  {Yoshida} N.,  {Takahashi} R.,    {Sugiyama}
  N.,  2008, \apjl, 676, L93

\bibitem[\protect\citeauthoryear{{Page}, {Barnes}, {Hinshaw}, {Spergel},
  {Weiland}, {Wollack}, {Bennett}, {Halpern}, {Jarosik}, {Kogut}, {Limon},
  {Meyer}, {Tucker} \& {Wright}}{{Page} et~al.}{2003}]{Page03}
{Page} L.,  {Barnes} C.,  {Hinshaw} G.,  {Spergel} D.~N.,  {Weiland} J.~L.,
  {Wollack} E.,  {Bennett} C.~L.,  {Halpern} M.,  {Jarosik} N.,  {Kogut} A.,
  {Limon} M.,  {Meyer} S.~S.,  {Tucker} G.~S.,    {Wright} E.~L.,  2003, \apjs,
  148, 39

\bibitem[\protect\citeauthoryear{{Panek}}{{Panek}}{1992}]{Panek92}
{Panek} M.,  1992, \apj, 388, 225

\bibitem[\protect\citeauthoryear{{Puchades}, {Fullana}, {Arnau} \&
  {S{\'a}ez}}{{Puchades} et~al.}{2006}]{Puchades06}
{Puchades} N.,  {Fullana} M.~J.,  {Arnau} J.~V.,    {S{\'a}ez} D.,  2006,
  \mnras, 370, 1849

\bibitem[\protect\citeauthoryear{{Quilis}, {Ibanez} \& {Saez}}{{Quilis}
  et~al.}{1995}]{Quilis95}
{Quilis} V.,  {Ibanez} J.~M.,    {Saez} D.,  1995, \mnras, 277, 445

\bibitem[\protect\citeauthoryear{{Quilis} \& {Saez}}{{Quilis} \&
  {Saez}}{1998}]{Quilis98}
{Quilis} V.,  {Saez} D.,  1998, \mnras, 293, 306

\bibitem[\protect\citeauthoryear{{Rees} \& {Sciama}}{{Rees} \&
  {Sciama}}{1968}]{Rees68}
{Rees} M.~J.,  {Sciama} D.~W.,  1968, \nat, 217, 511

\bibitem[\protect\citeauthoryear{{Rubi{\~n}o-Mart{\'{\i}}n},
  {Hern{\'a}ndez-Monteagudo} \& {En{\ss}lin}}{{Rubi{\~n}o-Mart{\'{\i}}n}
  et~al.}{2004}]{Rubino-Martin04}
{Rubi{\~n}o-Mart{\'{\i}}n} J.~A.,  {Hern{\'a}ndez-Monteagudo} C.,
  {En{\ss}lin} T.~A.,  2004, \aap, 419, 439

\bibitem[\protect\citeauthoryear{{Rudnick}, {Brown} \& {Williams}}{{Rudnick}
  et~al.}{2007}]{Rudnick07}
{Rudnick} L.,  {Brown} S.,    {Williams} L.~R.,  2007, \apj, 671, 40

\bibitem[\protect\citeauthoryear{{Sachs} \& {Wolfe}}{{Sachs} \&
  {Wolfe}}{1967}]{Sachs67}
{Sachs} R.~K.,  {Wolfe} A.~M.,  1967, \apj, 147, 73

\bibitem[\protect\citeauthoryear{{Saez}, {Arnau} \& {Fullana}}{{Saez}
  et~al.}{1993}]{Saez93}
{Saez} D.,  {Arnau} J.~V.,    {Fullana} M.~J.,  1993, \mnras, 263, 681

\bibitem[\protect\citeauthoryear{{Sanchez}, {Crocce}, {Cabre}, {Baugh} \&
  {Gaztanaga}}{{Sanchez} et~al.}{2009}]{Sanchez09}
{Sanchez} A.~G.,  {Crocce} M.,  {Cabre} A.,  {Baugh} C.~M.,    {Gaztanaga} E.,
  2009, ArXiv e-prints

\bibitem[\protect\citeauthoryear{{Smith}, {Hern{\'a}ndez-Monteagudo} \&
  {Seljak}}{{Smith} et~al.}{2009}]{Smith09}
{Smith} R.~E.,  {Hern{\'a}ndez-Monteagudo} C.,    {Seljak} U.,  2009, \prd, 80,
  063528

\bibitem[\protect\citeauthoryear{{Springel}, {White}, {Jenkins} \& et
  al.}{{Springel} et~al.}{2005}]{Springel05}
{Springel} V.,  {White} S.~D.~M.,  {Jenkins} A.,    et al. 2005, \nat, 435, 629

\bibitem[\protect\citeauthoryear{{Sunyaev} \& {Zeldovich}}{{Sunyaev} \&
  {Zeldovich}}{1972}]{Sunyaev72}
{Sunyaev} R.~A.,  {Zeldovich} Y.~B.,  1972, Comments on Astrophysics and Space
  Physics, 4, 173

\bibitem[\protect\citeauthoryear{{Tomita} \& {Inoue}}{{Tomita} \&
  {Inoue}}{2008}]{Tomita08}
{Tomita} K.,  {Inoue} K.~T.,  2008, \prd, 77, 103522

\bibitem[\protect\citeauthoryear{{Tuluie} \& {Laguna}}{{Tuluie} \&
  {Laguna}}{1995}]{Tuluie95}
{Tuluie} R.,  {Laguna} P.,  1995, \apjl, 445, L73

\bibitem[\protect\citeauthoryear{{Tuluie}, {Laguna} \& {Anninos}}{{Tuluie}
  et~al.}{1996}]{Tuluie96}
{Tuluie} R.,  {Laguna} P.,    {Anninos} P.,  1996, \apj, 463, 15

\bibitem[\protect\citeauthoryear{{Verde} \& {Spergel}}{{Verde} \&
  {Spergel}}{2002}]{Verde02}
{Verde} L.,  {Spergel} D.~N.,  2002, \prd, 65, 043007

\bibitem[\protect\citeauthoryear{{Vielva}, {Mart{\'{\i}}nez-Gonz{\'a}lez},
  {Barreiro}, {Sanz} \& {Cay{\'o}n}}{{Vielva} et~al.}{2004}]{Vielva04}
{Vielva} P.,  {Mart{\'{\i}}nez-Gonz{\'a}lez} E.,  {Barreiro} R.~B.,  {Sanz}
  J.~L.,    {Cay{\'o}n} L.,  2004, \apj, 609, 22

\bibitem[\protect\citeauthoryear{{Zhang} \& {Huterer}}{{Zhang} \&
  {Huterer}}{2009}]{Zhang09}
{Zhang} R.,  {Huterer} D.,  2009, ArXiv e-prints

\end{thebibliography}
\bibliographystyle{mn2e}
\appendix
\end{document}